\documentclass[12pt,a4]{article}

\renewcommand{\baselinestretch}{1.22}
\usepackage{amsthm,amsmath,latexsym,amssymb,amsfonts,amscd}
\usepackage{graphics,lscape,fancyhdr,array,stmaryrd,euscript}
\usepackage{empheq}
\usepackage{dsfont}
\usepackage{verbatim}
\usepackage{color,tikz,tikz-cd}
\usetikzlibrary[snakes]
\usepackage{relsize,slashed}
\numberwithin{equation}{section}
\usepackage{hyperref}
\newcommand{\pl}{\partial}

\usepackage[all]{xy}

\newcolumntype{w}[1]{%
>{\raggedright\hspace{0pt}}m{#1}}%
\newcolumntype{z}[1]{%
>{\raggedleft\hspace{0pt}}m{#1}}%

\usepackage{hyperref}

\newtheorem{theorem}{Theorem}[section]

\usepackage[numbers,sort&compress]{natbib}
\usepackage[nottoc]{tocbibind}

\newcommand{\aA}{{\ensuremath{\mathcal{A}}}}
\newcommand{\aB}{{\ensuremath{\mathcal{B}}}}

\newcommand{\ga}{\alpha}
\newcommand{\gb}{\beta}
\newcommand{\gc}{\gamma}

\newcommand{\gad}{{\dot{\alpha}}}
\newcommand{\gbd}{{\dot{\beta}}}

\newcommand{\bry}{{{\bar{y}}}}

\newcommand{\brk}{{{\bar{k}}}}

\newcommand{\fud}[2]{{}^{#1}{}_{#2}\,}
\newcommand{\fdu}[2]{{}_{#1}{}^{#2}\,}

\newcommand{\besubeqs}{\begin{subequations}}
\newcommand{\esubeqs}{\end{subequations}}

\newcommand{\momega}{{\boldsymbol{\omega}}}
\newcommand{\mxi}{{\boldsymbol{\xi}}}
\newcommand{\mC}{{\boldsymbol{C}}}

\newcommand{\mg}{{\boldsymbol{g}}}

\newcommand{\hs}{{\ensuremath{\mathfrak{hs}}}}

\newcommand{\Tr}{\mathrm{Tr}}
\newcommand{\Str}{\mathrm{Str}}
\newtheorem{proposition}[theorem]{Proposition}
\begin{document}
\pagenumbering{gobble}
\hfill
\begin{flushright}
    {}
\end{flushright}
\vskip 0.01\textheight
\begin{center}
{\Large\bfseries 
Characteristic Cohomology and Observables \\

\vspace{0.3cm}
in Higher Spin Gravity 
\vspace{0.4cm}}

\vskip 0.03\textheight

Alexey \textsc{Sharapov}${}^{1}$ \&  Evgeny \textsc{Skvortsov}${}^{2,3}$

\vskip 0.03\textheight

{\em ${}^{1}$Physics Faculty, Tomsk State University, \\Lenin ave. 36, Tomsk 634050, Russia}\\
\vspace*{5pt}
{\em ${}^{2}$ Albert Einstein Institute, \\
Am M\"{u}hlenberg 1, D-14476, Potsdam-Golm, Germany}\\
\vspace*{5pt}
{\em ${}^{3}$ Lebedev Institute of Physics, \\
Leninsky ave. 53, 119991 Moscow, Russia}\\

\end{center}

\vskip 0.02\textheight

\begin{abstract}
We give a complete classification of dynamical invariants in $3d$ and $4d$ Higher Spin Gravity models, with some comments on arbitrary $d$. These include holographic correlation functions, interaction vertices, on-shell actions, conserved currents, surface charges, and some others. Surprisingly, there are a good many conserved $p$-form currents with various $p$. The last fact, being  in tension with `no nontrivial conserved currents in quantum gravity' and similar statements, gives an indication of hidden integrability of the models. Our results rely on a systematic computation of Hochschild, cyclic, and Chevalley--Eilenberg cohomology for the corresponding  higher spin algebras. A new invariant in Chern--Simons theory with the Weyl algebra as gauge algebra is also presented. 
\end{abstract}
\newpage
\renewcommand{\baselinestretch}{0.8}\normalsize
\tableofcontents
\renewcommand{\baselinestretch}{1.22}\normalsize
\newpage
\section{Introduction}
\pagenumbering{arabic}
\setcounter{page}{2}
The idea of Higher Spin Gravities (HSGRA) is to construct viable models of Quantum Gravity by looking for extensions of classical gravity with massless higher spin fields. Conceptually, `masslessness' is important for modelling the high energy behaviour of massive fields. Technically, it also plays a crucial role in constraining interactions by the gauge symmetries associated with massless spinning fields. An extension of gravity by higher spin states is one of the key features of string theory and it also seems to be unavoidable in AdS/CFT scenarios since any CFT${}^d$ in $d\geq3$ has single-trace operators of any spin. Higher spin symmetry is supposed to leave no room for relevant counterterms and thus constructing a classical HSGRA is almost sufficient for having a quantum theory. However, there are many no-go-type results that prevent this simple scenario from happening, leaving us at present with only a handful of HSGRA's that avoid them.  

Let us mention some Higher Spin Gravities wherein various observables and quantum corrections can be studied. (Partially-)massless \cite{Blencowe:1988gj,Bergshoeff:1989ns,Campoleoni:2010zq,Henneaux:2010xg,Grigoriev:2020lzu} and conformal \cite{Pope:1989vj,Fradkin:1989xt,Grigoriev:2019xmp} higher spin gravities in $3d$ have to have the Chern--Simons form \cite{Grigoriev:2020lzu}, which facilitates the study thereof, but there are no local degrees of freedom. There is a higher spin extension of the $4d$ conformal gravity \cite{Segal:2002gd,Tseytlin:2002gz,Bekaert:2010ky}. The $4d$ chiral theory \cite{Metsaev:1991mt,Metsaev:1991nb,Ponomarev:2016lrm,Ponomarev:2017nrr} is the smallest higher spin theory with massless propagating fields; it is shown to be one-loop finite \cite{Skvortsov:2018jea,Skvortsov:2020wtf,Skvortsov:2020gpn}. In this paper, we address a class of as  yet hypothetical $4d$ HSGRA's that should be AdS/CFT dual of free/critical vector models \cite{Sezgin:2002rt,Klebanov:2002ja,Sezgin:2003pt,Leigh:2003gk} and, generally, of Chern--Simons Matter theories \cite{Giombi:2011kc}.

As long as the dynamics of higher spin fields are supposed to be entirely fixed by the higher spin symmetry, many important questions can be approached from the symmetry point of view even before a complete theory is constructed (symmetry, if powerful enough, can even eliminate the need for a theory). In the present paper we classify  invariant and covariant functionals of fields in $4d$ HSGRA, $3d$ HSGRA and with some remarks on HSGRA's in higher dimensions as well. The functionals are constructed to the lowest nontrivial order in fields. Various obstructions to extend them to higher orders are also given and discussed. 

The holographic HSGRA's, we are interested in, are known to exhibit a certain degree of nonlocality \cite{Bekaert:2015tva,Maldacena:2015iua,Sleight:2017pcz} that goes beyond the usual field theory approach, but have a perfectly local closed subsector represented by the chiral theory in $4d$.  Indeed, free and weakly-coupled CFT's, like free and large-$N$ critical vector models, do not have a large gap in the dimensions of single-trace operators. Therefore, the holographic duals face infinitely many fields from the onset and infinitely many derivatives starting from the quartic order \cite{Bekaert:2015tva}. This is not yet a concern for the paper since the observables we classify are consistent with the equations of motion to the lowest order, where such nonlocalities do not arise.

Of interest are either unintegrated observables that are strictly  gauge-invariant or observables that are gauge-invariant up to an exact differential (hence, one can  integrate the these latter over appropriate cycles to produce genuine gauge invariants).  On-shell closed forms (conserved currents) represent another class of dynamical invariants. In fact the last two classes coincide \cite{Barnich:2009jy}. As long as covariant, rather than just invariant, functionals go, cocycles with values in the higher spin algebra determine the interaction ambiguity in the higher spin equations themselves or represent possible deformations of extensions thereof with higher-degree forms, see e.g.  \cite{Boulanger:2011dd,Bonezzi:2016ttk}. They can also be contracted with global symmetry parameters to give conserved $p$-form currents. All these possibilities are studied in detail. The spectrum of $p$-form currents indicates that HSGRA's can give examples of higher form symmetries \cite{Gaiotto:2014kfa}.

What is surprising is that there exist a great many higher spin invariant and covariant functionals, as will be seen from our classification. This should be confronted with the lower spin gauge theories, e.g. Yang--Mills theory or Gravity, and more generally, with the Weinberg--Witten theorem \cite{Weinberg:1980kq}. This also contradicts a naive expectation that a bigger symmetry should be more restrictive. The study of invariants of higher spin symmetry and of observables in HSGRA was pioneered in \cite{Sezgin:2011hq}, where several different classes of invariants/observables were discovered, with an important prelude in \cite{Sezgin:2005pv}. See also \cite{Colombo:2010fu,Colombo:2012jx,VASILIEV2017219} for closely related results and alternative approaches. As different from \cite{Sezgin:2011hq}, we ignore all global aspects of the problem, non-local invariants like Wilson loops and off-shell invariants; instead, we give a complete classification of local invariants. 

More technically, the problem of classification of covariant and invariant functionals can be reduced to that of computing Chevalley--Eilenberg cohomology. The latter can further be related to the cyclic and, eventually, to the Hochschild cohomology of a given higher spin algebra: 
\begin{align}\notag
    & \text{Chevalley--Eilenberg} && \Longrightarrow && \text{Cyclic} && \Longrightarrow && \text{Hochschild}
\end{align}
The relationship owes its existence to a very special form of the Lie algebra associated to a given (associative) higher spin algebra. It is the Lie algebra of big matrices with elements belonging to the higher spin algebra. The matrix extension of higher spin symmetries is well-motivated by the AdS/CFT correspondence: one can always add global symmetries to the free (and critical) vector models, i.e., to the CFT duals of HSGRA's. 

The free higher spin fields on a maximally-symmetric backgrounds can be described by a simple system of equations 
\begin{align}
    d\momega&=\momega\star \momega+\mathcal{O}(\mC)\,, &
    d\mC&= \momega\star \mC-\mC\star \momega+\mathcal{O}(\mC^2)
\end{align}
for a one-form $\momega$ and a zero-form $\mC$ both taking values in a higher spin algebra \cite{Vasiliev:1988sa}, upon neglecting the higher-order corrections in $\mC$. Here, $\star$ is the product in a properly extended higher spin algebra. As is seen, the zeroth-order approximation is given by a flat connection $\momega$ associated with a higher spin algebra. The free fields are then described by a covariantly constant zero-form $\mC$. 

Therefore, the problem we solve is to classify all nontrivial multi-linear  forms
\begin{align}
    J(\underbrace{\momega,\ldots ,\momega}_m,\underbrace{\mC,\ldots,\mC}_n)
\end{align}
of $\momega$ and $\mC$ that are on-shell closed or gauge-invariant up to an exact differential. A part of our results on $4d$ HSGRA can be summarized by the following table with derivation and explicit expressions left to the main text. 
\begin{table}[h]
    \centering
\begin{tabular}{c|c|c|c|c|c|c|c|c|}
    & $\mC^0$ & $\mC^1$ &$\mC^2$ &$\mC^3$ &$\mC^4$ &$\mC^5$ &$\mC^6$ &$\mC^7$  \tabularnewline\hline
    $\momega^0$ & $\empty$ & $1$ & $1$& $1$& $1$& $1$& $1$& $1$\tabularnewline 
    $\momega^1$ & $1$& $0$& $0$& $0$& $0$& $0$& $0$& $0$\tabularnewline 
    $\momega^2$ & $0$ & $2$ & $2$ & $2$ & $2$ & $2$ & $2$ & $2$\tabularnewline 
    $\momega^3$ & $3$ & $0$& $0$& $0$& $0$& $0$& $0$& $0$\tabularnewline 
    $\momega^4$ & $0$ &  $1$ & $1$& $1$& $1$& $1$& $1$& $1$ \tabularnewline 
    $\momega^5$ & $4$ &  $0$ & $0$& $0$& $0$& $0$& $0$& $0$ \tabularnewline 
    $\momega^6$ & $0$ &  $0$ & $0$& $0$& $0$& $0$& $0$& $0$ \tabularnewline 
\end{tabular}
    \caption{Observables in $4d$ HSGRA to the lowest order (both in the physical and unphysical sectors). The number in each cell is the number of independent observables that are of $\momega^m \mC^n$-type. The table extends trivially in both directions; here, we kept a part that shows some irregularity (for higher $m$ and $n$ everything stabilizes). The multiplicities along the diagonal $m+n=\text{const}$ add up to $2,1,6,3,8,4,8,4,\ldots$.  }
    \label{tab:my_label}
\end{table}

We also make some comments on HSGRA's in $d>4$ and in $3d$, even though the latter are somewhat special in not having propagating degrees of freedom in the higher spin sector. One interesting finding in $3d$ is a new invariant of Chern--Simons theory that is based on the Weyl algebra. This type of invariant does not exist for $SU(N)$, for example. 

The outline of the paper is as follows. We begin in Section \ref{sec:formalequations} with the concept of formal equations of motion that any field-theoretic system can be fit in. In Section \ref{sec:basics}, we briefly discuss the problem of HSGRA in this language. In Section \ref{sec:observables}, we present and discuss our classification results, the mathematical details being left to the Appendices \ref{app:A}, \ref{app:B}.

\section{FDA approach to classical field theory} 
\label{sec:formalequations}
In this paper, we will study field theories formulated in terms of Free Differential Algebras.
These algebras became popular after the seminal paper by  D. Sullivan \cite{Sullivan77}; in physics, they were introduced in \cite{DAURIA1982101,vanNieuwenhuizen:1982zf}. Closely related topics include $Q$-manifolds, $L_\infty$-algebras and AKSZ-models \cite{Alexandrov:1995kv}.

\subsection{\texorpdfstring{FDA, ${L_\infty}$, ${Q}$-manifolds and all that}{FDA, L-infinity, Q-manifolds and all that}} Recall  that a Differential Graded Algebra $(A,\delta)$ is a graded $k$-vector space $A=\bigoplus A_n$ endowed with an associative (dot) product and a differential $\delta$ such that 
\begin{equation}
    A_n\cdot A_m\subset A_{n+m}\,,\qquad \delta A_n\subset A_{n+1}\,,\quad\mbox{and}\quad \delta ^2=0\,.
\end{equation}
The two operations are related by the Leibniz rule 
\begin{equation}
    \delta (a\cdot b)=\delta a\cdot b+(-1)^{|a|}a\cdot \delta b\,,
\end{equation}
where $|a|=n$ stands for the degree of a homogeneous element $a\in A_n$. The cohomology of $\delta$ is denoted by $H(A,\delta)=\bigoplus H^p(A,\delta)$. 

Below we are interested in DGA's that are free, commutative, and non-negatively graded.  This implies the 
existence of a finite (or countable) set  of generators $w^A\in A$ obeying no relations other than  graded commutativity:
\begin{equation}
    w^Aw^B=(-1)^{|A||B|}w^Bw^A\,, \qquad\qquad |A|=|w^A|\geq 0\,.
\end{equation}
 The general element of $A$ is given then by a polynomial $a(w)$ in $w$'s.  
In physics, the DGA's of this type are usually called Free Differential Algebras (FDA). Given a set of free generators,   the  
structure of FDA is completely determined by the differentials of its generators
\begin{equation}
    \delta w^A=Q^A(w)\,,\qquad\qquad Q^A(w)\in A\,.
\end{equation}
The r.h.s. of these relations can further be expanded into homogeneous polynomials as 
\begin{equation}\label{Ql}
    Q^A=Q^A_1+Q^A_2+Q^A_3+\cdots\,, \qquad\qquad Q_n^A=l_{A_1\cdots A_n}^Aw^{A_1}\cdots w^{A_n}\,,
\end{equation}
for some structure constants $l^A_{A_1\cdots A_n}$. 

By definition,  a FDA $(A,\delta)$ is called {\it linear} if $Q_k^A=0$ for all $k> 1$. If in addition $H(A,\delta)\simeq H^0(A,\delta )\simeq k$, then the algebra is said to be {\it linear contractible}.   An algebra $(A,\delta)$ is called {\it minimal} if $Q^A_1=0$. 

The main theorem on the structure of FDA's states that each algebra is decomposed into the product $A=A_{\mathrm{lc}}\otimes A_{\mathrm{min}}$ of a linear contractible algebra $A_{\mathrm{lc}}$ and a minimal algebra $A_{\mathrm{min}}$, see e.g. \cite[Sec. 4.5.1]{Kontsevich:1997vb}.  This implies the existence of a generating set $w^A=(u^\alpha, v^\alpha, \bar w^a)$ such that 
\begin{equation}\label{uv}
    \delta u^\alpha=v^\alpha\,,\qquad \delta v^\alpha=0\,,\qquad \delta \bar w^a=Q^a_2(\bar w)+Q_3^a(\bar w)+\cdots\,.
\end{equation}
Here $u$'s and $v$'s generate the linear contractible part $A_{\mathrm{lc}}$, while $\bar w$'s are generators for $A_{\mathrm{min}}$. In particular, this implies the isomorphism $H(A,\delta)\simeq H(A_{\mathrm{min}},\delta)$. 

The nilpotency condition $\delta^2=0$ for the differential imposes a sequence of quadratic relations on the structure constants $l$'s. These read
\begin{equation}\label{ll}
   \sum_{n+m=k} l^C_{(A_1\cdots A_n}l^B_{CB_{1}\cdots B_m)}=0\,, \qquad k=1,2,\ldots\,,
\end{equation}
where the round brackets stand for the (graded) symmetrization of the indices enclosed. The concept of { $L_\infty$-algebras} (aka {\it strongly homotopy Lie algebras}) delivers an alternative interpretation of these relations as generalized Jacobi identities for a system of multi-brackets \cite{StasheffLada}. The multi-brackets are naturally defined on the graded vector space  $V=A^\ast[1]$, that is, the space dual to $A$ and with the degree of homogeneous subspaces  shifted by one.\footnote{Upon the shift the even subspaces go odd and vice versa; hence, symmetrization in (\ref{ll}) becomes  anti-symmetrization from the viewpoint of $V$. } In terms of the dual basis $e_A\in V$, $|e_A|=|w^A|+1$, the structure of an $L_\infty$-algebra on $V$ is given by the sequence of multi-brackets 
\begin{equation}\label{linf}
    [e_{A_1}, \ldots, e_{A_k} ]_k=l_{A_1\cdots A_k}^Ae_A\,, \qquad k=1,2,\ldots. 
\end{equation}
It follows from the first identity in (\ref{ll}) that the unary bracket $[-]_1$ defines a differential $d: V_n\rightarrow V_{n+1}$ making $V$ into a cochain complex. Then, the second identity  ($k=2$) tells us that $d$ differentiates the binary bracket $[-,-]_2$ by the Leibniz rule.   For minimal algebras the unary bracket is absent and the first nontrivial relation in (\ref{ll}) identifies $[-,-]_2$ as a graded Lie bracket with the structure constants $l^A_{BC}$.  Hence, with each minimal FDA one can associate a graded Lie algebra\footnote{By a graded Lie algebra we mean that the bracket is graded antisymmetric and not just the fact that the underlying space is graded. More pedantic would be a ``graded Lie superalgebra''. } $\mathcal{L}=\mathcal{L}(A)$ on the dual vector space $V$. 
This graded Lie algebra is the starting point for and the most crucial part of the entire $L_\infty$-algebra structure  dual to $(A,\delta)$. In particular, the next ternary bracket $[-,-,-]_3$ appears to be nothing else but a $3$-cocycle of the Chevalley--Eilenberg cohomology of the graded Lie algebra $\mathcal{L}$ with coefficients in the adjoint representation. We will return to this point in Sec. \ref{HPT}.

There is also a nice geometric approach to the algebraic structures above. It treats the FDA $(A,\delta)$ as the algebra of smooth functions on a graded manifold $N$ `coordinatized' by the generators $w^A\in A$. Upon this interpretation the differential $\delta$ gives rise to an odd vector field $Q$ on $N$, whose coordinate expression  is as follows:
\begin{equation}\label{Q}
    Q=\sum_{n=1}^\infty w^{A_1}\cdots w^{A_n}l_{A_1\cdots A_n}^A\frac{\partial }{\partial w^A}\,.
\end{equation}
Then the whole set of Jacobi identities (\ref{ll}) is compactly encoded by a single equation $[Q,Q]=2Q^2=0$. In mathematics, such an odd vector field $Q$ is called {\it homological} and a pair $(N,Q)$ is referred to as a {\it $Q$-manifold} \cite{schwarz1993}, \cite{Alexandrov:1995kv}. The terminology is justified by the fact that the cohomology of the operator (\ref{Q}) acting in the space of smooth functions $C^\infty(N)=A$ coincides with the cohomology of the initial FDA $(A,\delta)$. 

\subsection{Formal Dynamical Systems}\label{FDS}
In order to formulate a field theory on a space-time manifold $M$ with local coordinates $x^\mu$ one introduces the total space of the `shifted' tangent bundle $T[1]M$, i.e., the tangent bundle of $M$ with tangent space coordinates $\theta^\mu$ assigned degree $1$.  The algebra $C^\infty({U})$ of `smooth functions' on a trivializing coordinate chart ${U}\subset T[1]M$ is then a linear contractible FDA with the generators $(x^\mu, \theta^\mu)$ and the differential  
\begin{equation}
    dx^\mu=\theta^\mu\,,\qquad d\theta^\mu=0\,.
\end{equation}
This FDA is clearly isomorphic to the algebra of exterior differential forms on ${U}\cap M$ upon the 
identification 
\begin{equation}\label{ff}
    f(x,\theta)\quad \Leftrightarrow \quad f(x,dx)\,.
\end{equation}
Geometrically, one can regard $d=\theta^\mu\frac{\partial}{\partial x^\mu}$ as a canonical homolological vector field on $T[1]M$ that mimics the de Rham differential. 

Let $N$ be another $Q$-manifold with coordinates $w^A$ and a homological vector field $Q$.  Given a triple $(M,N,Q)$, we define classical fields to be smooth 
maps 
\begin{equation}\label{W}
W: T[1]M\rightarrow N
\end{equation}
of degree zero from the source $Q$-manifold $T[1]M$ to the target $Q$-manifold $N$. The true field configurations (=maps) are, by definition, those relating the homological vector field on $T[1]M$ to that on $N$; in symbols $W_\ast(d)=Q$. In terms of local coordinates the last condition amounts to the differential equations
\begin{equation}\label{FDA}
    dW^A=Q^A(W)\,,
\end{equation}
where the functions $w^A=W^A(x,\theta)$ provide the coordinate description of the map (\ref{W}). 
With identification (\ref{ff}) the l.h.s. of (\ref{FDA}) is given by the de Rham differential of the form fields $W^A$, while
the r.h.s. involves their exterior products. It is known that any system of PDE's can be reformulated in the form (\ref{FDA}), perhaps,  at the expense of introducing an infinite collection of the forms $W^A$, see e.g. \cite{Barnich:2004cr}. Numerous examples of the field equations (\ref{FDA}) are provided by the AKSZ-models \cite{Alexandrov:1995kv}. 
In the context of HSGRA, bringing field's dynamics into the form (\ref{FDA}) is known as an {\it unfolded representation} \cite{Vasiliev:1988sa,Bekaert:2005vh}. 

Besides apparent general covariance the system (\ref{FDA}) enjoys the gauge symmetry of the form 
\begin{equation}\label{gtr}
    \delta_\varepsilon W^A=d\varepsilon^A +\varepsilon^B\partial_BQ^A \,,
\end{equation}
$\varepsilon^A$ being infinitesimal gauge parameters (the forms of appropriate degrees). Applying the de Rham differential to both sides of (\ref{FDA}) one can also see that the system is formally consistent whenever $Q^2=0$. 

It follows from the decomposition theorem (\ref{uv}) that the contractible pairs of fields associated with the generators of $A_{\mathrm{lc}}$ completely decouple from the system (\ref{FDA}). Moreover, the subsystem $dU^\alpha=V^\alpha$, $dV^\alpha=0$ appears to be  tautological, and hence, dynamically empty: the first equation just introduces the fields $V^\alpha$ as names for the differentials $dU^\alpha$ and the second equation trivially follows from the first. Therefore, without loss in generality, one may always assume the FDA underlying the field equations (\ref{FDA}) to be minimal, in which case
\begin{equation}\label{MFDA}
    dW^A=l^A_{BC}W^B\wedge W^C+l^A_{BCD}W^B\wedge W^C\wedge W^D+\cdots\,.
\end{equation}

A word of caution is needed about  these equations:  Even though any PDE can be brought into the form  (\ref{MFDA}), only specific infinite-dimensional FDA's and only very particular choices of coordinates therein lead to well-defined differential equations. In practice, most of the fields  $W^A$ appear to  be auxiliary, being expressible via derivatives of a small (or even finite) subset of dynamical fields. If the number of auxiliary fields is infinite, they can encode an infinite number of derivatives. As a result, the effective dynamics may happen to be far from those described by genuine PDE's. 
In this paper, we leave aside all subtle  analytical issues related to the field equations (\ref{MFDA}) focusing upon their formal consistency. 

\subsection{\texorpdfstring{$Q$-cohomology vs characteristic cohomology}{Q-cohomology vs. characteristic cohomology}}

Suppose we are given a collection of fields $\phi^i$ living on a space-time manifold $M$ and obeying a set of PDE's  
\begin{equation}\label{T}
T_a(\phi,\partial_\mu\phi, ...)=0\,.\end{equation} 
Apart from the equations of motion in themselves, there are many other quantities of interest (like  actions, Lagrangians, conserved currents,  surface charges etc.) that can be attributed to a given field-theoretic system. All these quantities are defined in terms of differential forms that look like  
\begin{equation}\label{J}
    J=J_{\mu_1\cdots \mu_p}(x,\phi,\partial_\mu\phi, ...)dx^{\mu_1}\wedge \cdots\wedge dx^{\mu_p}\,.
\end{equation}
Here the coefficients of a $p$-form $J$ on $M$ are assumed to be smooth functions of the space-time coordinates $x^\mu$, the fields $\phi^i$ and a finite number of their derivatives. On integrating the form $J$ over a $p$-cycle $\Sigma\subset M$ one gets a local functional 
of fields of the form 
\begin{equation}\label{O}
    \mathcal{O}_\Sigma[\phi]=\int_\Sigma J\,.
\end{equation}
For want of a better term we refer to (\ref{J}) as a {\it current} of degree $p$ ($p$-current for short)\footnote{Conventional currents correspond to forms of degree $\dim M-1$. With a metric tensor one can dualize them to get vector fields of the form $J^\mu(x,\phi,\partial\phi, \ldots)$.} and call the functional (\ref{O}) an {\it observable} associated to the current $J$. In general, the value of $\mathcal{O}_\Sigma[\phi]$ depends both on a particular field configuration $\phi^i$ and a chosen cycle $\Sigma$. This value, however, does not change if we add to  $J$ the differential $dI$ of any other current $I$  of degree $p-1$, since $\partial \Sigma=0$.  Therefore, if we are interested in observables alone, it makes sense  to consider currents (\ref{J}) modulo the equivalence relation 
\begin{equation}
    J\sim J+dI\,.
\end{equation}
A typical example of a current of top degree is a Lagrangian density; the corresponding observable is an action functional. A $p$-current $J$ is said to be conserved if 
\begin{equation}
    dJ\approx 0\,,
\end{equation}
where $\approx$ means ``equal when the equations of motion hold". Clearly, observables (\ref{O}) associated to conserved $p$-currents depend only on homotopy classes  of $p$-cycles $\Sigma$ whenever evaluated on solutions to the field equations (\ref{T}). Such observables are usually called {\it charges}.  
On restricting to the subspace of conserved currents, a stronger equivalence relation can be imposed:  two conserved $p$-currents $J$ and $J'$ are considered equivalent if  
\begin{equation}
    J-J'\approx dI
\end{equation}
for some $(p-1)$-current $I$. Clearly, equivalent currents result in equal charges for all solutions of the field equations. 
The equivalence classes of conserved currents associated to PDE's were intensively studied in mathematics under the name of {\it characteristic cohomology} \cite{Bryant1995}, \cite{Verbovetsky:1997fi}, \cite[Sec. 6.2]{Barnich:2000zw}. 

In the presence of gauge symmetries the above definition of an observable should be modified a little. 
To justify its name the functional (\ref{O}) must be invariant under the gauge transformations $\delta_{\varepsilon}\phi^i$ of fields and this implies that
\begin{equation}\label{deJ}
\delta_\varepsilon J\approx dI
\end{equation}
for some $(p-1)$-current $I$ that depends linearly on the gauge parameters $\varepsilon^\alpha$. Then $\delta_\varepsilon \mathcal{O}_{\Sigma}=0$ for any cycle $\Sigma$. Notice that all conserved currents are automatically gauge invariant in the sense of (\ref{deJ}). Indeed, by definition, $\delta_\varepsilon T_a\approx 0$ and $d\delta_\varepsilon=\delta_\varepsilon d$; whence it follows that 
\begin{equation}
    \delta_\varepsilon dJ=d\delta_\varepsilon J\approx 0\quad \Rightarrow\quad \delta_\varepsilon J\approx dI\,.
\end{equation}
For a more systematic discussion of currents and observables in (non-)Lagrangian gauge theories we refer the reader to \cite{Barnich:2000zw}, \cite{Kaparulin:2011xy}.

When dealing with formal dynamical systems (\ref{FDA}) it seems natural to start with  currents and observables associated with certain elements of the underlying FDA $(A,\delta)$. To any  element $j(w)\in A_p$ or, what is the same, homogeneous function on  the $Q$-manifold $N$, we can assign  the $p$-current $J=j\circ W$:
\begin{equation}\label{Jalg}
    J=\sum_{n}\sum_{|A_1|+\cdots +|A_n|=p} j_{A_1\cdots A_n}W^{A_1}\wedge \cdots \wedge W^{A_n}\,.
\end{equation}
As distinguished from the general expression (\ref{J}) these currents do not involve the partial derivatives of the fields  $W^A$ and for this reason one may call them {\it algebraic}.   Checking the gauge invariance (\ref{gtr}), we find 
\begin{equation}
    \delta_{\varepsilon} J=d\varepsilon^A\partial_{A}J+\varepsilon^A\partial_AQ^B\partial_BJ\approx d(\varepsilon^A\partial_{A}J) +\varepsilon^B\partial_B(Q^A\partial_A J)
\end{equation}
(by abuse of notation we omit the wedge product sign). The current $J$ is seen to be gauge invariant iff the last term vanishes.  
The forms $\varepsilon^A$ being arbitrary, this means $QJ=const$. Since $|QJ|>0$ and there are no constants of positive degree, we are lead to conclude that $QJ=0$. In other words, the current $J$ is gauge invariant\footnote{We would like to stress that for a current `gauge invariance' means gauge-invariance up to 
an exact differential. } iff the corresponding function $j(w)$ on $N$ is $Q$-invariant.  
On the other hand, for any algebraic current $I$
\begin{equation}
    dI\approx QI\,,
\end{equation}
and hence, all $Q$-invariant conserved currents of the form $J=QI$ are trivial and any gauge invariant current $J$ is conserved. Let us summarize the above considerations  as follows:
\begin{itemize}
    \item the equivalence classes of gauge invariant $p$-currents of the form (\ref{Jalg}) are in one-to-one correspondence with the elements of the cohomology group $H^p(A,\delta)$;
    \item all gauge-invariant algebraic currents are conserved and all conserved currents are gauge invariant. 
\end{itemize}
Thus, the gauge invariant currents (\ref{J}) constitute a part of characteristic cohomology. The characteristic cohomology of the PDE's of the form (\ref{FDA}) was studied in \cite{Barnich:2009jy}.  In that paper, the authors argue that, under  mild assumptions, each gauge invariant current 
is equivalent to an algebraic one. This reduces the study of characteristic cohomology and observables to a pure algebraic problem of computing the cohomology groups $H(A,\delta)$.

\subsection{Homological perturbation theory in action}\label{HPT}

Generally computation of cohomology groups $H(A,\delta)$ is a rather nontrivial problem for infinite-dimensional algebras. 
What alleviates the problem significantly for FDA's is the use of the natural filtration on $A$ by the polynomial degree of its  elements. 
At this point it is convenient to switch  to the dual picture considering $A$ as the algebra of functions on some $Q$-manifold $N$.  
The homological vector field associated to a minimal FDA  decomposes then into the sum $Q=Q_2+Q_3+\cdots$ of homogeneous components (\ref{Q}) and the same decomposition for the equation $Q^2=0$ yields the sequence of relations 
\begin{equation}\label{QQ}
    [Q_2,Q_2]=0\,,\qquad [Q_2,Q_3]=0\,,\qquad [Q_3,Q_3]=-2[Q_2,Q_4]\,,\quad \ldots\,,
\end{equation}
which are perfectly equivalent to the generalized Jacobi identities (\ref{ll}) for the $L_\infty$-algebra (\ref{linf}).  As we have mentioned earlier,  the leading term  $Q_2=w^Aw^Bl_{AB}^C{\partial}/{\partial w^C}$ is a homological vector field by itself, the expansion coefficients $l_{AB}^C$ being the structure constants of some graded Lie algebra $\mathcal{L}$. Then the action of $Q_2$ on $C^\infty(N)$ models the  Chevalley--Eilenberg complex computing the cohomology of the graded Lie algebra $\mathcal{L}$ with trivial coefficients. This action extends naturally from scalar functions to arbitrary tensor fields on $N$ through the operation of Lie derivative $L_{Q_2}$. 
The corresponding complex  of tensor fields splits into the direct sum $\mathcal{T}=\bigoplus \mathcal{T}^{p,q}$ of various subcomplexes composed of $(q,p)$-type tensors. 
Moreover, all natural tensor operations (the tensor product, contraction and permutation of indices, commutator of vector fields, etc.) pass through the cohomology, so that we may speak of the 
tensor algebra of $Q_2$-cohomology  $H(\mathcal{L}, \mathcal{T})$. The scalar functions on $N$ give rise to a graded commutative subalgebra $H(\mathcal{L}, \mathcal{T}^{0,0})$ that we are most interested in. 

Second in importance is the $Q_2$-cohomology  with coefficients in vector fields on $N$. The space $H(\mathcal{L},\mathcal{T}^{1,0})$ carries the natural structure of a graded Lie algebra, induced by the commutator of vector fields, along with the structure of $H(\mathcal{L}, \mathcal{T}^{0,0})$-module, cf. Appendix \ref{Lie-op}. Notice that Eqs.(\ref{QQ}) hold true if one rescales 
the homogeneous components  as $Q_n\rightarrow g^{n-2}Q_n$, $g$ being an arbitrary  parameter. 
This results in a family of homological vector fields $Q(g)$, which can be viewed as a one-parameter deformation of the quadratic vector field $Q_2=Q(0)$. In particular, the first-order deformation is determined by a vector field $Q_3$ that must be a  $Q_2$-cocycle as suggested by Eq.(\ref{QQ}). If this cocycle happens  to be trivial, i.e., $Q_3=[Q_2, X]$, then one can `gauge it out' by the formal diffeomorphism $e^{gX}: N\rightarrow N$. 
Thus, all the nontrivial first-order deformations are in one-to-one correspondence with the degree-$1$ elements  of the cohomology group $H^3(\mathcal{L},\mathcal{T}^{1,0})$.\footnote{\label{f5} It is conceivable that  all $Q_3,\ldots, Q_{k-1}$ vanish and the deformation  starts in effect with the $k$-th order. Then the leading term $Q_k$ defines and is defined by an element of $H^k(\mathcal{L}, \mathcal{T}^{1,0})$. Formal deformations with $k>3$ are called on occasion {\it irregular}. }\label{f5}  To study higher-order deformations it is convenient to rewrite Eqs.(\ref{QQ}) in the following form:
\begin{equation}
\begin{array}{c}
    L_{Q_2}Q_n=\Delta_n(Q_3,\ldots, Q_{n-1})\,,\\[3mm] 
   \displaystyle  \Delta_n=-\frac12\sum_{m=3}^{n-1} [Q_m,Q_{n-m+2}]\,,\qquad n=3,4,\ldots \,.
    \end{array}
\end{equation}
These equations can be handled effectively by the
`step-by-step obstruction' method of {\it homological perturbation theory } \cite[Sec.7]{Barnich:2000zw}, \cite{stasheff1997}. Proceeding by induction on $n$, one can see that the vector field $\Delta_n$ is a $Q_2$-cocycle provided all previous equations for  $Q_3,\ldots, Q_{n-1}$ are satisfied.  The existence of the $n$-th order deformation $Q_n$ amounts then to the triviality of the cocycle $\Delta_n$. This allows one to interpret the degree-$2$ elements of the cohomology groups   $H^n(\mathcal{L},\mathcal{T}^{1,0})$, $n>2$, as potential obstructions  to deformation of $Q_2$.  If all these groups are happen to be trivial, then each first-order deformation $Q_3$ extends to all higher orders giving a family $Q(g)$. 

In the context of formal dynamical systems of Sec. \ref{FDS}, one may regard the deformation procedure above as inclusion of a consistent  interaction, with $g$ playing the role of a coupling constant.   The three- and four-dimensional  HSGRA's that we discuss in Sec. \ref{sec:observables}  provide major  examples of such interactions.   

A similar perturbative analysis applies to the cohomology groups $H(\delta, A)$ associated with the algebraic conserved currents (\ref{Jalg}).
Writing $j=j_k+j_{k+1}+\cdots$ for a $j\in A$, we find that the function $j$ is $Q$-invariant  if
\begin{equation}\label{Lj}
    L_{Q_2} j_n =-\sum_{m=3}^{n-k+2} L_{Q_m}j_{n-m+2}\,,\qquad n=k,k+1,\ldots\,.
\end{equation}
Again, the r.h.s. of the $n$-th equation is a $Q_2$-cocycle providing all previous equations are satisfied.
The triviality of this cocycle ensures the existence of a function $j_n$ obeying (\ref{Lj}). The sequence (\ref{Lj}) starts with the equation 
$L_{Q_2}j_k=0$ identifying the leading term $j_k$ as a $Q_2$-cocycle. This may be assumed nontrivial: 
otherwise one could remove it by the equivalence transformation $j\rightarrow j-\delta f$ with $j_k=Q_2 f$, so that Taylor's expansion for $j$ would start actually with $j_{k+1}$. In homological algebra, such a perturbative approach to computation of $Q$-cohomology is known as a {\it spectral sequence technique} \cite[Ch. XI]{MacLane}.  

All in all, much of the  structure of a minimal FDA  $(A,\delta)$ as well as its cohomology groups are controlled by the cohomology of the associated Lie algebra $\mathcal{L}(A)$.

\section{Basics of Higher Spin Gravity}
\label{sec:basics}
We setup the problem of Higher Spin Gravity in the language of formal equations of motion \eqref{FDA}, as was first suggested in \cite{Vasiliev:1988sa}. An immediate benefit is that we can capture  relevant algebraic structures that are hard to see, if at all,  within any standard perturbative approach like Noether procedure. A drawback is that there is a long way to any such standard field theory approach. For instance,  the problem of extracting proper interaction vertices appears to be highly nontrivial and may require further structures that are not yet fully understood. Fortunately, the problem starts to be effective at the second order and does not affect any result of the present paper. 

According to \cite{Vasiliev:1988sa} the formal equations of motion for a HSGRA  are formulated in terms of a one-form field $\omega$ and a zero-form field $C$, both taking values in a higher spin algebra $\hs$, i.e., $W^A=(\omega, C)$. At this point, $\hs$ may be any associative algebra. By the form-degree counting the equations should have the following general form:
\besubeqs\label{problem}
\begin{align}\label{problemAA}
d\omega&=\omega\star \omega + \mathcal{V}_3(\omega,\omega,C)+\mathcal{V}_4(\omega,\omega,C,C)+\mathcal{O}(C^3)=Q^\omega(\omega,\omega|C)\,,\\ \label{problemAB}
d C&=\omega\star C-C\star \pi(\omega)+\mathcal{V}_3(\omega,C,C)+\mathcal{O}(C^3)=Q^C(\omega|C)\,.
\end{align}
\esubeqs
Here, $\star$ combines the wedge product of forms with the associative product in $\hs$ and $\pi$ is a certain automorphism of $\hs$.\footnote{In a more general context, where $\hs$ is just an associative algebra, the automorphism represents a natural option to change the adjoint action of $\hs$ on itself (assuming that the field content is already fixed to be $\omega$ and $C$ valued in $\hs$). It does not have to be nontrivial, in general, but it is so for the higher spin problem.  } Thus, the bilinear vertices are completely determined by the pair $(\hs,\pi)$, which is a starting point for the deformation problem. As was explained in the previous section, a formal consistency of the field equations (\ref{problem}) implies the r.h.s. to be given by some homological vector field $Q=(Q^\omega,Q^C)$ on the target space of fields $(\omega, C)$. {\it A central  problem of formal HSGRA is to construct interaction vertices $\mathcal{V}_k$ for a given $(\hs,\pi)$ respecting formal consistency. }

The first particular solution of this problem was given in \cite{Vasiliev:1990en}. The general solution applicable to any higher spin (or arbitrary associative) algebra was proposed in \cite{Sharapov:2019vyd}. There are a number of such and very similar models in the literature \cite{Vasiliev:1990en,Vasiliev:2003ev,Bekaert:2013zya,Neiman:2015wma,Bonezzi:2016ttk,Bekaert:2017bpy,Arias:2017bvi,Grigoriev:2018wrx,Sharapov:2019vyd,Sharapov:2019pdu,Alkalaev:2019xuv}. 

\paragraph{Quasi-holographic interlude. } We refer to \cite{Bekaert:2005vh,Didenko:2014dwa,Sharapov:2017yde} for clarifications on the general structure of Eqs. (\ref{problem}). Let us make just a few AdS${}_{d+1}$/CFT${}^d$-inspired comments that hopefully explain the origin of the basic ingredients. 

The first and foremost fact about general HSGRA's  is that the graded Lie algebra $\mathcal{L}$ governing the quadratic part of equations (\ref{problem}) always comes from an associative algebra $\hs$ of infinite dimension in $d\geq3$.  The easiest way to perceive this fact is through the perspective of AdS/CFT correspondence. All HSGRA's are thought to be holographically dual to free CFT's \cite{Sundborg:2000wp,Sezgin:2002rt, Klebanov:2002ja,Sezgin:2003pt,Leigh:2003gk,Beisert:2004di} for an appropriate choice of boundary conditions. The simplest of such CFT's are free vector models, i.e., free fields valued in vector representations of some weakly-gauged symmetry group.\footnote{An ambiguity in the choice of boundary conditions for fields in $AdS_{d+1}$ -- Neumann vs. Dirichlet -- allows for more interesting dualities, e.g. the large-$N$ critical vector models \cite{Klebanov:2002ja,Sezgin:2003pt}.} Free CFT's, e.g. free scalar theory $\square\phi=0$, always feature an infinite-dimensional algebra of symmetries. Here, by symmetries we understand \cite{Eastwood:2002su} just differential operators $S=S(x,\pl)$ that map any solution  to a solution again.\footnote{This is equivalent to $E S=S' E$, where $E$ is the l.h.s. of the equations of motion ($E=\square$ in the example above) and $S'$ is some other operator that pops up when $S$ is pushed through the equations of motion. Indeed, $ES\phi=S'E\phi=0$ whenever $E\phi=0$. Two symmetries $S_1$ and $S_2$ are equivalent if $S_1=S_2+RE$ for some differential operator $R$. The symmetries of the form $RE$ annihilate any solution and for this reason they   are called trivial. } In particular, the generators of conformal symmetries $so(d,2)$ are realized by $S$ that are operators of the first order at most. The  composition of differential operators endows the symmetries with the structure of an associative algebra. Moreover, the symmetry algebra most likely to be infinite-dimensional.  Indeed, each symmetry operator $S$
gives rise to an infinite sequence of higher-order symmetries $S^n$, most of which may happen to be nontrivial.  In this respect linear theories differ greatly from nonlinear ones: in the latter case, the symmetries form a `weaker' structure of a Lie algebra, which is normally finite-dimensional.

The same symmetries can also be understood as the usual Noether symmetries associated with conserved tensors of schematic form\footnote{In the AdS/CFT context one needs to take $\phi$ to be vectors of some global symmetry group, e.g. $O(N)$ or $U(N)$, and take $N$ large. We will ignore this additional complication in a heuristic explanation below. }
\begin{align}\label{Jff}
    J_{i_1\cdots i_s}&= \phi \pl_{i_1}\cdots \pl_{i_s}\phi+\cdots\,. 
\end{align}
From the AdS/CFT point of view $J_{i_1\cdots i_s}$, together with the degenerate member of this family $J=\phi^2$, are single-trace operators. According to the standard AdS/CFT lore, they should be dual to the fields of the AdS/CFT-dual gravitational theory. Since the tensors are conserved, the fields have to be gauge fields, and hence,  massless.

As with any associative algebra, we may consider the Lie algebra $L(\hs)$ generated by commutators in $\hs$. In the AdS/CFT context, this Lie algebra always contains the conformal algebra $so(d,2)$ as a finite-dimensional subalgebra.  The existence of an associative structure on symmetries allows one to define the higher spin algebra as the quotient of the universal enveloping algebra of $so(d,2)$ by a certain two-sided ideal \cite{Eastwood:2002su}. Denoting the universal enveloping by $\mathcal{U}$, we can write  $\hs=\mathcal{U}/\mathcal{J}$. The ideal $\mathcal{J}$, called the Joseph ideal,\footnote{The relation to the Joseph ideal was clarified e.g. in \cite{Fernando:2009fq,Joung:2014qya}. The term higher spin algebra was introduced in \cite{Fradkin:1986ka}.} is known to be primitive, meaning the existence of a representation $\rho: \mathcal{U}\rightarrow \mathrm{End}(V)$ such that 
$\ker \rho= \mathcal{J}$. The representation $\rho$  is given by a  Verma module $V=\mathrm{span}\{|\phi\rangle\}$ for the Lie algebra $so(d,2)$.  The corresponding lowest weight vector $|\Delta, \mathbf{s}\rangle$ is specified by a conformal weight $\Delta$ -- the eigenvalue of the dilatation operator $D$ -- and the weight ${\bf s}=(s_1,\ldots, s_{[d/2]})$ of the Lorentz subalgebra $so(d-1,1)$ generated by $L_{ij}$. By definition, $|\Delta, {\mathbf{s}}\rangle$ is annihilated by the conformal boosts $K_i$, so that each vector $|\phi\rangle\in V$ is obtained from  $|\Delta,\mathbf{s}\rangle$ by applying translations $P_i$.\footnote{Let $T_{AB}=-T_{BA}$ the standard basis for $so(d,2)$, $A,B=0,\ldots ,d+1$. In the conformal basis, we split the indices as $A=i,+,-$ and use the light-cone form of the metric: $X^AX^B\eta_{AB}=2X^+X^- +X^iX^j\eta_{ij}$. Then, $L_{ij}=T_{ij}$, $D=-T_{d,d+1}$, $P_i=T_{i,d+1}-T_{i,d}$, $K_i=T_{i,d+1}+T_{i,d}$.} For the free scalar CFT above, $\Delta=(d-2)/2$ and ${\mathbf{s}}=0$, i.e., $L_{ij}|\Delta, {\mathbf{0}}\rangle =0$.

The higher spin algebra $\hs$ associated with the Verma module $V$ is just the algebra of all (local) operators $\rho(\mathcal{U})$ that act on $V$. In particular, the conformal algebra generators belong to $\hs$. As usual, we can represent (some of) operators from  $\mathrm{End}(V)$ in the form $|\phi\rangle\langle \phi'|\in V\otimes V^\ast$. The single-trace operators, being bilinear in $\phi$, belong to the tensor product $V\otimes V=\mathrm{span}\{|\phi\rangle |\phi'\rangle\}$. Normalizable on-shell bulk fields form, basically, the same representations as the currents (\ref{Jff}).\footnote{See e.g. \cite{Gunaydin:1998jc,Iazeolla:2008ix} for a careful treatment of the real forms and basis adapted to the $so(d,2)$ representation theory at the boundary and in the bulk.}

Since the higher spin symmetry $\hs$ is a global symmetry on the CFT side, it has to be gauged in the bulk. To this end, one should begin with a one-form connection $\omega$ taking values in $\hs$, that is, in $\mathrm{End}(V)$. The flatness condition \eqref{problemAA} describes then the maximally symmetric backgrounds with $\hs$ symmetry. The bulk degrees of freedom can be described by taking a zero-form $C$ with values in $V\otimes V$ and requiring it to be covariantly constant w.r.t. $\omega$, as in \eqref{problemAB}. One small tweak in \eqref{problemAB} is 
that the $\hs$-modules  $V\otimes V$, $V\otimes V^\ast$, and $\mathrm{End}(V)$ are formally isomorphic to each other.\footnote{Of course,  care is required in dealing with such tensor products, see e.g. \cite{Iazeolla:2008ix,Basile:2018dzi} for some subtleties.} Therefore, one can choose $C$ to assume values in the algebra $\hs$ itself and account for the difference between $|\phi\rangle$ and $\langle\phi|$ with the help of a certain automorphism $\pi$. In the conformal basis, $\pi$ is related to the inversion: $\pi(D)=-D$, $\pi(K_i)=P_i$, $\pi(P_i)=K_i$, and $ \pi(L_{ij})=L_{ij}$. In the $AdS_{d+1}$ setup $\pi$ acts as $\pi(P_\aA)=-P_\aA$, $\pi(L_{\aA\aB})=L_{\aA\aB}$, \cite{Vasiliev:1988sa,Bekaert:2005vh,Didenko:2014dwa}.\footnote{In order to relate the two basis we choose $\eta_{AB}=(-,+,\cdots,+,-)$. In the $AdS_{d+1}$ basis we split $A$ as $A=\{\aA,5\}$, etc., where $\aA,\aB=0,\ldots ,d$ are the indices of the AdS-Lorentz algebra $so(d,1)$ and $5$ is an extra dimension, $\eta^{55}=-1$. Then, the Lorentz and translation generators are $L^{\aA\aB}=T^{\aA\aB}$, $P^\aA=T^{\aA5}$, so that $[P^\aA,P^\aB]=L^{\aA\aB}$. } 

Though not rigorous enough,  the above consideration should hopefully explain, at least heuristically, why both $\omega$ and $C$ are needed, why either takes values in $\hs$, and what is the role  of the automorphism $\pi$ in  \eqref{problemAB}. Now, the problem is to look for nonlinear deformations of \eqref{problem} and to relate them with $\hs$ and $\pi$. $\blacksquare$

We note that one can construct certain higher-spin Lie (super)algebras with the help of (anti-)automorphisms, see e.g. \cite{Konstein:1989ij}. These algebras are genuine Lie (super)algebras, where the Lie bracket does not result from the commutator in an associative algebra. Nevertheless, they are still obtained as truncations of certain associative algebras. In these cases, the relation is similar to that between $gl_N$ and $o(N)$, the latter resulting from truncation $\tau(a)=-a$ of the former, where $\tau(a)=a^T$ is an anti-automorphism. Therefore, one always begins with an associative algebra $\hs$. At the end, possible truncations of the nonlinear equations \eqref{problem} can be studied, if needed.
 
In accordance with Sec. \ref{HPT} all relevant quantities, e.g. vertices and invariants, should correspond to the Chevalley--Eilenberg cohomology of the Lie algebra associated with  the higher spin algebra. Nevertheless, the fact that the Lie algebra comes  from an associative algebra is very helpful in computing the cohomology of the former. 

As a side remark, let us mention that the problem of completing \eqref{problem} with interaction vertices makes sense for any associative algebra, with $\pi$ being e.g. the identity automorphism. This leads to a new class of integrable models as the classical equations of motion \eqref{problem} can be solved explicitly via an auxiliary Lax pair \cite{Sharapov:2019vyd}.  

\subsection{Higher Spin Cohomology}
\label{sec:HScohomology}
Let us explain in a few words, avoiding  technical details, how the problem of classifying all relevant deformations and observables can be approached. The initial data are given by a higher spin algebra $\hs$, which is always an associative algebra, and by an automorphism $\pi$. These data allow us to write down  the free equations of motion\footnote{It is worth stressing, and this is one of the crucial advantages of our approach, that the equations are not free in the usual sense. They describe propagation of free fields $C$ in the background given by a flat connection $\omega$ of a higher spin algebra. In the simplest case when only the spin-two subsector of $\omega$ is turned on, the equations reduce to the usual Bargmann--Wigner type equations on the Maxwell tensor, Weyl tensor, and higher spin generalizations thereof (usually called higher-spin Weyl tensors) as well as to the Klein--Gordon equation for the scalar field. A generic flat $\omega$ describes a collection of topological background Fronsdal fields and `free' equations \eqref{freeeqA} result in equations for the Maxwell-, Weyl-tensors and the scalar field that are highly nonlinear in the background fields. Probing different backgrounds is a useful way to probe interactions. Eqs. \eqref{freeeqA} probe the maximally-symmetric backgrounds, of which $AdS_{d+1}$ is the simplest case. Within the Noether procedure the higher spin algebra becomes visible at the quartic level only, \cite{Boulanger:2013zza}.}
\begin{align}\label{freeeqA}
    d\omega&=\omega\star \omega\,, &
    dC&= \omega\star C-C\star \pi(\omega)\,, && \omega\,, C \in \hs\,.
\end{align}
Now, suppose we are interested in deformations of this free system 
or in the conserved currents that are $d$-closed on-shell to the leading order. For example, we can be looking for an algebraic current
\begin{equation}
   J(\omega,\ldots,\omega,C,\ldots,C)\in \mathbb{C}\,,
\end{equation}
or for an interaction vertex
\begin{equation}
    \mathcal{V}(\omega,\omega,C,\ldots,C)\in \hs
\end{equation}
that can deform the r.h.s. of the first equation in \eqref{freeeqA}. In either case, we see that $(i)$ the free equations take advantage of the associated Lie algebra $L(\hs)$; $(ii)$ the second equation in (\ref{freeeqA}) suggests that the field $C$ takes values in the $\pi$-twisted adjoint representation $\hs_\pi$ of $L(\hs)$.  Therefore, we face quite a difficult  problem of computing Chevalley--Eilenberg cohomology of $L(\hs)$ with nontrivial coefficients. One may think of multi-linear functions of $C$ as functions with values in the symmetric tensor product $S(\hs_\pi^*)$ of the module $\hs_\pi^*$ that is conjugate to $\hs_\pi$. We will reduce the problem to a much simpler one in a few steps:

\begin{itemize}
    \item Firstly, we would like to consider the matrix extension of the higher spin algebra $\mathrm{Mat}_n(\hs)=\hs\otimes \mathrm{Mat}_n$, i.e., $n\times n$-matrices with entries in $\hs$. The associated Lie algebra $gl_n(\hs)\equiv L(\mathrm{Mat}_n(\hs))$ is much more handy: for $n$ big enough the  Chevalley--Eilenberg cohomology of $gl_n(\hs)$ can be effectively reduced to the Hochschild cohomology of $\hs$, see Appendix \ref{AR}. Apart from being just a simplifying assumption, the matrix extension is well-motivated by the AdS/CFT correspondence. Indeed, with appropriate boundary conditions each HSGRA is dual to a certain free CFT and in a free CFT there is always an option to add global symmetries.\footnote{For example, one can take the $U(N)$-vector model with free complex conjugate fields $\bar\phi^i$ and $\phi_i$. In the simplest case the higher spin fields are dual to the conserved  currents $J_s=\bar\phi^i \pl^s \phi_i$. The bulk coupling constant is of order $1/\sqrt{N}$, which justifies taking the large-$N$ limit (even though the CFT is free and not much can change with $N$). The dual HSGRA gauges the higher spin symmetry algebra $\hs$ that is induced by $J_s$. There is always a simple extension: one takes vector fields $\bar\phi^{I,i}$ and $\phi_{J,j}$ of $U(M)\times U(N)$, then $U(N)$ is to be weakly gauged, while $U(M)$ is left as a global symmetry. Therefore, the higher spin currents $J_s{}\fud{I}{J}=\bar\phi^{I,i} \pl^s \phi_{J,i}$ are $U(M)$ valued and  the higher spin algebra is $gl_M(\hs)=\hs \otimes \mathrm{Mat}_M$ (its appropriate real form, to be precise). Now the dual HSGRA has to gauge $gl_M(\hs)$.}
    
    \item Secondly, we absorb $\pi$ by extending the higher spin algebra to the smash product algebra $A=\hs \rtimes \mathbb{Z}_2$, where the cyclic group $\mathbb{Z}_2$ is generated by $\varkappa$ and imitates the action of the involutive automorphism $\pi$: $\varkappa f \varkappa =\pi(f)$, $\varkappa^2=1$, \cite{Iazeolla:2008ix}. The multiplication in $A$ is given by formula (\ref{smpr}). Upon such an extension  $\pi$ becomes an inner automorphism of $A$. The fields are, however, doubled
    \begin{align}\label{doubledfields}
        \mC&=K+C\varkappa\,, & \momega&=\omega+ \omega' \varkappa\,.
    \end{align} 
    The free equations (\ref{freeeqA}) can be rewritten as
    \begin{align}\label{freeeqAA}
    d\momega&=\momega\star \momega\,, &
    d\mC&= \momega\star \mC-\mC\star \momega\,, && \momega\,, \mC \in A
    \end{align}
    without any mentioning of $\pi$. The problem of completing \eqref{problem} with vertices is now reduced to a more concise problem of completing
    \besubeqs\label{problemB}
    \begin{align}\label{problemBAA}
    d\momega&=\momega\star \momega + \mathcal{V}_3(\momega,\momega,\mC)+\mathcal{V}_4(\momega,\momega,\mC,\mC)+\mathcal{O}(\mC^3)\,,\\ \label{problemBAB}
    d \mC&=\momega\star \mC-\mC\star \momega+\mathcal{V}_3(\momega,\mC,\mC)+\mathcal{O}(\mC^3)\,,
    \end{align}
    \esubeqs
    (no $\pi$ in the second equation). Once we solve any higher spin problem for $A$, we get the sought for solution for $\hs$ by setting $\omega'=0$ and $K=0$. As we discuss below, $K$ from \eqref{doubledfields} can also be useful sometimes.

    \item Finally, we need to take into account that \eqref{freeeqAA} comprises two fields, $\momega$ and $\mC$. As was discussed at length in Sec. \ref{sec:formalequations}, the quadratic vertices of any formal dynamical system \eqref{MFDA} define and are defined by a graded Lie algebra $\mathcal{L}$. It is easy to see that the graded Lie algebra at hand can be modelled by taking the tensor product of $A$ with the Grassmann algebra $\Lambda$ on one generator, $\Lambda=\mathbb{C}[\theta]$, $\theta^2=0$. This allows us to combine the fields $\momega$ and $\mC$ into a single `superfield' $\Phi=\momega+\theta \mC$ subject to the equation
    \begin{equation}
        d\Phi=\Phi\star \Phi+\mathcal{V}_3(\Phi,\Phi,\Phi)+\mathcal{O}(\Phi^4)\,.
    \end{equation}
    The superfield $\Phi$ being odd, the star-product reduces in fact to the commutators of component fields inducing thus the graded Lie algebra structure $\mathcal{L}$.
\end{itemize}

We may now summarize that the majority of HSGRA problems boil down to computing  the Chevalley--Eilenberg cohomology of the graded Lie algebra $\mathcal{L}=gl_n(A\otimes \Lambda)$, where $A=\hs \rtimes \mathbb{Z}_2$. The matrix extension allows one to reduce the complicated Chevalley--Eilenberg problem to a much simpler problem of Hochschild cohomology;\footnote{An important question is how big the matrices are, i.e., what is the lower bound on $n$ such that the Chevalley--Eilenberg cohomology can be reduced to the Hochschild one. Theorem \ref{A1} implies that $n$ has to be greater or equal to the number of arguments that a given cocycle has. However, our preliminary study of the Chevalley--Eilenberg cohomology of the Weyl algebra $A_1$ indicates that all the results in the paper are true all the way down to $n=1$, i.e., without any matrix extension (the whole Chevalley--Eilenberg cohomology of $A_1$ is induced from the Hochschild one). } the details are left to Appendices \ref{app:A}, \ref{app:B}. The following general arguments will be used repeatedly below:
\begin{itemize}
   
    \item Given that the form-degree should not exceed the space-time dimension, it is important to decompose the  cohomology groups  $H(\mathcal{L},\mathcal{T})$ into representatives $\momega^m\mC^n$ that have a definite degree in $\mC$ and $\momega$.

     \item If a relevant structure corresponds to a cohomology class of $H(\mathcal{L},\mathcal{T})$ in form-degree $p$, then the same cohomology group in degree ${p+1}$ contains possible obstructions to extending this structure  to higher orders in deformation. Depending on the situation the same cohomology class can give either a nontrivial $(p+1)$-current or an obstruction to the deformation of some other  $p$-current.

    \item As we compute the cohomology of the extended higher spin algebra, $A=\hs \rtimes \mathbb{Z}_2$, there is an extra information that may be helpful. Both $\momega$ and $\mC$ consist of the physical and unphysical parts \eqref{doubledfields}. Upon setting $\omega'=0$, so that $\momega=\omega$, we see that the second equation in  \eqref{freeeqAA} reduces to\footnote{So far we have not been able to take advantage of one-forms $\omega'$ in the twisted-adjoint representation. They do not seem to have any physical interpretation.}
    \begin{align}\label{freeeqAAAX}
        dC&= \omega\star C-C\star \pi(\omega)\,, &
        dK&= \omega\star K-K\star \omega\,.       
    \end{align}
    $C$ appears as expected and  $K$ takes values in the adjoint representation of $L(\hs)$. Therefore, $K$ corresponds to a global symmetry parameter in the $\omega$-background.\footnote{Indeed, suppose we have a flat connection $\omega$ (maximally-symmetric background). Global symmetries are the gauge transformations that leave $\omega$ invariant,  $\delta_\xi \omega=0$. This leads to the equation $d\xi=\omega\star \xi-\xi\star \omega$, which is identical to the second equation in \eqref{freeeqAAAX}. }
    
    First of all, it is important to see if a given cocycle survives upon projecting onto the physical sector:  $\omega'=0$ and $K=0$. If it does so, we have a physically relevant $p$-form current. If it does not, then setting $\omega'=0$ to zero, all but one $\mC$ to $C$, and the remaining one to $K$ gives a nonvanishing form of the type
    \begin{align}
        J(\omega,\ldots,\omega,C,\ldots,C,K)\,.
    \end{align}
    This can be interpreted as a collection of conserved $p$-currents parameterized by a global symmetry parameter $K$. One can also project several $\mC$'s onto $K$'s, which gives a conserved $p$-current that depends on a global symmetry with values in the tensor product of several adjoint representations of the higher spin algebra.\footnote{In the actual examples we find below, the nonlinearity in $\mC$ is of simple form $\mC^k$. }  
    
    \item Instead of scalar $p$-form currents one may be interested in the Chevalley--Eilenberg cocycles with values in a given higher spin algebra. They correspond either to possible deformations of the free equations or  to introducing  higher-degree forms with values in the higher spin algebra and adding some sources there, see e.g. \cite{Boulanger:2015kfa,Bonezzi:2016ttk}. In order to get such cocycles, one can simply remove one $\mC$ factor from the scalar ones, i.e., by representing them as $\Tr(f(\momega,\ldots,\mC)\star \mC)$. This agrees, of course, with the honest computation of cohomology in the present paper. 
\end{itemize}
Supersymmetric extensions of the bosonic $4d$ HSGRA can also be of some interest, see \cite{Sezgin:2002rt,Sezgin:2012ag} for discussion. In practice, these are obtained \cite{Sezgin:2012ag} by appending $\hs$ with the generators of the Clifford algebra. This is equivalent to taking the tensor product with an additional matrix factor, which does not change the cohomology. Therefore, all the results on the higher spin cohomology below apply to various supersymmetric extensions as well.

\subsection{Algebraic structure of interactions}
The general solution of the problem on how to construct the interaction vertices in \eqref{problem} was given in \cite{Sharapov:2019vyd}. Let us recall a few details from that work as we are going to discuss not only the observables of the free theory, but also possible obstructions to their deformation by interaction. 

The problem of constructing the vertices $\mathcal{V}_n$ in \eqref{problemB} can be solved whenever the algebra $A$ which determines the bilinear terms in \eqref{problemB} admits a  nontrivial deformation as an associative algebra. This means that there exists  a one-parameter family of associative algebras with the product  
\begin{align}\label{fullproduct}
    a\ast b&= a\star b+\sum_{k>0} \phi_k(a,b) u^k\,, 
\end{align}
 $u$ being a formal deformation parameter. As usual associativity requires  the  first-order deformation $\phi_1$ to be  a  Hochschild two-cocycle. According to the general definition (\ref{HD}) it obeys 
\begin{align}\label{untwistedcocycle}
   (\partial\phi_1)(a,b,c)= a\star \phi_1(b,c)-\phi_1(a \star b,c)+\phi_1(a,b\star c) -\phi_1(a,b)\star c&=0\,.
\end{align}
Given the $\phi_k$'s, one can write down immediately all the vertices. For example,
\begin{align}\label{firstV}
    \mathcal{V}_3(\momega,\momega,\mC)&= \phi_1(\momega,\momega)\star \mC 
\end{align}
and
\begin{align}\label{ffour}
    \mathcal{V}_4(\momega,\momega,\mC,\mC)&=\phi_2(\momega,\momega)\star \mC\star \mC +\phi_1(\phi_1(\momega,\momega),\mC)\star \mC\,.
\end{align}
Of course, the vertices are determined up to an isomorphism  of the corresponding $L_\infty$-algebra (or FDA), which corresponds to field redefinitions in the field theory language or to a change of coordinates in the language of $Q$-manifolds. 

Eq. \eqref{untwistedcocycle} and $\phi_1$ is the simplest example of the higher spin cohomology and the size of the cohomology group has a clear physical meaning:\footnote{The superscript $n$ in $H^n(\mathcal{L},\mathcal{T}^{1,0})$ denotes the number of arguments that representative cocycles have. The deformation of \eqref{problemB} has to be cubic, therefore $n=3$. We can prove, under the assumptions above, that this cohomology is equivalent to the second Hochschild cohomology of $A$, that is, $HH^2(A,A)$. }
\begin{align}
    &\dim H^3(\mathcal{L}, \mathcal{T}^{1,0})=\begin{cases}\text{AdS-side:\,\, \# independent coupling constants}\,,\\ \text{CFT-side: \# marginal deformations}+1\,. \end{cases}
\end{align}
It is worth stressing that the deformation of equations \eqref{problemB}, i.e., the condition for $\mathcal{V}_3$, leads to a more complicated Chevalley--Eilenberg problem for the Lie algebra $\mathcal{L}=gl_n(A\otimes \Lambda)$. Under the assumptions discussed briefly above and in more detail in Appendices \ref{app:A}, \ref{app:B}, this problem can be reduced to a much simpler problem of Hochschild cohomology. Therefore, in many instances we expect that the higher spin cohomology is obtained from the Hochschild cohomology $HH^\bullet(A,A)$ in the sense that the knowledge of $HH^\bullet(A,A)$ and applying the standard cohomological operations allows one to reconstruct $H^\bullet(\mathcal{L}, \mathcal{T})$. 

This  means that the number of independent deformations of the star-product on $A$ is equal to the number of independent coupling constants in the HSGRA that gauges $A$.\footnote{It is important to check whether these deformations survive on physical configurations $\momega=\omega$ and $\mC=C\varkappa$, i.e., whether the deformation is along $\varkappa$ or it is the higher spin algebra $\hs$ itself  that gets deformed. In the latter case, deformation has nothing to do with interaction, see the $3d$ example below. } On the dual CFT side the same number is equal to the number of marginal deformations plus one. Indeed, one coupling constant in the bulk just counts the degree in $C$ and it is proportional to $1/\sqrt{N}$ on the CFT side. The rest of the couplings should parameterize the space of CFT's, i.e., the marginal deformations (at least in the large-$N$ limit). 

In general $\dim HH^2(A,A)=1$, so that there is a single bulk coupling constant of order $1/\sqrt{N}$. An interesting example with $\dim HH^2(A,A)=2$  is provided by  Chern--Simons Matter theories that feature one more coupling constant -- the Chern--Simons level $k$ -- and the corresponding effective t'Hooft coupling $\lambda=N/k$ becomes a continuous parameter in the large-$N$ limit. Accordingly, the extended higher spin algebra $A$ admits a two-parameter deformation. For more detail we refer to \cite{Sharapov:2018kjz,Sharapov:2017lxr}.

That all nonlinear terms in \eqref{problemB} follow, up to equivalence, in a simple way from a one-parameter family of associative algebras is an enormous simplification of the problem. There is a variety of nonlinear equations of type \eqref{problemB} in the literature that do not seem to use this structure at all \cite{Bekaert:2013zya,Bekaert:2017bpy,Grigoriev:2018wrx}, use it implicitly \cite{Vasiliev:1990en,Vasiliev:2003ev,Neiman:2015wma,Bonezzi:2016ttk,Arias:2017bvi}, and use it explicitly \cite{Sharapov:2019vyd,Sharapov:2019pdu}.\footnote{There are two more models: the collective dipole approach, see e.g. \cite{deMelloKoch:2018ivk}, which we discuss below, and the higher spin IKKT model of \cite{Sperling:2017dts}, whose relation to the above is yet to be clarified. } There are some general methods to deform associative algebras in a constructive way, e.g. deformation quantization \cite{Fedosov:1994zz, Kontsevich:1997vb} and the injective resolution technique  \cite{2019LMaPh.109..623S,Sharapov:2018ioy}, which help to construct models of this kind.

\section{Characteristic cohomology and observables }
\label{sec:observables}
The main case of interest for us is $4d$ HSGRA wherein the higher spin algebra is based on the Weyl algebra $A_2$, i.e., the noncommutative algebra of functions in two pairs of creation and annihilation operators.  Nevertheless, all the observables turn out to be made out of the building blocks that exist already for the smallest Weyl algebra $A_1$. This is due to the obvious isomorphism $A_2\simeq A_1\otimes A_1$. Therefore, we first go through the case of $A_1$, which can be of some interest by itself, e.g. in the context of $3d$ HSGRA or specific Chern--Simons theories.

\subsection{\texorpdfstring{Three-dimensional HSGRA: warming up with $A_1$}{Three-dimensional HSGRA: warming up with A-one} }
\label{sec:Aone}
Let us consider the algebra of complex  polynomials in creation and annihilation operators: 
\begin{align}
    [\hat a,\hat a^\dag]&=1 && \Longleftrightarrow && [\hat y_\ga,\hat y_\gb]=-2\epsilon_{\ga\gb}\,, \qquad \hat y_\ga=(\sqrt{2}\hat a^\dag,\sqrt{2}\hat a)\,.
\end{align}
We prefer to work with $f(\hat y_\ga)$ instead of $f(\hat a,\hat a^\dag)$. As one more improvement, one can consider commuting variables $y_\ga$ and endow the algebra of polynomial functions in $y_\ga$ with the Weyl--Moyal star-product:
\begin{align}\label{starsucks}
    (f\star g)(y)&=\exp \left[ p_1\cdot p_2 \right]f(y+y_1)g(y+y_2)\Big|_{y_{1,2}=0}\,,
\end{align}
where $p_0\equiv y$, $p_i^\ga=\pl/\pl y_{i,\ga}$, and the symplectic inner product is defined as $p_1\cdot p_2\equiv - p_1^\ga \epsilon_{\ga\gb} p_2^\gb$. Note that the derivatives are defined to act as $p_1^\ga y_1^\gb=\epsilon^{\ga\gb}$ and the indices are raised and lowered according to $y^\ga=\epsilon^{\ga\gb}y_\gb$, $y_\gb=y^\ga \epsilon_{\ga\gb}$.\footnote{The same rule applies to derivatives. Starting from the canonical $\pl_\ga y^\gb=\delta\fdu{\ga}{\gb}$ we get $\pl^\ga y^\gb=\epsilon^{\ga\gb}$, which is not the same as to lower the index on $y_\ga$ and then take the derivative. Note, $\epsilon_{12}=1$, $\epsilon_{\ga\gb}\epsilon^{\gc\gb}=\delta\fdu{\ga}{\gc}.$} The star-product \eqref{starsucks} can be rewritten in a more concise form as\footnote{The only trick here is that $\exp[p_0\cdot p_1]f(y_1)\equiv \exp[y^\nu  (\frac{\pl}{\pl y_1})_\nu]f(y_1)\equiv f(y+y_1)$.}
\begin{align}\label{starcool}
    (f\star g)(y)&=\exp \left[p_0\cdot p_1+p_0\cdot p_2+ p_1\cdot p_2 \right]f(y_1)g(y_2)\Big|_{y_{1,2}=0}\,.
\end{align}
In general, any operator $R$ (cochain) that takes a number of elements from the Weyl algebra -- complex polynomial functions $f(y)$ -- can be written as
\begin{align}\label{operatorR}
    R(f_1,\ldots,f_n)(y)&= R(p_0,p_1,\ldots,p_n) f_1(y_1)\cdots f_n(y_n)\Big|_{y_1=\cdots =y_n=0}\,.
\end{align}

Let us assume for a moment that the higher spin algebra of interest is $\hs=A_1$. It is well known that the algebra $A_1$ is rigid, i.e., it cannot be  deformed nontrivially as an associative algebra. This is due to the fact that $HH^2(A_1,A_1)=0$, so that Eq. \eqref{untwistedcocycle} has only  trivial solutions $\phi_1=\partial\psi$. The only nonzero cohomology group of $HH^\bullet(A_1,A_1)$ is $HH^0(A_1,A_1)\simeq\mathbb{C}$. It is naturally identified with the center of the Weyl algebra $\mathbb{C}\subset A_1$, see Appendix  \ref{PWA}.

A more interesting case is the groups $HH^\bullet(A_1,A_1^*)$. This time  $\dim HH^2(A_1,A_1^*)=1$ and most of nontrivial cohomology considered  in the present paper takes its origin in this fact. This means that there is a nontrivial two-cocycle with values in the module dual to $A_1$. It is this cocycle that will lead to the deformation of the extended higher spin algebra in the case of $4d$ HSGRA. Therefore, it is important to discuss it in more detail. The dual module $A_1^\ast$ can be realized in two different ways. 

\textbf{1.} With the help of the supertrace\footnote{It is a supertrace in the sense that even/odd functions $f(y)$ are considered to be even/odd elements of the superalgebra $A_1$, which is the same $A_1$, but understood as a $\mathbb{Z}_2$-graded associative algebra. } $\Str f=f(0)$ one can define a nondegenerate  pairing $\Str(f\star g)$, which identifies $A_1^*$ with $A_1$. The action of $A_1$ on $A_1^*$ can be derived from the equalities $\Str(f\star g)=\Str(\pi({g})\star f)=\Str (g\star \pi({f}))$, where $\pi$ is the involutive  automorphism  of $A_1$ defined by
\begin{equation}\label{-y}
\pi(f)(y)=f(-y)\,.
\end{equation}
The group $HH^2(A_1,A_1^*) $ is now generated by a nontrivial solution to
\begin{align}\label{twistedcocycle}
    a\star \phi(b,c)-\phi(a \star b,c)+\phi(a,b\star c) -\phi(a,b)\star \pi(c)&=0\,, \qquad a,b,c\in A_1\,.
\end{align}
It is illuminating to see how far from the star-product \eqref{starcool} such $\phi$ is. It can be written in two different forms. First, there is an explicit expression of the form \eqref{operatorR}:
\begin{align}\label{ffs}
    \phi_{FFS}(y\equiv p_0,p_1,p_2) a_1(y_1) a_2(y_2) \Big|_{y_{1,2}=0}\,,
\end{align}
where 
\begin{align}\label{FFS-symb}
   \phi_{FFS}&= \frac{z \left[(x-z) e^{x+y+z}+(y+z) e^{-x-y+z}-(x+y) e^{x-y-z}\right]}{4 (x+y) (x-z) (y+z)}\,.
\end{align}
Note that $\phi_{FFS}$ is nonsingular despite the way it is written. The cocycle condition \eqref{twistedcocycle} 
can be rewritten in terms of the symbols of operators as
\begin{align*}
    \phi(p_0\!+\!p_1,p_2,p_3) e^{p_{01}}\!-\!\phi(p_0,p_1\!+\!p_2,p_3) e^{p_{12}}\!+\!\phi(p_0,p_1,p_2\!+\!p_3) e^{p_{23}}\!-\!\phi(p_3\!+\!p_0,p_1,p_2) e^{-p_{03}}\!=0\,.
\end{align*}
For reference, the symbol of the  star-product is just $e^{x+y+z}$. Second, the cocycle $\phi_{FFS}$ can be rewritten as an integral over a two-simplex,\footnote{It follows from the Kontsevich--Shoikhet--Tsygan Formality \cite{FFS}, but implicitly was found for $A_1$ even earlier in \cite{Vasiliev:1988sa}. } namely,
\begin{align}
    \phi_{FFS}&= z\int\limits_{0<t_1<t_2<1} \exp{\left[x(1-2t_1)+y(1-2t_2)+z(1+2(t_1-t_2))\right]}\,.
\end{align}

\textbf{2.} Operators with values in $A_1^*$ can be thought of as linear functionals on $A_1$:
\begin{align}\label{ffs1}
    \phi(a_0,a_1,a_2)&=\phi(\tfrac{\pl}{\pl y_0}\equiv p_0,p_1,p_2) a_0(y_0)a_1(y_1) a_2(y_2) \Big|_{y_{0,1,2}=0}\,.
\end{align}
The first argument $a_0$ accounts for taking values in $A_1^*$. The cocycle condition now reads
\begin{align}\label{cyclicClosure}
    \phi(a\star b,c,d)-\phi(a,b\star c,d)+\phi(a,b,c\star d)-\phi(d\star a,b,c)=0\,.
\end{align}
With the help of the pairing between $A_1$ and $A_1^*$ via $\Str$, we can express the cocycle with three-arguments (still  called two-cocycle since the first argument is of a different nature) as
\begin{align}\label{noncyclic}
    \phi(a,b,c)&= \Str (a\star \phi(b,c))\,.
\end{align}
The last formula links representations (\ref{ffs}) and (\ref{ffs1}).

Associated to the Hochschild cocycle (\ref{ffs1}) is a cyclic cocycle, which is a representative of cyclic cohomology, see Appendix \ref{CC} for precise definitions. The cyclic complex is a subcomplex of the Hochschild one, where the cochains are invariant under the cyclic permutations of arguments up to sign. In particular, our cyclic two-cocycle has to obey
\begin{align}\label{cycle-pr}
    \phi_C(a,b,c)&= \phi_C(c,a,b)\,.
\end{align}
Note that \eqref{noncyclic} is not a cyclic cocycle. Moreover, one cannot obtain $\phi_C$ by averaging \eqref{noncyclic} over the cyclic permutations. Nevertheless, both the cocycles are known to be cohomologous to each other. In our case, it is possible to find a cyclic representative  directly:
\begin{align}\label{fc}
    \phi_C(x,y,z)=\frac{\left(y^2-x^2\right) e^{x-y-z}+\left(x^2-z^2\right) e^{x+y+z}+\left(z^2-y^2\right) e^{-x-y+z}}{8 (x+y) (x-z) (y+z)}\,.
\end{align}
One can also represent it  as the sum of integrals over two- and one-simplices:
\begin{align}\label{2-1}
    \phi_C&=\phi_{FFS}+\frac14\int\limits_{0<t<1} \Big\{e^{\left[(1-2t)(y+z)+x\right]}-e^{\left[(1-2t)(x+y)+z\right]}+e^{\left[(1-2t)(x-z)-y\right]}\Big\}\,.
\end{align}

\paragraph{Extended higher spin algebra.} Let $\pi$ be the involutive automorphism (\ref{-y}) of our higher spin algebra $A_1$ and we would like to deform the free equations \eqref{freeeqA}.\footnote{An alternative  point of view is that we would like to `gauge' the discrete $\mathbb{Z}_2$-symmetry and to project onto the $\pi$-invariant subalgebra, i.e., to consider $\omega$ and $C$ assuming  values in the even subalgebra $A_1^e\subset A_1$.} According to the general recipe we should introduce a new algebra generator $\varkappa$ such that  $$\varkappa^2=1 \,,\qquad \varkappa \star f(y)\star \varkappa=f(-y)\quad \Leftrightarrow\quad \varkappa\star y_\alpha=-y_\alpha\star \varkappa\,.$$ Now, the extended algebra $\mathcal{A}$ is $A_1\rtimes \mathbb{Z}_2$ and its generic element reads $f=f_1(y)+f_2(y)\varkappa$.\footnote{It makes sense to omit $\star$ between $y$ and $\varkappa$ whenever no confusion arises. Such 
$\varkappa$'s  are called the Klein operators \cite{Ohnuki:1973jb} sometimes.} From now on, somewhat strange structures on $A_1$ that we defined above -- the supertrace $\Str$ and the two-cocycle $\phi$ with values in $A_1^*$ -- acquire a simple interpretation for $\mathcal{A}$ (or for its $\pi$-invariant subalgebra $A_1^e$).

A trace $\Tr$ on  $\mathcal{A}$ is given by a cyclic zero-cocycle, which may be viewed either as an  element of $\mathcal{A}^*$ or a linear functional on $\mathcal{A}$. The cocycle condition takes the familiar form  $\Tr[a\star b]=\Tr[b\star a]$. The supertrace on $A_1$ gives rise to the usual trace on $\mathcal{A}$ defined by
\begin{align}
    \Tr[f]&=\Str[f\star \varkappa]=f_2(0)\,, && f=f_1(y)+f_2(y)\varkappa\in \mathcal{A}\,.
\end{align}
That the group $HH^2(A_1,A_1^*)$ is nonzero translates into the fact that $\dim HH^2(\mathcal{A},\mathcal{A})=1$ and the algebra $\mathcal{A}$ admits a nontrivial deformation. The second cohomology group is generated by the two-cocycle $\phi_1(a,b)=\phi(a,b)\star \varkappa$, where $\phi$ from \eqref{twistedcocycle} can be taken to be $\phi_{FFS}$. Now, $\phi_1$ obeys the  condition
\begin{align}\label{untwistedcocycleB}
   a\star \phi_1(b,c)-\phi_1(a \star b,c)+\phi_1(a,b\star c) -\phi_1(a,b)\star c&=0\,, && a,b,c\in \mathcal{A}\,,
\end{align}
and determines an infinitesimal deformation of $\mathcal{A}$. Since $HH^3(\mathcal{A},\mathcal{A})=0$, the deformation is not obstructed and there exist all higher-order corrections $\phi_{k}(-,-)$ that define an associative $\ast$-product \eqref{fullproduct}. It is also true that the $\pi$-invariant subalgebra, $A_1^e$, admits a one-parameter deformation.\footnote{The one-parameter family of algebras that goes through $A_1^e$ is not mysterious and can be obtained as the  quotient of the universal enveloping algebra $\mathcal{U}(sl_2)$ by the two-sided ideal associated with the quadratic Casimir operator $C_2-(-3/4-u)$ \cite{Feigin1987}. The even subalgebra $A_1^e$ corresponds to $C_2=-3/4$. This family was dubbed $gl_\lambda$ in \cite{Feigin1987} as it interpolates between $gl_n$ in some sense.} 

\paragraph{Relation to HSGRA in three dimensions.} Although the structures discussed above are intended to be building blocks for a more complex case of 4d HSGRA, it is also of interest to adapt them to 3d HSGRA. As is well-known, due to the isomorphisms $so(2,2)\sim sp_2\oplus sp_2$, $sp_2\sim sl_2$ the Einstein--Hilbert action for $3d$ gravity can be rewritten as the difference of two $sl_2$ Chern--Simons actions \cite{Witten:1988hc}:
\begin{align}
    S_{EH}&= \frac{k}{4\pi}S_{CS}[A_L]-\frac{k}{4\pi}S_{CS}[A_L]\,, &&S_{CS}[A]=\int \mathrm{Tr}\Big(A\wedge dA-\frac{2}3A\wedge A\wedge A\Big)\,.
\end{align}
This admits a straightforward generalization to the case where $sl_2$ is replaced by any Lie algebra bigger than $sl_2$ \cite{Blencowe:1988gj,Bergshoeff:1989ns,Campoleoni:2010zq,Henneaux:2010xg} with a specific embedding of $sl_2$ such that decomposition into $sl_2$-modules contains representations bigger than the adjoint one. In particular, one may consider the even  subalgebra $A_1^e\subset A_1$. The associated  Lie algebra $L(A_1^e)$ contains $sp_2\sim sl_2$ as a subalgebra generated by quadratic polynomials $\{y_\ga, y_\gb\}_\star$. Decomposition into $sp_2$-modules looks as 
\begin{align}
    A_1^e&= V_0\oplus V_1\oplus V_2\oplus \cdots \,, & \dim V_j&= 2j+1 \,.
\end{align}
The singlet $V_0$ corresponds to the unit of $A^e_1$ and results in the direct $u(1)$-factor in $L(A_1^e)$. $V_1$ is given by the embedding of $sp_2$. All HSGRA's with massless, partially-massless, and conformal higher spin fields in $3d$ must be of the Chern--Simons form \cite{Grigoriev:2020lzu}. One can try to look for more complicated theories with matter fields (scalars or fermions). As was shown in \cite{Vasiliev:1996hn,Prokushkin:1998bq}, the matter fields (we consider the scalar field for simplicity) can be described by the equations
\begin{align}
    dA_L&= A_L\star A_L\,, & dA_R&= A_R\star A_R\,, &dC&=A_L\star C-C\star A_R\,,
\end{align}
where $A_L$, $A_R$, and $C$ take values in $A_1^e$. This system can be reduced to the standard one
\eqref{freeeqAA}  with the help of an additional extension by $2\times 2$-matrices \cite{Vasiliev:1996hn,Prokushkin:1998bq}.\footnote{Representing $\mathrm{Mat}_2$ as the Clifford algebra $\gamma_0^2=\gamma_1^2=1$, $\gamma_0\gamma_1+\gamma_1\gamma_0=1$, one can associate $A_{L,R}$ with the entries on the diagonal and $C$ as a component along the matrix $[(0,1),(1,0)]$. The remaining components are similar to the those considered around \eqref{doubledfields}.} Finally, the relevant algebra is $\mathcal{A}=(A_1\rtimes \mathbb{Z}_2)\otimes \mathrm{Mat}_2$, where $\mathbb{Z}_2$ is to be gauged to reduce the equations to even functions of $y_\ga$. As was discussed in Sec. \ref{sec:HScohomology}, all our results are obtained in conditions where the matrix extension is always possible.\footnote{There might exist some sporadic Chevalley--Eilenberg cocycles for $L(\hs\otimes \mathrm{Mat}_n)$ with small $n$'s, but we are not aware of any example. } Therefore, we can go back to $\mathcal{A}=A_1\rtimes \mathbb{Z}_2$ at any time as the matrix factor does not make any difference for the classification of observables.

\subsubsection*{Conserved currents and interactions}
In this section, we present a complete classification of conserved (and hence, gauge-invariant up to an exact differential) currents for the free system
\begin{align}\label{freeeqAAA}
    d\momega&=\momega\star \momega\,, &
    d\mC&= \momega\star \mC-\mC\star \momega\,, && \momega\,, \mC \in \mathcal{A}\,,
\end{align}
associated with the extended higher spin algebra $\mathcal{A}=A_1\rtimes \mathbb{Z}_2$. The results can be summarized by the following table, which can be extracted from Table \ref{table1} in Appendix \ref{app:B} and where we do not list those currents that have more than five $\momega$'s. 
\begin{table}[h]
    \centering
\begin{tabular}{c|c|c|c|c|c|}
    & $\mC^0$ & $\mC^1$ &$\mC^2$ &$\mC^3$ &$\mC^4$   \tabularnewline\hline
    $\momega^0$ & $\empty$ & $1$ & $1$& $1$& $1$\tabularnewline 
    $\momega^1$ & $1$& $0$& $0$& $0$& $0$\tabularnewline 
    $\momega^2$ & $0$ & $1$ & $1$ & $1$ & $1$\tabularnewline 
    $\momega^3$ & $2$ & $0$& $0$& $0$& $0$\tabularnewline 
    $\momega^4$ & $0$ &  $0$ & $0$& $0$& $0$\tabularnewline 
    $\momega^5$ & $2$ & $0$& $0$& $0$& $0$
\end{tabular}
    \caption{On-shell closed/gauge-invariant currents in the $A_1$ toy model. The table shows  the number of independent currents that are of $\momega^m \mC^n$-type, i.e., involve $m$ one-forms $\momega$ and $n$ zero-forms $\mC$.  }
    \label{tab:table3d}
\end{table}

\noindent Now, we give an account of the expressions that explain the multiplicities here-above. 
\paragraph{Class $\boldsymbol{\mC^n}$. } This is the simplest class comprising the zero-form invariants\footnote{Note that the supertrace $\Str$ on $A_1$ reduces to a trace on $A_1^e$.}
\begin{align}
    J_{0,n}&=\Tr[\mC^{n}]\equiv\Tr[\overbrace{\mC\star\cdots \star \mC}^{n}]
\end{align}
that are obviously gauge invariant and $d$-closed on-shell. For an appropriately chosen wave-function $\mC$ it can give correlation functions of $O=\chi^2$, where $\chi$ is a generalized free scalar field on the $2d$ CFT boundary of $AdS_3$. The significance of this is unclear since the dual CFT is expected to have a much bigger $W$-symmetry.\footnote{There are a number of important differences between HSGRA in $d\geq4$ and $d<4$. In the latter case higher spin fields do not have propagating degrees of freedom, while the higher spin algebra does not seem to play much role on the CFT side since the conformal symmetry gets extended to the Virasoro algebra and even further to $W$-algebras \cite{Campoleoni:2010zq,Henneaux:2010xg,Gaberdiel:2012uj}.} These invariants are unobstructed.

\paragraph{Class $\boldsymbol{\omega^{2k+1}}$, Chern--Simons forms. } This class consists of the on-shell closed Chern-Simons forms
\begin{align}
    \tilde J_{2k+1,0}&=\Tr[\momega^{2k+1}]\equiv\Tr[\,\overbrace{\momega\star \cdots \star \momega}^{2k+1}\,]\,.
\end{align}
It is easy to see that they are gauge invariant up to an exact differential. In fact, the Chern--Simons forms with $k>0$ are all generated from $\Tr[\momega]$ by the standard $S$-operation, see Appendix  \ref{PM}. The $k=0$ member is obstructed, but the others are not.

\paragraph{Class $\boldsymbol{\phi_C(\omega,\omega,\omega)}$. } The importance of the cyclic cocycle $\phi_C$ found in Sec. \ref{sec:Aone} is that it allows us to construct a new $3$-current, namely, 
\begin{align}\label{j30}
    J_{3,0}(\momega)&=\phi_C(\momega,\momega,\momega)\,.
\end{align}
It is closed on-shell:
\begin{align}
    dJ_{3,0}(\momega)&\approx 3\phi_C(\momega\star \momega,\momega,\momega)=0\,.
\end{align}
Here we used the equations of motion \eqref{freeeqAAA}, the cyclic property \eqref{cycle-pr}, and the cocycle condition \eqref{cyclicClosure}. One should also take into account that $\momega$ is a one-form. Equivalently, we can show that the current \eqref{j30} is gauge-invariant on-shell up to an exact differential:
\begin{align}
    \tfrac13\delta_\mxi J_{3,0}(\momega)&=\phi_C(d\mxi,\momega,\momega)-\phi_C([\momega,\mxi]_\star,\momega,\momega)\approx d\phi_C(\mxi,\momega,\momega)\,.
\end{align}
Therefore, $J_{3,0}(\momega)$ can be integrated to give a gauge-invariant observable in Chern--Simons theory.
This current is specific to the gauge algebra being the Weyl algebra (its even subalgebra). For the classical Lie algebras we have just $\Tr[\momega^3]$. We can use this additional observable to modify the Chern--Simons action as follows:
\begin{align}\label{ggg}
    \frac{k}{4\pi}\int \mathrm{Tr}\Big(\momega\wedge d\momega-\frac{2}3\momega\wedge \momega\wedge \momega\Big) +g\int \phi_C(\momega,\momega,\momega)+\mathcal{O}(g^2)\,.
\end{align}
The new current $J_{3,0}$ owes its existence to the fact that the algebra $A_1^e$ admits a one-parameter deformation and the same deformation  gives the one-parameter family (\ref{ggg}) of Chern--Simons theories. The cyclic cocycle $\phi_C$ determines the first-order deformation at the action level, while the corresponding Hochschild cocycle $\phi$ does the same at the level of algebra. Lastly, $J_{3,0}$ is unobstructed.   

\paragraph{Class $S^k\boldsymbol{\phi_C(\omega,\omega,\omega)}$. } With the help of the $S$-operation on cyclic cohomology one can generate a family of forms of the type $\momega^{2k+3}$ by applying $S$ to $\phi_C$. Applied ones, it gives
\begin{align}
    J_{5,0}=(S\phi_C)(\momega,\momega,\momega,\momega,\momega)&=3\phi_C(\momega^3,\momega,\momega)-\phi_C(\momega^2,\momega^2,\momega)\,.
\end{align}
In order to check the gauge-invariance of the resulting current directly one may use
the following identities, which are obtained by splitting $\momega^6$ into the four arguments that enter the cocycle condition:
\begin{align}
    2\phi_C(\momega^4,\momega,\momega)-\phi_C(\momega^3,\momega^2,\momega)+\phi_C(\momega^3,\momega,\momega^2)=0\,,\\
    \phi_C(\momega^2,\momega^2,\momega^2)-\phi_C(\momega^3,\momega,\momega^2)+\phi_C(\momega^3,\momega^2,\momega)+\phi_C(\momega^4,\momega,\momega)=0\,.
\end{align}
As the $S$-operation can be applied repeatedly, cyclic cohomology is unbounded from above as distinct from the case of Hochschild cohomology. We do not discuss invariants whose form degree is significantly higher than the space-time dimension. All members of this class, including the case $k=0$, are unobstructed. 

\paragraph{Class $\boldsymbol{\phi_C(\omega,\omega,C^k)}$. } All members of this class are two-forms. The first its representative is obtained by replacing one $\momega$ with $\mC$ in $\phi_C(\momega,\momega,\momega)$. This yields 
\begin{align}
    J_{2,1}&=\phi_C(\momega,\momega,\mC)\,.
\end{align}
It can also be rewritten through the Hochschild two-cocycle as
\begin{align}
    J_{2,1}&\sim\Tr[\phi(\momega,\momega)\star \mC]
\end{align}
and the cyclic property \eqref{cycle-pr} is not needed anymore to check that $dJ_{2,1}\approx 0$. The other members of this family are obtained by replacing $\mC$ with $\mC^k$:
\begin{align}
    J_{2,k}&=\phi_C(\momega,\momega,\mC^k)\equiv\phi_C(\momega,\momega,\overbrace{\mC\star\cdots \star \mC}^{k})\,.
\end{align}
The family contains infinitely many two-forms and all of them are unobstructed. Canonically, on-shell closed two-forms in $3d$ lead to conserved charges and we seem to have too many of them. The physically relevant charges should be finite. However, quite generally one can show that the charges built out of $J_{2,k}$ are divergent. We leave the discussion of possible applications of these observables to future publications.

Lastly, all of the here-above $p$-forms serve as the generators of the algebra of scalar invariants, i.e., the exterior product thereof gives new invariants. 

\paragraph{$\boldsymbol{H(\mathcal{L}, \mathcal{T}^{1,0})}$, class $\boldsymbol{\phi(\omega,\omega) C^k}$. } So far we discussed the scalar observables, i.e., $H(\mathcal{L}, \mathcal{T}^{0,0})$, but cocycles with values in the higher spin algebra are also of interest as they determine the deformation of equations. They can be obtained by undressing one $\mC$ in the observables above. Since this is rather a trivial exercise, we consider just one example -- the leading deformation of the equation
\begin{align}\label{deformed3d}
    d\momega&= \momega\star \momega +\phi(\momega, \momega)\star \mC^k+\mathcal{O}(\mC^{k+1})\,.
\end{align}
If $k=0$ the r.h.s. comes form the deformation of the algebra itself. This case is also covered by the  deformation with $k=1$ as the field  $\mC$ can be set constant in space and proportional to the unit of $\mathcal{A}$. This is a major difference with the case of $4d$ HSGRA. The $3d$ higher spin algebra $A_1^e$ admits a nontrivial  deformation and this is the only deformation of the extended algebra $\mathcal{A}$ as well. Therefore, one can tell straight away that the end result of this deformation is a system of the form
\begin{align}
    d\momega&= \momega\ast \momega\,, & d\mC&=\momega\ast \mC-\mC\ast \momega\,,
\end{align}
where the only difference is that the fields assume their values in the deformed algebra with the $\ast$-product \eqref{fullproduct}, see also \cite{Prokushkin:1998bq}. This system is clearly inconsistent with AdS/CFT since there is no backreaction from matter fields $\mC$ to the gravitational sector of $\momega$. The backreaction cannot be captured by the formal consideration and requires locality to be taken into account.\footnote{One obvious difference between the formal scheme and the physical one is that the stress-tensors are formally exact, i.e., do not correspond to nontrivial cocycles. Indeed, the stress-tensor contribution has to be a cocycle of type $\momega^2\mC^2$. The exactness of such stress-tensors for $\momega$ being the $AdS_3$ vacuum was shown in \cite{Prokushkin:1999xq} and discussed further in \cite{Kessel:2015kna}. If \eqref{deformed3d} begins with the $k=1$ deformation and the $\momega^2\mC^2$-deformation is also added, the latter can be obtained via $\mC\rightarrow \mC+\mC\star \mC$ redefinition from the $\momega^2\mC$-deformation and is no longer a nontrivial deformation.   Vertices with $k>1$ are examples of irregular deformations mentioned in footnote \ref{f5}.}

Everything we  said about the observables above holds true for any generic point of the one-parameter family of algebras that passes through $A_1^e$ or $\mathcal{A}=A_1\rtimes \mathbb{Z}_2$. Indeed, at any generic $u$ there is a Hochschild two-cocycle $\phi_u(a,b)=\pl_u (a\ast b)$. This $\phi_u$ can be used in place of $\phi$ above provided the $\star$-product is  replaced with $\ast$ at $u$. One only needs to be careful about whether or not a given observable descends to the physical sector.

\subsection{HSGRA in four dimensions}\label{4dim}
As was known already to Dirac \cite{Dirac:1963ta}, the algebra that naturally acts on the free $3d$ massless scalar field is the even subalgebra $A_2^e$ of the next to the smallest Weyl algebra $A_2=A_1\otimes A_1$. This algebra is the same as in \cite{Fradkin:1986ka, Eastwood:2002su}. Also, this is the same as the algebra of symmetries of the free $3d$ massless fermion field, which is a necessary condition for the $3d$ bosonization duality \cite{Giombi:2011kc, Maldacena:2012sf, Aharony:2012nh,Aharony:2015mjs,Karch:2016sxi,Seiberg:2016gmd} to work at least in the large-$N$ limit. To realize $A_2$ we simply double the variables, as compared to the previous section, and introduce the star-product language from the onset:
\begin{align}
   [y_\ga,y_\gb]_\star&=-2\epsilon_{\ga\gb}\,, &
   [y_\gad,y_\gbd]_\star&=-2\epsilon_{\gad\gbd}\,.
\end{align}
The higher spin algebra $\hs$ is the star-product algebra of even functions, $f(y,\bry)=f(-y,-\bry)$. Note that both $A_2$ and $A_2^e$ are rigid and do not have nontrivial deformations as associative algebras (cf. $A_1$ vs. $A_1^e$, the latter being deformable). 

The $\pi$-automorphism is defined to flip the sign of the $AdS_4$-translation generators $P_{\ga\gad}$ \cite{Vasiliev:1988sa,Bekaert:2005vh,Didenko:2014dwa}. Given that $P_{\ga\gad}\sim y_\ga y_\gad$, one can realize $\pi$ either as $\pi_1(f)(y,\bry)=f(-y,\bry)$ or as $\pi_2(f)(y,\bry)=f(y,-\bry)$. Both $\pi_1$ and $\pi_2$ are equivalent on $A_2^e$, but not on $A_2$.   
Since $A_2=A_1\otimes A_1$, the extended higher spin algebra $\mathfrak{A}$ is constructed by taking the tensor product of two copies of $\mathcal{A}=A_1\rtimes \mathbb{Z}_2$. Therefore, we introduce  the pair of generators $k$ and $\brk$ such that $k^2=\brk^2=1$ and\footnote{\label{alternativeK}Instead of taking the tensor product one could define a single $\varkappa$ satisfying $\varkappa^2=1$ and  $\varkappa\star P_{\ga\gad}\star \varkappa=-P_{\ga\gad}$. However, there are general theorems that allows one to compute the cohomology of the tensor product $A\otimes A$ whenever  the cohomology of $A$ is known. The higher spin cohomology corresponds to $\mathfrak{A}_0=A^e_2\rtimes \mathbb{Z}_2$ and, as is shown in Appendix \ref{EHSA}, it coincides with that of $\mathfrak{A}=\mathcal{A}\otimes \mathcal{A}$, which is more natural to compute. Note that the algebra with $k$ and $\brk$ was introduced in \cite{Vasiliev:1986qx} as $\mathcal{N}=2$ SUSY higher spin algebra, which is not an interpretation we use here. }
\begin{align}
    \{y_\ga, k\}&=0\,, & [\bry_\gad,k]&=0\,, & \{\bry_\gad,\brk\}&=0\,, &[y_\ga,\brk]&=0\,.
\end{align}
As for $A_1$, there is a unique supertrace $\Str$ on $A_2$. It induces the usual trace $\Tr$ on $A_2^e$.  Projection onto the subspace spanned by $k\brk$ gives a trace on $\mathfrak{A}$. 

In order to extract the physical sector of fields associated with $(A^e_2,\pi)$ one has to gauge the diagonal $\mathbb{Z}_2$ in $\mathcal{A}\otimes \mathcal{A}$. In practice, one takes $$\momega= \tfrac12(1+k \brk)\omega(y,\bry)\quad \mbox{and}\quad \mC=\tfrac12(k+\brk) C(y,\bry)\,, $$ where $\omega$ and $C$ are even functions. The relation to footnote${}^{\ref{alternativeK}}$ is that the element  $E=\tfrac12(1+k \brk)$ is central on the subalgebra of even functions and $\varkappa=\tfrac12(k+\brk)$ plays the role of $\varkappa$ on this subalgebra, $\varkappa^2=E$. If one wishes to keep the global symmetry parameters, then $\mC= E K(y,\bry)$ for any even $K$.

\subsubsection*{Conserved currents and interactions}
The conserved currents associated with  the free system
\begin{align}\label{freeeqAAAA}
    d\momega&=\momega\star \momega\,, &
    d\mC&= \momega\star \mC-\mC\star \momega\,, && \momega\,, \mC \in \mathfrak{A}\,,
\end{align}
are summarized in Table \ref{tab:table4d}, which can be extracted from Table \ref{table2} of Appendix \ref{app:B}. The structure of cyclic cohomology here is such that, translating it to the physics language, there are representatives of arbitrarily high form degree. Those that have the form degree much above the space-time dimension are hardly relevant for HSGRA studies. 
\begin{table}[h]
    \centering
\begin{tabular}{c|c|c|c|c|c|c|c|c|}
    & $\mC^0$ & $\mC^1$ &$\mC^2$ &$\mC^3$ &$\mC^4$ &$\mC^5$ &$\mC^6$ &$\mC^7$  \tabularnewline\hline
    $\momega^0$ & $\empty$ & $\boldsymbol{0}+1$ & $\boldsymbol{1}+0$& $\boldsymbol{0}+1$& $\boldsymbol{1}+0$& $\boldsymbol{0}+1$& $\boldsymbol{1}+0$& $\boldsymbol{0}+1$\tabularnewline 
    $\momega^1$ & $\boldsymbol{1}+0$& $0$& $0$& $0$& $0$& $0$& $0$& $0$\tabularnewline 
    $\momega^2$ & $0$ & $\boldsymbol{2}+0$ & $\boldsymbol{0}+2$ & $\boldsymbol{2}+0$ & $\boldsymbol{0}+2$ & $\boldsymbol{2}+0$ & $\boldsymbol{0}+2$ & $\boldsymbol{2}+0$\tabularnewline 
    $\momega^3$ & $\boldsymbol{1}+2$ & $0$& $0$& $0$& $0$& $0$& $0$& $0$\tabularnewline 
    $\momega^4$ & $0$ &  $\boldsymbol{0}+1$ & $\boldsymbol{1}+0$& $\boldsymbol{0}+1$& $\boldsymbol{1}+0$& $\boldsymbol{0}+1$& $\boldsymbol{1}+0$& $\boldsymbol{0}+1$ \tabularnewline 
    $\momega^5$ & $\boldsymbol{2}+2$ &  $0$ & $0$& $0$& $0$& $0$& $0$& $0$ \tabularnewline 
    $\momega^6$ & $0$ &  $0$ & $0$& $0$& $0$& $0$& $0$& $0$ \tabularnewline 
\end{tabular}
    \caption{On-shell closed/gauge-invariant currents in $4d$ Higher Spin Gravity to the lowest order. The sum in each cell gives the number of independent currents that are of $\momega^m \mC^n$-type, i.e., involve  $m$ one-forms $\momega$ and $n$ zero-forms $\mC$, cf. Table \ref{table2}. The (first) boldface number is the number of currents that do not vanish upon projection to the physical sector.  }
    \label{tab:table4d}
\end{table}

\paragraph{$\boldsymbol{H(\mathcal{L}, \mathcal{T}^{1,0})}$.} We will also discuss (obstructions to) the extension of observables to higher orders. Before doing that it is important to understand what are the options to deform the free equations \eqref{freeeqAAAA}. Since either  factor  in $\mathfrak{A}=\mathcal{A}\otimes \mathcal{A}$ admits a nontrivial deformation, it is not surprising that the there are two independent cubic vertices\footnote{These vertices were found in \cite{Vasiliev:1988sa}, but the proof of uniqueness was not given there. The uniqueness is important in view of the $3d$ bosonization duality conjecture. Had we found more nontrivial deformations, Chern--Simons Matter theories would have had more free parameters. The deformations can be extracted from Table \ref{table3} of Appendix \ref{app:B}.}
\begin{align}\label{deformed4d}
    d\momega&= \momega\star \momega +g_1\phi(\momega, \momega)\star \mC+g_2\bar\phi(\momega, \momega)\star \mC+\cdots\,.
\end{align}
The notation is that the  cocycle $\phi$ entering in $\mathcal{V}_3$ acts on $y$'s, while the Weyl--Moyal star-product is taken over $\bry$. More explicitly,
\begin{align}
    \mathcal{V}_3(a\otimes \bar a, b\otimes \bar b, c\otimes \bar c)&= \phi(a,b)\star c\otimes \bar a \star \bar b \star \bar c\,,
\end{align}
where $a,b,c$ depend on $y$ and $\bar a,\bar b,\bar c$ depend on $\bry$. Both the vertices descend to the physical sector and give $\mathcal{V}_3=\phi(\omega,\omega)\star \pi(C)$ and likewise for $\bar{\mathcal{V}}_3$. There are two independent coupling constants $g_{1,2}$. Unless we need one of the two deformations, $\phi$ or $\bar\phi$, we will assume that $\Phi=g_1\phi+g_2\bar\phi$ is the most general one.

The $4d$ case should be confronted with the $3d$ one. Here, the higher spin algebra does not have any deformations. Only its $\mathbb{Z}_2$-extension does; the deformation happens along any of the $k$ or $\brk$ directions, that is,  along $C$ in the $\pi$-twisted representation. Therefore, we do not include bare $\phi(\momega,\momega)$ in \eqref{deformed4d} since its projection onto the physical sector vanishes.

The cocycles $\phi$ and $\phi_C$, being inherited from $A_1$, can always be made to respect the $sp_2$-subalgebra. Identifying the Lorentz algebra with $sl_2(\mathbb{C})$,  one can realize it  by the complex conjugate generators $y_\ga y_\gb$ and $y_\gad y_\gbd$. In particular, the spin-connection is represented by its self-dual $\varpi^{\ga\gb}$ and anti-self-dual $\varpi^{\gad\gbd}$ parts. The Hochschild two-cocycle $\phi$ can be adjusted such that $\phi(\varpi,\omega)+\phi(\omega,\varpi)=0$ whenever $\varpi\in sp_2$ and $\omega$ is arbitrary. This guarantees that the equations are manifestly Lorentz-covariant, i.e., the spin-connection appears only inside the Lorentz-covariant differential $D= d-[\varpi+\bar\varpi,-]_\star$ and its curvature two-form $R^{\ga\gb}=d \varpi ^{\ga\gb}-\varpi\fud{\ga}{\gc} \wedge \varpi^{\gc\gb}$, \textit{idem.} $\bar R^{\gad\gbd}$. This is a form of the Equivalence Principle.

\noindent Now we explain the multiplicities in the table with concrete expressions. 

\paragraph{Class $\boldsymbol{C^n}$, holographic correlation functions. } This is the simplest class of invariants and the only class of gauge-invariant functionals that are strictly gauge-invariant:
\begin{align}
    I_{0,n}&=\Tr[\mC^{n}]\equiv\Tr[\overbrace{\mC\star\cdots\star \mC}^n]\,.
\end{align}
Therefore, these are the only observables that do not have to be integrated. $I_{0,n}$ are of physical significance since, for appropriately chosen wave-functions $C$, they compute the correlation functions of the single-trace operators in the free CFT duals (these are the free $3d$ scalar or free $3d$ fermion) \cite{Colombo:2012jx,Didenko:2012tv,Didenko:2013bj,Bonezzi:2017vha,Neiman:2017mel,Neiman:2018ufb}.\footnote{The correlation functions of the higher spin currents ($J_s$, $s>0$) in the critical vector model and in the Gross--Neveu model are given by the same formulas in the large-$N$ limit. Those with insertions of the scalar operator $J_0$ differ structurally between free and critical models. For three-point functions those with two and three scalar operators $J_0$ are fixed by the conformal symmetry up to a constant.} Invariants of this type were first proposed in \cite{Sezgin:2005pv} to distinguish between different solutions, were related to topological string theory in \cite{Engquist:2005yt} and were explored further in \cite{Colombo:2010fu,Sezgin:2011hq,Colombo:2012jx}.

We note, however, that what is plugged in are zero-forms $\mC= \Pi K$ in the adjoint representation that are obtained from those in the twisted-adjoint as $K=C\star \delta$, where $\delta$ is the star-product delta-function. This transformation between the adjoint and the twisted-adjoint representations takes the form of the Fourier transform and is  ill-defined in general. Nevertheless, the fact that it needs to be employed to compute the correlation functions and is well-defined for certain physical configurations indicates that there is no sharp difference between the two representations. Therefore, the above splitting into the physical and unphysical configurations, as in Table \ref{tab:table4d}, has to be treated with care.

Another important property of the invariants $I_{0,n}$ is that they do not depend on $x$. Indeed, $dI_{0,n}=0$ by construction. It is also instructive to see how the $x$-dependence gets washed away. In general, $\mC$ can be solved as $\mg^{-1}\star \mC_0\star \mg$, where $\momega=-\mg^{-1}\star d\mg$ and $C_0$ is an $x$-independent element of the (extended) higher spin algebra. Upon taking the trace the $x$-dependence of $I_{0,n}$, which is in $\mg$, gets erased. As a consequence, all $I_{0,n}$ are diffeomorphism invariant. This is an example of `gravitationally dressed' observables, see e.g. \cite{Donnelly:2015hta}. In this case, the dressing is via Wilson's line.

Being $x$-independent, the invariants $I_{0,n}$ are implicitly nonlocal. The relation to the correlation functions implies that $I_{0,n}$ can be represented as an integral of a certain $n$-point `contact vertex' in $AdS_{d+1}$ \cite{Bekaert:2015tva}. This is not in contradiction with $I_{0,n}$ being seemingly local as they are obtained by sewing several $\mC(x)$, all taken at the same point $x$. Indeed, $\mC$ is a generating function of auxiliary fields that are expressed by virtue of the equations of motion as derivatives (of unbounded order) of the higher-spin Weyl tensors. A knowledge  of all on-shell derivatives of the higher spin fields allows one to compute certain space-time integrals as star-products (and other functionals in the $Y$-space), i.e., on the higher spin algebra.   

The zero-form invariants are also very close to the vertices of the collective dipole action, see e.g. \cite{deMelloKoch:2018ivk}, which by definition gives the correlation functions of higher spin currents, see \cite{Neiman:2017mel,Neiman:2018ufb} for details. The collective dipole action should correspond to the sum of all $I_{0,n}$ with certain numerical prefactors. Other ideas to employ $I_{0,n}$ as building blocks of an action can be found e.g. in \cite{Boulanger:2012bj}.

The second raw in Table \ref{tab:table4d} shows that there are no obstructions to the extension of $I_{0,n}$ in the nonlinear theory. The deformed $I_{0,n}$ are expected to give correlation functions in  Chern--Simons Matter theories \cite{Sharapov:2018kjz}. This was conjectured in \cite{Sharapov:2018kjz}, while the new results prove the uniqueness of these invariants. Therefore, if the slightly-broken higher spin symmetry \cite{Maldacena:2012sf} is realized in  Chern--Simons Matter theories, then the deformed observables $I_{0,n}$ seem to provide a unique answer for the correlation functions.

\paragraph{Class $\boldsymbol{\omega^{2k+1}}$, Chern--Simons forms. } This class consists of the on-shell closed Chern--Simons forms
\begin{align}
    I_{2k+1,0}&=\Tr[\momega^{2k+1}]\equiv\Tr[\,\overbrace{\momega\star \cdots \star \momega}^{2k+1}\,]\,.
\end{align}
Clearly, $I_{1,0}=\Tr[\momega]$ is just the spin-one gauge potential $A_\mu\, dx^\mu\equiv \omega_\mu (y=0,\bry=0)\, dx^\mu$. 
This first member of the family does not survive the inclusion of interaction. Indeed, a simple computation with the help of \eqref{deformed4d} gives
\begin{align}
    d\Tr[\momega]&\approx  \Tr[\Phi(\momega,\momega)\star \mC]\,,
\end{align}
which is a nontrivial cocycle.\footnote{It is a nontrivial cocycle for the free equations of motion. Here, we used the first-order deformation of the free equations.} The next current $I_{3,0}$ is not obstructed at the first order
\begin{align}
    d\Tr[\momega\star\momega\star\momega +\Phi(\momega,\momega)\star \mC\star \momega+\mathcal{O}(\mC^2)]&\approx0\,,
\end{align}
where the `expansion parameter' is $\mC$; hence, we use \eqref{deformed4d} for $d\momega$ and \eqref{freeeqAAAA} for $d\mC$. However, it is obstructed at the second order for $g_1g_2\neq0$ in  \eqref{deformed4d}. The obstruction is of $\momega^4\mC^2$-type, which is clear from Table \ref{tab:table4d} and is discussed below.

The invariant $I_{3,0}$,  being a three-form, looks like a conventional conserved current. It is felt that $I_{3,0}$ can appear as a boundary term. The projection of $I_{3,0}$ onto the physical sector can be uplifted to an off-shell invariant, which is the Chern--Simons action for the higher spin algebra $\hs=A_2^e$: 
\begin{align}
    \frac{k}{4\pi}\int \mathrm{Tr}\left[\omega\wedge d\omega-\frac{2}3\omega\wedge \omega\wedge \omega\right]\,.
\end{align}
It gives the action of one of the $3d$ conformal higher spin gravities \cite{Pope:1989vj,Fradkin:1989xt,Grigoriev:2019xmp}. The higher Chern--Simons forms can lead to exotic higher spin theories discussed recently in \cite{Fuentealba:2019bgb}. The higher Chern-Simons forms are unobstructed.

\paragraph{Classes $\boldsymbol{\phi_C(\omega,\omega,\omega)}$ and $\boldsymbol{\bar\phi_C(\omega,\omega,\omega)}$. } In addition to $\Tr[\momega^3]$ there are two more three-forms. They can be constructed by applying the cyclic cocycle \eqref{fc} to either $\mathcal{A}$-factor of $\mathfrak{A}=\mathcal{A}\otimes \mathcal{A}$ and tracing over the rest of the variables:
\begin{align}
    J_{3,0}&=\phi_C(\momega,\momega,\momega)\,, && \bar J_{3,0}=\bar\phi_C(\momega,\momega,\momega)\,.
\end{align}
According to Table \ref{tab:table4d}, there is only one obstructing cocycle. Therefore, depending on the values of $g_{1,2}$ in \eqref{deformed4d} or which linear combination of $\phi_C$ and $\bar\phi_C$ is taken, we can face an obstruction. The obstruction  is of type $\phi\bar\phi$, see below. Therefore, at $g_2=0$ the cocycle $\phi_C$ is unobstructed (the interaction vertex contains only $\phi$), but at $g_1=0$ it is obstructed. We are lead to conclude that there is a linear combination of $J_{3,0}$ and $\bar J_{3,0}$ that survives switching on interaction. Together with $\Tr[\momega^3]$ we have three  candidates for global symmetry currents, but only $\Tr[\momega^3]$ projects to the physical sector. 

\paragraph{Classes $\boldsymbol{S^k\phi_C(\omega,\omega,\omega)}$ and $\boldsymbol{S^k\bar\phi_C(\omega,\omega,\omega)}$. } With the help of the $S$-operation on cyclic cohomology one can generate two families of type $\momega^{2k+3}$ by applying  $S$ to $\phi_C$ and $\bar\phi_C$. For example,
\begin{align}
     J_{5,0}&=S\phi_C[\momega]=3\phi_C(\momega^3,\momega,\momega)-\phi_C(\momega^2,\momega^2,\momega)\,,\\
    J_{7,0}&=S^2\phi_C[\momega]=2\phi_C(\momega^5,\momega,\momega)-\phi_C(\momega^3,\momega^2,\momega^2)\,,
\end{align}
and likewise for $\bar\phi_C$. They vanish upon the physical projection and are  unobstructed.

\paragraph{Classes $\boldsymbol{\phi_C(\omega,\omega,\mC^k)}$ and $\boldsymbol{\bar\phi_C(\omega,\omega,\mC^k)}$. } It is easy to see that the following families of two-forms are closed on-shell:
\begin{align}
    J_{2,k}&=\phi(\momega,\momega,\mC^k)\,, && \bar J_{2,k}=\bar\phi(\momega,\momega,\mC^k)\,.
\end{align}
They can also be rewritten in an equivalent, if not identical, way
\begin{align}
    &\Tr[ \phi(\momega,\momega)\star \mC^k]\,, && \Tr[\bar\phi(\momega,\momega)\star \mC^k]\,.
\end{align}
These are candidates for the role of surface currents and there are infinitely many of them. We do not expect those with $k>1$ to give a finite charge in general. All these currents are unobstructed and can easily be continued to higher orders. Those with $k$ odd do not vanish when projected onto the physical sector.  For example, the current $\Tr[ \bar\phi(\momega,\momega)\star \mC]$ gives $\Tr[ \bar\phi(\omega,\omega)\star \pi(C)]$. To the lowest order we obtain $h\fud{\ga}{\gad}\wedge h^{\gb\gad} C_{\ga\gb}$, which is the self-dual part of the Maxwell tensor $F_{\mu\nu}$ and likewise for the anti-self-dual part and $\phi$. See also \cite{Sezgin:2011hq,Vasiliev:2015mka} for other discussions and proposals of two-forms.

\paragraph{Class $\boldsymbol{\phi_C(\momega,\momega,\momega,\momega,\momega)}$.} The fourth Hochschild cohomology group $HH^4(A_2,A_2^\ast)$ of the second Weyl algebra $A_2$ is one-dimensional. This fact implies that the corresponding cyclic cohomology group $HC^4(A_2)$ is one-dimensional too. It is quite difficult to find an explicit expression for the Hochschild four-cocycle of $A_2$. It was first obtained in \cite{FFS} as a consequence of the Kontsevich--Shoikhet--Tsygan formality theorem. An alternative construction was proposed in our paper \cite{Sharapov:2017lxr}.  The cocycle can be represented as a function of four arguments in $A_2$ with values in $A_2$. The cocycle condition reads
\begin{equation}
\begin{array}{l}
    a\star \phi(b,c,d,e) -\phi(a\star b,c,d,e)+\phi(a, b\star c,d,e)+\\[3mm]
   -\phi(a, b,c\star d,e) +\phi(a, b,c,d\star e)-\phi(a,b,c,d)\star \Pi(e)=0\,.
   \end{array}
\end{equation}
Here, $\Pi[f(y,\bry)]=f(-y,-\bry)$, the effect that can also be achieved with $k\bar k$. The automorphism $\Pi\equiv \pi_1\pi_2$ takes into account the difference between $A_2$ and $A_2^*$ understood as $A_2$ bimodules. 

The Hochschild four-cocycle $\phi$ is known to be cohomologous to a cyclic cocycle $\phi_C$. 
The latter is defined by the conditions 
\begin{equation}
    \phi_C(a,b,c,d,e)=-\phi_C(e,a,b,c,d)\,,
\end{equation}
\begin{equation}
\begin{array}{l}
    \phi_C(a\star b,c,d,e,f)-\phi_C(a, b\star c,d,e,f)+\phi_C(a, b,c\star d,e,f)+\\[3mm]
    -\phi_C(a, b,c,d\star e,f)+\phi_C(a, b,c,d,e\star f)-\phi_C(f\star a,b,c,d,e)=0\,.
    \end{array}
\end{equation}
Since $A_2=A_1\otimes A_1$, the cyclic four-cocycle $\phi_C$ can be obtained as the cup product of the two-cocycles $\phi_C$ and $\bar\phi_C$, see Eq. (\ref{cup2prod}). However, this representative enjoys the $sp_2$ symmetry only. A manifestly $sp_4$-invariant cocycle can be written as a sum of integrals over $4$-, $3$-, and $2$-simplicies, cf. (\ref{2-1}). With the help of the cyclic four-cocycle we define the five-form
\begin{align}
    J_{5,0}&=\phi_C(\momega,\momega,\momega,\momega,\momega)\,,
\end{align}
which is on-shell closed and nontrivial. This form is a well-defined observable in the five-dimensional Chern--Simons theory based on $\hs=A_2^e$ whenever we consider solutions to the zero-curvature equation $F=0$ rather than $F\wedge F=0$. The five-form descends to the physical sector and is unobstructed.

\paragraph{Class $\boldsymbol{S^k\phi_C(\omega,\omega,\omega,\omega,\omega)}$}
Applying the $S$-operation to the four-cocycle above one can generate $\momega^{5+2k}$-forms. For example, applying $S$ once we get a nontrivial seven-form 
\begin{align}
    J_{7,0}&=5 \phi_C \left(\momega ^3,\momega ,\momega ,\momega ,\momega \right)+\phi_C \left(\momega ^2,\momega ,\momega ^2,\momega ,\momega \right)-3 \phi_C \left(\momega ^2,\momega ^2,\momega ,\momega ,\momega \right)\,.
\end{align}
All these forms survive the projection onto the physical subsector and are unobstructed. 

\paragraph{Class $\boldsymbol{\momega^4\, \mC}$.} This class is obtained by replacing one $\momega$ in the cyclic four-cocycle with $\mC$:
\begin{align}\label{jjj}
    J_{4,1}&=\phi_C(\momega,\momega,\momega,\momega,\mC)\,.
\end{align}
Its physical projection vanishes as $\mC$ has to be in the adjoint representation of the higher spin algebra. At the same time, this implies that there is an on-shell closed four-form that depends on a global symmetry parameter. In this case, there is no need to use a cyclic representative and we can 
replace (\ref{jjj}) with the equivalent current\footnote{In checking $dJ_{4,1}=0$ one should remember that $\Pi$ leaves $A_2^e$ invariant. Therefore, $\phi$ also determines a representative of $HH^4(A_2^e,A_2^e)$. }
\begin{align}
    J_{4,1}&=\Tr[K\star \phi(\omega,\omega,\omega,\omega)]\,.
\end{align}
 This four-form current does not have any physical interpretation at the moment.

There is a general property of FFS cocycles \cite{FFS}, which we have already used for the $A_1$ case. In the case being considered it says that  
\begin{align}\label{sp-b}
    \phi(\Omega,\omega,\omega,\omega)+\phi(\omega,\Omega,\omega,\omega)+\phi(\omega,\omega,\Omega,\omega)+\phi(\omega,\omega,\omega,\Omega)=0
\end{align}
whenever $\Omega$ belongs to the $sp_4$-subalgebra generated by $y_\alpha y_\beta$, $y_{\dot \alpha}y_{\dot \beta}$, and $y_\alpha y_{\dot\beta}$. As a consequence, $J_{4,1}=0$ on purely gravitational backgrounds. The flat connection $\Omega$ of $sp_4\sim so(3,2)$ describes the empty $AdS_4$ space provided that the vierbein component of $\Omega$ is nondegenerate. To make $J_{4,1}$ nontrivial $\omega$ is required to contain genuine higher spin components. Another comment is that similar to the zero-form invariants, one can map certain field configurations $C$ to the adjoint representation and still take the advantage of $J_{4,1}$. This class is unobstructed.

\paragraph{Class $\boldsymbol{\omega^4\, C^2}$.} This is obtained by inserting the product  $\mC\star \mC$ into the cyclic four-cocycle:
\begin{align}
    J_{4,2}&=\phi_C(\momega,\momega,\momega,\momega,\mC\star \mC)\,.
\end{align}
The importance of this current is that it is the first four-form that does not vanish on generic physical configurations. Upon projecting onto the physical sector it can also be written as
\begin{align}
    J_{4,2}&=\Tr[\phi(\omega,\omega,\omega,\omega)\star C\star \pi(C)]\,.
\end{align}
In view of the property (\ref{sp-b}), $J_{4,2}$ vanishes whenever $\omega$ is purely gravitational, i.e., for $\omega\in sp_4$. This four-form is unobstructed. 

\paragraph{Class $\boldsymbol{\omega^4\, C^k}$.} The two classes above are the first members of the family
\begin{align}
    J_{4,k}&=\phi_C(\momega,\momega,\momega,\momega,\mC^k)\,,
\end{align}
of which we make no special comments, except we do not expect those with $k>2$ to lead to convergent integrals in general. 

The expression for the cyclic four-cocycle $\phi_C$ in terms of its two-cocycle constituents is quite cumbersome. Nevertheless, there is a much simpler expression for $J_{4,k}$ in terms of the Hochschild two-cocycle $\phi$, which is equivalent to
\begin{align}
    J_{4,k}&=\Tr[\momega \star\phi(\momega, \boldsymbol{X})+\phi(\momega,\boldsymbol{X}\star \momega)]\,,
\end{align}
where the two-form $\boldsymbol{X}$ can be chosen simply as $\mC^n\star \phi(\momega,\momega)$ or $\mC^{n-k}\star \phi(\momega,\momega) \star \mC^k$, $k=0,\ldots ,n$.

A general comment about the invariants  $J_{4,k}$ that do not vanish identically on physical configurations is that they are natural building blocks for the on-shell action. In particular, we see that there is no analog of the cosmological term of $\omega^4$-type in $4d$ HSGRA. Also there is no tadpole of $\omega^4C$-type. All $J_{4,k}$ are unobstructed and extend to higher orders.

\paragraph{Final note.} It is worth stressing that all the invariants presented above form a multiplicative basis for the graded-commutative algebra of invariants under the exterior product of differential forms, see Proposition \ref{P1}. Therefore, we have even more invariants. 

Coming back to the deformation of the equations of motion \eqref{problemB} by interaction. According to our  classification the two-form currents $\Tr[\phi(\momega,\momega)\star \mC^k]$ with one $\mC$ removed determine all the possible deformations of the free equations. Those that remain nontrivial upon the physical projection are of the form  \begin{equation}\phi(\omega,\omega)\star \pi(C)\star (C\star \pi(C))^n\end{equation} and the same for $\bar\phi$. This interaction ambiguity was first observed in \cite{Vasiliev:1992av}. The terms with $n>0$ are similar to the zero-form invariants\footnote{The $y=\bry=0$ component of $(C\star \pi(C))^n$ is just $\Tr[(C\star \pi(C))^n]$; hence, it does not even depend on $x$, which is not what is expected from an interaction vertex.} and are too nonlocal, as interaction vertices, even for HSGRA \cite{Boulanger:2015ova,Skvortsov:2015lja}.

Lastly, after the complete classification of invariants is presented, it is worth commenting on which are the new ones. All invariants that contain the cyclic cocycles $\phi_C$, $\Phi_C$ are new and have never been mentioned in the literature. In particular, this leads to a new invariant of Chern--Simons theory that is based on the Weyl algebra. The zero-form invariants are definitely not new, but they do not {\it a priory} coincide with those in \cite{Sezgin:2005pv,Sezgin:2011hq},\footnote{We note that there is a doubly-infinite series of zero-form invariants introduced in \cite{Sezgin:2011hq}. The classification in the present paper reveals that not all of them are independent and/or reduce to finite zero-form invariants in terms of $C$. } except for the leading order where one or another prescription adopted in \cite{Colombo:2012jx,Bonezzi:2017vha} gives the same results. Surprisingly, what we call the Chern--Simons forms are also new, at least in this context. In addition, there are infinite series of invariants generated via the $S$-operation. All these are new as well. Nevertheless, there are other types of invariants proposed in \cite{Sezgin:2011hq}, which are not captured by the present analysis: off-shell invariants, (decorated) Wilson lines and observables that are invariant under the Lorentz subalgebra of the higher spin algebra. It would be interesting to obtain a complete classification of these classes as well.

\section{Discussion}
After the classification is presented, it is high time to glance over Table \ref{tab:table4d} and notice that we have many more observables than we expected to find for a theory of quantum gravity, see e.g. \cite{Harlow:2018jwu}. Surprisingly, the higher spin symmetry, being much bigger than the gauge symmetries in  Yang--Mills theory or GR, enjoys a wealth  of conserved $p$-form currents that the low spin theories do not normally have.  

For example, the surface charges are known to correspond to reducibility parameters, that is, those gauge parameters that leave the field configuration invariant on-shell, see e.g. \cite{Barnich:2005bn} in the higher spin context. The surface charges are of no surprise in free gauge theories, but it is hardly probable to find them in nonlinear models. In GR and YM, the surface charges can be defined only in the asymptotic region, where they correspond to the symmetries of the  asymptotically free background. However, most of the conserved $p$-form currents we found are not obstructed and their nonlinear completion does exist in the full theory. This might indicate that the nonlinear theory is still free in a sense. Indeed, this is just what AdS/CFT duality between HSGRA's and free CFT's suggests. Interacting higher spin fields, if not free in themselves, are dual to `mesons' $J_{i_1\cdots i_s}=\phi \pl_{i_1}\cdots\pl_{i_s}\phi+\cdots$ that are bilinear in the free `parton' fields $\phi$ of the corresponding CFT. 

This also reincarnates in the bulk as hidden integrability of any formal HSGRA. As was shown in \cite{Sharapov:2019vyd} all solutions to the nonlinear equations \eqref{problemB} can be obtained from a `free system' of the same form
\begin{align}\label{Lax}
    d\hat\momega&= \hat\momega\ast \hat\momega\,, & d\hat\mC&=\hat\momega\ast \hat\mC-\hat\mC\ast \hat\momega\,.
\end{align}
The only difference is that $\hat\momega$ and $\hat \mC$ take now  values in the deformed higher spin algebra \eqref{fullproduct} and depend on the deformation parameter $u$. There is an explicit map that  takes solutions of Eqs. \eqref{Lax} to those of Eqs. \eqref{problemB},  see also \cite{Prokushkin:1998bq}.  The existence of such a map is due to an inevitable nonlocality of HSGRA that goes far beyond what is allowed in conventional  field theory. The presence of infinitely many conserved surface charges also agrees with general conclusions of \cite{VASILIEV2017219}. 

Another comment is about local gauge invariant operators: there are plenty of such operators in YM and none in GR. In GR, however, we find infinitely many local scalars (they are scalars under diffeomorphisms) that are built out of the Riemann tensor and its covariant derivatives. Our results imply that there are no local `curvature invariants' in HSGRA. The only true scalars are $\Tr[\mC^n]$, but they are $(i)$ nonlocal and $(ii)$ do not actually depend on the point and thus are diffeomorphism invariant. The reason is that the diffeomorphisms are actually part of gauge transformations (\ref{gtr}).

It is possible to foresee the higher dimensional generalization of Table \ref{tab:table4d}. Generically, there is just one Hochschild two-cocycle $\phi$ on a higher spin algebra \cite{Sharapov:2018kjz}. Associated to it is a cyclic two-cocycle $\phi_C$. Together with the trace $\Tr$ they generate the following families: 
$$\Tr[\mC^k]\,,\qquad  \Tr[\momega^{2k+1}]\,,\qquad  S^k\phi_C(\momega^{3+2k})\,,\qquad  \phi_C(\momega,\momega, \mC^k)\,.$$ 
In addition, there should be a Hochschild cocycle $\psi$ of degree $d-2$ (in $4d$ this results in that we have $\phi$ and $\bar\phi$), which also gives rise to families similar to those originating from $\phi$, i.e., $S^k\psi(\momega^{d+2k-1})$ and $\psi(\momega^{d-2},\mC^k)$. The cup products $\phi\sqcup \cdots\sqcup \phi$ do not seem to generate  nontrivial classes. Nevertheless, we expect $\chi=\phi\sqcup \psi$ to give a nontrivial $d$-cocycle and the corresponding two families: $S^k\chi(\momega^{d+2k+1})$ and $\chi(\momega^{d},\mC^k)$.

Finally, all conventional  gauge theories do not have any characteristic cohomology in degree less than $d-2$ provided that gauge symmetries are irreducible \cite{Barnich:2000zw}. On the contrary in HSGRA we always have zero-form invariants and for $d>4$ there is an infinite family of two-form currents at the very least. Therefore, HSGRA's may give interesting examples of higher form symmetries \cite{Gaiotto:2014kfa}.

\section*{Acknowledgments}
\label{sec:Aknowledgements}
We are grateful to Xavier Bekaert, Maxim Grigoriev, Yasha Neiman, David de Filippi, Ergin Sezgin and Stefan Theisen for useful discussions. We also benefited a lot from illuminating discussions with Dennis Sullivan, Xiang Tang and Boris Tsygan on various mathematical aspects of the present work. The work of E.~Skvortsov was supported by the Russian Science Foundation grant 18-72-10123 in association with the Lebedev Physical Institute. The work of A.~Sharapov was supported by the Ministry of Science and Higher Education of the Russian Federation, Project No. 0721-2020-0033.

\appendix
\section{The essentials of Hochschild, cyclic, and Lie algebra cohomology}\label{app:A}
This appendix collects, in a highly condensed form, some basic facts on different cohomology theories for graded algebras. A systematic exposition of the material can be found in  \cite{Loday}, \cite{Co}, \cite{Feigin1987}.
For a quick introduction to the subject we refer the reader to \cite{Kassel}.  

Throughout, $k$ is a ground field of characteristic zero and $M$ is a bimodule  over a graded associative $k$-algebra $A$, that is,
$A=\bigoplus_{n\in \mathbb{Z}}A_n$ and $M=\bigoplus_{n\in \mathbb{Z}} M_n$ such that 
$$
A_n\cdot A_m\subset A_{n+m} \quad \mbox{and}\quad A_m\cdot M_n\subset M_{m+n}\supset M_n\cdot A_m\,.
$$
The degrees of homogeneous elements $a\in A_n$ and $m\in M_n$ will be denoted by $|a|=n$, $|m|=n$.   
Both the algebra and bimodule are supposed to be unital. All unadorned tensor products $\otimes$ and Hom's are taken over $k$. 
Given two graded algebras $A$ and $B$, we define the tensor product $A\otimes B$ to be a graded algebra with 
$$
(A\otimes B)_n=\bigoplus_{p+q=n}A_p\otimes B_q\quad\mbox{and}\quad (a\otimes b)\cdot (a'\otimes b')=(-1)^{|b||a'|}aa'\otimes bb'\,.
$$
In a similar way one can define the tensor product of graded bimodules over $A$ and $B$. 
Notice that the sign in the definition of the product above agrees with the {\it Koszul sign convention}:  if in manipulation with
monomial expressions involving graded objects an object $a$ jumps over an object $b$, then the
sign $(-1)^{|a||b|}$ appears in front of the expression. For example, if $f$ and $g$ is a pair of homomorphisms of graded modules, then $(f\otimes g)(a\otimes b)=(-1)^{|g||a|}f(a)\otimes g(b)$ and the dual bimodule $M^\ast=\mathrm{Hom}(M,k)$ is defined by the relation 
\begin{equation}
    (afb)(c)=(-1)^{|a|(|f|+|b|+|c|)}f(bca)\,, \qquad \forall f\in M^\ast \quad \forall a,b,c\in A\,.
\end{equation}
We follow this sign convention throughout the paper. Many formulae below are simplified significantly if one uses the shifted degree $\bar a=|a|-1$ that results from the desuspension $A\rightarrow A[-1]$, where $A[-1]_n=A_{n-1}$.

Finally, we will use $\mathbb{Z}_2$ to denote the cyclic group $\mathbb{Z}/2\mathbb{Z}$ and $Sp_{2n}(k)$ to denote the symplectic group $Sp(n,k)$ since this is 
notation prevalent in the physical literature.

\subsection{Hochschild cohomology} 

The Hochschild cohomology $HH^\bullet(A,M)$ of a graded algebra $A$  with coefficients in $M$ is the cohomology 
of the Hochschild cochain complex $C^\bullet(A,M)$:
$$
\xymatrix{0\ar[r]   &C^0\ar[r]^{\partial}& C^1\ar[r]^{\partial}& C^2\ar[r]^{\partial} &\cdots}
$$
with 
$$
C^p=\mathrm{Hom}(A^{\otimes p}, M)\,,\qquad A^{\otimes p}=\underbrace{A\otimes \cdots \otimes A}_p\,, 
$$
and the differential 
\begin{equation}\label{HD}
(\partial f)(a_1,\ldots, a_{p+1})=(-1)^{(\bar a_1+1)(\bar f+1)}a_1 f(a_2,\ldots,a_{p+1})  - (-1)^{\bar a_1+\cdots +\bar a_p}f(a_1,\ldots,a_{p})a_{p+1}
\end{equation}
$$
+\sum_{k=1}^{p}(-1)^{\bar a_1+\cdots +\bar a_k} f(a_1,\ldots,a_ka_{k+1},\ldots, a_{p+1})\,.
$$
The complex $C^
\bullet(A,M)$ contains a large subcomplex $\bar C^\bullet(A,M)$ of cochains that vanish whenever at least one of their arguments is equal to unit $e\in A$. The latter is called the {\it normalized Hochschild complex}. It is not hard to see that the inclusion map $i: \bar C(A,M)\rightarrow C(A,M)$ induces an isomorphism in cohomology. On restricting to normalized cochains, one concludes immediately that $HH^\bullet(k,M)\simeq HH^0(k,M)\simeq M$, where the ground field $k$ is regarded as a one-dimensional unital  $k$-algebra.   

Of particular interest are two special cases of bimodules: $M=A$ and $M=A^\ast$. The cohomology groups $HH^\bullet(A,A)$ arise naturally in deformation theory of the algebra $A$, while the groups $HH^\bullet(A,A^\ast)$ behave functorialy in $A$: each $k$-algebra homomorphism $h:A\rightarrow B$ indices a homomorphism in cohomology 
$h^\ast: HH^\bullet(B,B^\ast)\rightarrow HH^\bullet(A,A^\ast)$. 

Upon identification $C^p(A,A^\ast)\simeq\mathrm{Hom}(A^{\otimes(p+1)},k)$, formula  (\ref{HD}) takes the form   
\begin{equation}\label{HD1}
\begin{array}{rl}
(\partial g)(a_0, a_1,\ldots, a_{p+1})=&\displaystyle \sum_{k=0}^{p}(-1)^{\bar a_0+\cdots+\bar a_k} g(a_0, a_1,\ldots,a_ka_{k+1},\ldots, a_{p+1})\\[5mm]
+&(-1)^{(\bar a_0+1)(\bar a_1+\cdots +\bar a_{p+1})} g(a_1, \ldots, a_p,a_{p+1}a_0)\,,
\end{array}
\end{equation}
where $g(a_0,\ldots,a_{p-1},a_{p})=(-1)^{|a_{p}|}f(a_0,\ldots,a_{p-1})(a_{p})$.

\subsection{Cyclic cohomology}\label{CC}

The complex $ C^\bullet(A,A^\ast)$ contains a subcomplex $C_{\mathrm{cyc}}^\bullet(A)$ of {\it cyclic cochains}, i.e., 
cochains $g\in \mathrm{Hom}(A^{\otimes(p+1)},k)$ satisfying the symmetry condition 
\begin{equation}\label{cyclic}
g(a_0,a_1,\ldots, a_p )=(-1)^{\bar a_0(\bar a_1+\cdots+\bar a_p)}g(a_1,\ldots, a_{p},a_0)\,.
\end{equation}
The cohomology of the complex $C_{\mathrm{cyc}}^\bullet(A)$ is called the {\it cyclic cohomology} of $A$ and the corresponding cohomology groups are denoted by $HC^\bullet(A)$.  Upon restricting to cyclic cochains, one can bring the differential (\ref{HD1}) into a more familiar form\footnote{Notice that the signs in either form obey the Koszul rule if one shifts the degree of all $a$'s by $-1$, so that the dot product acquires degree $1$. Upon this interpretation the  dot product and cyclic permutation of $a$'s go first and the map $g$ after.}
\begin{equation}\label{CD}
\begin{array}{rl}
(\partial g)(a_0, a_1,\ldots, a_{p+1})=&\displaystyle \sum_{k=0}^{p}(-1)^{\bar a_0+\cdots+\bar a_k} g(a_0, a_1,\ldots,a_ka_{k+1},\ldots, a_{p+1})\\[5mm]
+&(-1)^{\bar a_{p+1}(\bar a_0+\cdots +\bar a_{p}+1)} g(a_{p+1}a_0, a_1, \ldots, a_p)\,.
\end{array}
\end{equation}

Let us consider the ground field $k$ as a one-dimensional algebra over itself. It follows from the cyclicity condition (\ref{cyclic}) that $C^{2n}_{\mathrm{cyc}}(k)\simeq k$ and $C^{2n+1}_{\mathrm{cyc}}(k)=0$. Hence, $HC_{\mathrm{cyc}}^{2n}(k)\simeq k$ and $HC_{\mathrm{cyc}}^{2n+1}(k)=0$.
On the other hand, $HH^\bullet(k,k^\ast)\simeq HH^0(k,k^\ast)\simeq k$. We thus conclude that $HH^\bullet(A,A^\ast)\neq HC^\bullet (A)$ in general. In particular, the cyclic subcomplex $C_{\mathrm{cyc}}^\bullet(A)$ is not a retract of $C^\bullet(A,A^\ast)$. 

There is a general construction to produce cyclic cocycles. It is based on Connes' notion of a cycle over an algebra  \cite[Ch. III]{Co}. 
Recall that each differential graded algebra $(\Omega, d)$
possesses  a canonical decreasing filtration $\Omega\supset F^1\Omega\supset F^2\Omega\supset\cdots $, where the differential subalgebra 
$F^p\Omega$ is given by the $p$-th power of the ideal generated by $d\Omega$. In other words,  the $k$-space $F^p\Omega$ is spanned by
elements of the form  $a=a_0da_1\cdots da_p$. 

Following Connes, we define a $p$-cycle to be a triple $(\Omega, d, \int)$ consisting of a differential graded algebra endowed with a closed trace $\int: \Omega\rightarrow k $ vanishing on $F^{p+1}\Omega$, that is,
$$
\int [a,b]=0\,,\qquad \int da=0\,,\qquad \int c=0\,, \qquad \forall a,b\in \Omega\,,\quad \forall c\in F^{p+1}\Omega\,. 
$$
By definition,  a {\it $p$-cycle over an algebra} $A$ is given by a $p$-cycle $(\Omega, d,\int)$ together with a homomorphism $\rho:A\rightarrow \Omega$ of graded algebras. Let us denote $\hat{a}=\rho(a)$ for all $a\in A$. 

Given a $p$-cycle over an algebra $A$, we define its {\it character} $\tau: A^{p+1}\rightarrow k$ by the rule
$$
\tau (a_0,a_1,\ldots, a_p)=\int \hat{a}_0d\hat{a}_1\cdots d\hat{a}_p\,.
$$
One can check that $\tau$ is a cyclic $p$-cocycle. 
When manipulating with cycles the following multiplication formulae are useful:
\begin{equation}
    \hat{a}(\hat{a}_0d\hat{a}_1\cdots d\hat{a}_p)=\widehat{aa}_0d\hat{a}_1\cdots d\hat{a}_p\,,
\end{equation}
$$
(\hat{a}_0d\hat{a}_1\cdots d\hat{a}_p)\hat{a}_{p+1}=\hat{a}_0d\hat{a}_1\cdots d\widehat{a_pa_{p+1}} -(-1)^{|a_p|}(\hat{a}_0d\hat{a}_1\cdots d\hat{a}_{p-1})\hat{a}_p d\hat{a}_{p+1}
$$
$$
=\sum_{k=1}^p (-1)^{|a_{k+1}|+\cdots+|a_p|+p-k} \hat{a}_0d\hat{a}_1\cdots d\widehat{a_{k}a_{k+1}}\cdots d\hat{a}_{p+1}+(-1)^{|a_1|+\cdots+|a_p|+p} \widehat{a_0a_1}d\hat{a}_2\cdots d\hat{a}_{p+1}\,.
$$
With the help of these relations one readily finds 
\begin{equation}
    (\partial \tau)(a_0,\ldots, a_{p+1})=(-1)^{\bar a_0+\cdots+ \bar a_p}\int [\hat a_0d\hat a_1\cdots d\hat a_p, \hat a_{p+1}]=0\,. 
\end{equation}
It can  be shown that each cyclic cocycle can be represented by a character of an appropriate cycle over $A$, see \cite[III.1.$\alpha$, Prop. 4]{Co}.

\subsection{Cohomology of Lie algebras}\label{LAC}
 Let $V=\bigoplus V^n$ be a $\mathbb{Z}$-graded module over a $\mathbb{Z}$-graded Lie algebra $L=\bigoplus L^n$.  We use the square brackets $[-,-]$ to denote both the Lie bracket in $L$ and the action (representation) of $L$ in $V$. The Chevalley--Eilenberg cochain complex consists of the sequence of groups $C^p(L,V)=\mathrm{Hom}(\Lambda^pL,V)$ endowed with a coboundary operator  $d: C^p(L,V)\rightarrow C^{p+1}(L,V)$. By definition, 
 \begin{equation}\label{transp}
     c(a_1,\ldots,a_k,a_{k+1},\ldots, a_{p})=(-1)^{\bar a_k\bar a_{k+1}}c(a_1,\ldots, a_{k+1}, a_k,\ldots,a_p)
 \end{equation}
 and
\begin{equation}\label{ChE}
\begin{array}{c}
\displaystyle (d c)(a_1, \ldots,  a_{p+1})=-\sum_{k=1}^{p+1}(-1)^{\epsilon_k }[c(a_1,\ldots,\hat{a}_k,\ldots,a_{p+1}), a_k]\\[6mm]
\displaystyle +\sum_{1\leq k<l\leq p+1}(-1)^{\epsilon_{kl}}c([a_k,a_l],  a_1, \ldots,\hat{a}_k,\ldots,\hat{a}_l,\ldots, a_{p+1})\,,
\end{array}
\end{equation}
where 
$$
\epsilon_k=\bar a_1+\cdots+{\bar a}_{k-1}+\bar a_{k+1}+\cdots+\bar a_{p+1}+\bar a_k(\bar a_{k+1}+\cdots +\bar a_{p+1})\,,
$$
$$
\epsilon_{kl}=\bar a_k+\bar a_k(\bar a_1+\cdots +\bar a_{k-1})+\bar a_l(\bar a_1+\cdots+\bar a_{k-1}+\bar a_{k+1}+\cdots + \bar a_{l-1})\,.
$$
The hats indicate omitting of the corresponding arguments.  For trivial coefficients $V=k$ the sum in the first line of (\ref{ChE}) is absent.  
By definition, the Lie algebra cohomology with coefficients in $V$ is the cohomology of the Chevalley--Eilenberg complex above.  The corresponding cohomology groups are denoted by $H^\bullet (L,V)$ or just $H^\bullet(L)$ for trivial coefficients.

\subsection{Cohomological operations}

\subsubsection{Cotrace map and Morita invariance} \label{cotrmap}
Let $\mathrm{Mat}_n(k)$ denote the algebra of $n\times n$-matrices over $k$. One may regard the tensor product $\mathrm{Mat}_n(A)=A\otimes \mathrm{Mat}_n(k)$ as the algebra of $n\times n$-matrices with  entries in $A$ and endow the $k$-space $\mathrm{Mat}_n(M)=M\otimes \mathrm{Mat}_n(k)$ with the natural  bimodule structure over $\mathrm{Mat}_n(A)$. Define the {\it cotrace map} 
\begin{equation}\label{cotr}
    \mathrm{cotr}: C^p (A, M) \rightarrow C^{p}(\mathrm{Mat}_n(A),\mathrm{Mat}_n(M))\,,\qquad p=0,1,\ldots,
\end{equation}
by the relation 
\begin{equation}\label{cotrace}
    \mathrm{cotr}(f)(a_1\otimes m_1,\ldots, a_p\otimes m_p)=f(a_1,\ldots, a_p)\otimes m_1\cdots m_p
\end{equation}
for $f\in C^p(A,M)$ and $a_i\otimes m_i\in A\otimes \mathrm{Mat}_n(k)$. It is stated that (\ref{cotr}) is a cochain map inducing an isomorphism 
\begin{equation}\label{cotrh}
    \mathrm{cotr}^\ast: HH^p(A,M)\rightarrow HH^{p}(\mathrm{Mat}_n(A),\mathrm{Mat}_n(M))\,.
\end{equation}
in Hochschild cohomology \cite[Sec. 1.5.6]{Loday}. In the case of cyclic $p$-cochains (\ref{cyclic}) the map (\ref{cotrace}) takes the form
\begin{equation}\label{cotrc}
    \mathrm{cotr}(g)(a_0\otimes m_0,\ldots, a_p\otimes m_p)=g(a_0,\ldots, a_p)\mathrm{tr}(m_0\cdots m_p)
\end{equation}
and gives rise to the isomorphism 
\begin{equation}\label{cotrch}
    \mathrm{cotr}^\ast: HC^p(A)\rightarrow HC^{p}(\mathrm{Mat}_n(A))
\end{equation}
of cyclic cohomology groups \cite[Sec. 2.4.6]{Loday}. 

The isomorphisms above are particular manifestations of the {\it Morita invariance} of Hoch\-schild and cyclic cohomology.  
By definition, two algebras $A$ and $B$ are said to be {\it Morita equivalent} ($A\sim B$) if there is an isomorphism between their categories of (left) modules.
In other words, there exist a $B$-$A$-bimodule $M$ and an $A$-$B$-bimodule $N$ such that 
$M\otimes_A N\simeq B$ and $N\otimes_B M\simeq A$.
For example, $A\sim \mathrm{Mat}_n(A)$ for any associative algebra $A$ and two commutative algebras are Morita equivalent iff they are isomorphic. Hence, the phenomenon of Morita equivalence is essentially noncommutative. It is known, see e.g. \cite[Sec. 1.2]{Loday}, that the Hochschild and cyclic cohomology functors are Morita invariant, that is, 
\begin{equation}
    HH^\bullet(A,A^\ast)\simeq HH^\bullet(B,B^\ast) \quad \mbox{and}\quad HC^\bullet (A)\simeq HC^\bullet (B)
\end{equation}
whenever $A\sim B$.

\subsubsection{\texorpdfstring{The Gerstenhaber algebra structure on $HH^\bullet (A,A)$}{The Gerstenhaber algebra structure on Hochschild cohomology}}\label{Gas}

Given a pair of cochains $f\in C^p(A,A)$ and $g\in C^q(A,A)$, we set 
\begin{equation}\label{cup}
    (f\cup g)(a_1,\ldots,a_{p+q})=(-1)^{(\bar g-1)(\bar a_1+\cdots +\bar a_p)}f(a_1,\ldots,a_p)g(a_{p+1},\ldots,a_{p+q})\,,
\end{equation}
\begin{equation}\label{1br}
    f\{ g\} = \sum_{k=0}^{p-1}(-1)^{\bar g(\bar a_1+\cdots+\bar a_k)} f(a_1,\ldots,a_k,g(a_{k+1},\ldots, a_{k+q}), a_{k+q+1},\ldots,  a_{p+q-1})\,,
\end{equation}
and 
\begin{equation}\label{g-com}
    [f,g]=f\{ g\} -(-1)^{\bar g\bar f}g\{f\}\,.
\end{equation}
The first operation (\ref{cup}), called the cup product, makes $C^\bullet (A,A)$ into a graded associative algebra, while the commutator (\ref{g-com}) equips the space $C^{\bullet}(A,A)$ with the structure of a graded Lie algebra. As is seen, the latter structure makes no use of the associative product on the $k$-space $A$. The product, however, gives rise to  a Maurer--Cartan element of  the graded Lie algebra. Setting
\begin{equation}
    m(a, b)=(-1)^{\bar a  }a b\,,\qquad \forall a,b\in A\,,
\end{equation}
one can easily see that $[m,m]=2m\{ m\}=0$. With the help of $m$ the Hochschild differential (\ref{HD}) for $M=A$ can be written as 
\begin{equation}
    \partial f=[f,m]\,,
\end{equation}
so that the condition $\partial^2=0$ becomes a simple consequence of the Jacobi identity. As another consequence, one finds  that the operator $\partial$
differentiates the Lie bracket (\ref{g-com}). A straightforward calculation shows 
that it is also a (right) derivation of the cup product:  
\begin{equation}
    \partial (f\cup g)= f\cup \partial g+(-1)^{|g|}\partial f\cup g\,.
\end{equation}
As a result both the multiplication operations pass through cohomology. Furthermore, at the level of cohomology, the cup product appears to be graded commutative and compatible with the bracket in the sense of the graded  Poisson relation. This follows from the identities
\begin{equation}
    f\cup g-(-1)^{|f||g|}g\cup f= -(\partial f)\{ g\}-(-1)^{| g|}\partial (f\{g\})+(-1)^{|g|}f\{\partial g\}\,,
\end{equation}

\begin{equation}
\begin{array}{l}
    [f, g\cup h]-[f, g]\cup  h-(-1)^{\bar f|g|} g\cup[f, h]\\[5mm]
    =-\partial (f)\{g,h\}\pm\partial (f\{g,h\}) \pm f\{\partial g,h\} \pm f\{g,\partial h\} \,.
\end{array}
\end{equation}
The r.h.s. of the second identity involves a higher cohomological operation, called a two-brace, whose definition can be found in \cite{GV}.  
 In such a way the space $HH^\bullet(A,A)$ is endowed with the structure of a {\it Gerstenhaber algebra} \cite{Gerst}.

\subsubsection{Operations in Lie algebra cohomology}\label{Lie-op}
There are various multiplicative structures on the Lie algebra cohomology depending on coefficients. For instance, the natural multiplication of scalar cochains $C^\bullet(L,k)$ as exterior forms on  $L^{\bullet}$ makes  $C^\bullet(L,k)$  a differential graded algebra with the same differential (\ref{ChE}). As a result, the cohomology space $H^{\bullet}(L)$ acquires the structure of graded commutative  algebra:
\begin{equation}\label{d-prod}
[c_1]\cdot [c_2]=[c_1\cdot c_2 ]=(-1)^{\bar c_1\bar c_2}[c_2\cdot  c_1]\,,
\end{equation}
$c_{i}$ being cocycles representing the cohomology classes $[c_{i}]$.

Denoting $H_+^\bullet (L)=\bigoplus_{p>0} H^p(L)$, we define the space of {\it indecomposable} elements of the algebra 
$H^{\bullet}(L)$
as the quotient  $\mathrm {Indec}\, H^{\bullet}(L)= H^\bullet (L)/H_+^\bullet (L)\cdot H_+^\bullet (L)$.

Another type of multiplication arises in the case of cohomology with coefficients in the adjoint representation.
By analogy with (\ref{1br}) we put 
\begin{equation}\label{compprod}
    c_1\langle c_2\rangle (a_1,\ldots, a_{q+p})=\sum_{_{\mbox{\tiny $ij$-unshuffles}}} (-1)^{\epsilon_{ij}} c_1(c_2(a_{i_1},\ldots, a_{i_q}), a_{j_1},\ldots, a_{j_p})\,.
\end{equation}
Here the sum is taken over all unshuffle sequences 
$$
i_1<i_2<\cdots<i_q\,,\quad j_1<j_2<\cdots< j_p\,,
$$
of $\{1,2,\ldots, q+p\}$ and $(-1)^{\epsilon_{ij}}$ is the Koszul sign of the permutation 
$$
(a_1,\ldots, a_{q+p})\mapsto (a_{i_1},\ldots, a_{i_q},a_{j_1},\ldots, a_{j_p})\,,
$$
where the degree of each homogeneous element $a_i\in A$ is shifted by  $-1$, cf. (\ref{transp}). Then the bracket 
\begin{equation}
    [c_1,c_2]=c_1\langle c_2\rangle -(-1)^{\bar c_1\bar c_2}c_2\langle c_1\rangle 
\end{equation}
gives  $C^{\bullet}(L,L)$ the structure of  a differential  graded Lie algebra with the differential (\ref{ChE}). Passing to cohomology, one can speak of the graded Lie algebra of cohomology  $H^\bullet(L,L)$.   

Using the above formulae for the dot (\ref{d-prod}) and composition  (\ref{compprod}) products and taking the algebra of functions and vector fields on a graded manifold as a model, one can easily guess how to endow  $H^\bullet(L,L)$ with the structure of $H^\bullet(L)$-module and turn $H^\bullet(L)$ into a module over the graded  Lie algebra $H^\bullet(L,L)$.

Observe that  multiplication  (\ref{d-prod}) makes sense  for $[c_1]\in H^\bullet (L)$ and $[c_2]\in H^{\bullet}(L,V)$, since  the $L$-module $V$ is simultaneously a  $k$-vector space. This allows one to regard $H^{\bullet}(L,V)$ as a (left) module over $ H^\bullet (L)$
and to identify  the space of {\it indecomposable} elements of $H^{\bullet}(L,V)$ with the quotient 
$\mathrm{Indec}\, H^{\bullet}(L,V)=H^{\bullet}(L,V)/
H_+^{\bullet}(L)\cdot H^{\bullet}(L,V)$.

\subsubsection{Cup products in Hochschild and cyclic cohomology}\label{cpcc} These cup products\footnote{Do not mix with the cup product of \ref{Gas}.} relate the cohomology groups of algebras $A$ and $B$ with those of the tensor product $A\otimes B$.   
The case of Hochschild cohomology appears to be simpler than the cyclic one as it is possible to define the desired product at the level of cochains. Let $M$ and $N$ be bimodules over algebras $A$ and $B$, respectively. Then one puts 
\begin{equation}\label{cp}
\begin{array}{c}
    (f\sqcup g)(a_1\otimes b_1,\ldots,a_{q+p}\otimes b_{q+p})\\[5mm]
    =(-1)^\epsilon f(a_1,\ldots,a_q)a_{q+1}\cdots a_{q+p}\otimes b_1\cdots b_{q}g(b_{q+1},\ldots, b_{q+p})
    \end{array}
\end{equation}
for any $f\in C^{q}(A,M)$ and $g\in C^{p}(B,N)$. Here $(-1)^\epsilon$ is the Koszul sign resulting from permutations of $a$'s, $b$'s, and $g$. By construction, $f\sqcup g\in C^{q+p}(A\otimes B, M\otimes N)$. 
The cup product (\ref{cp}) is differentiated by the Hochschild coboundary operator (\ref{HD}) thereby inducing a product in cohomology:
\begin{equation}\label{cup-hoh}
    \sqcup : HH^q(A,M)\otimes HH^p(B,N)\rightarrow HH^{q+p}(A\otimes B, M\otimes N)\,.
\end{equation}

By way of example let us consider the case wherein $B$ is the matrix algebra $\mathrm{Mat}_n(k)$ viewed as a bimodule over itself. Then $A\otimes B=\mathrm{Mat}_n(A)$
and $M\otimes N=\mathrm{Mat}_n(M)$. In view of the Morita equivalence $\mathrm{Mat}_n(k)\sim k$, we have
\begin{equation}
    HH^\bullet (\mathrm{Mat}_n(k),\mathrm{Mat}_n(k))\simeq HH^\bullet (k,k)\simeq HH^0(k,k)\simeq k\,.
\end{equation}
The group $HH^0 (\mathrm{Mat}_n(k),\mathrm{Mat}_n(k))$, being the center of the matrix algebra, is generated by the unit matrix $1\in \mathrm{Mat}_n(k)$.  
Then one can see that $f\sqcup 1=\mathrm{cotr}(f)$ for any $f\in C^{p}(A,M)$. Thus, the cotrace map (\ref{cotrh}) is induced by the cup product with the generator of 
the group $HH^\bullet (\mathrm{Mat}_n(k),\mathrm{Mat}_n(k))$. 

The simplest way to introduce the cup product for cyclic cohomology is through the notion of a cycle over an algebra, see Sec. \ref{CC}. Given a pair of cycles $\Omega$ and $\Omega'$ of dimensions $p$ and $p'$, one can define their tensor product $\Omega''=\Omega\otimes \Omega'$, which is a cycle of dimension $p+p'$. As a differential graded algebra $\Omega''$ is given by the tensor product of the algebras $\Omega$ and $\Omega'$ and the closed trace is defined by  
$$
\int \omega \otimes\omega'=(-1)^{|\omega||\omega'|}\int \omega\int\omega'\,,\qquad \forall \omega\in \Omega, \quad \forall \omega'\in \Omega'\,.
$$
This definition naturally extends to the tensor product of cycles over algebras inducing  
a cup product in cyclic cohomology:
\begin{equation}\label{ccup}
\sqcup: HC^q(A)\otimes HC^p(B)\rightarrow HC^{p+q}(A\otimes B)\,.
\end{equation}
Specifically, the cup product of cyclic cocycles is defined as the product of their characters: If 
$$
\phi (a_0, a_1, \ldots, a_q)=\int \hat{a}_0d\hat{a}_1\cdots d\hat{a}_q\,,\qquad
\psi (a_0, a_1, \ldots, a_p)=\int \hat{a}_0d\hat{a}_1\cdots d\hat{a}_p\,,
$$
then 
\begin{equation}\label{genform}
(\phi\sqcup \psi)(a_0\otimes b_0,\ldots, a_{p+q}\otimes a_{p+q})=\int (\hat{a}_0\otimes \hat{b}_0) d(\hat{a}_1\otimes\hat{b}_1)\cdots d(\hat{a}_{p+q} \otimes\hat{b}_{p+q})\,.
\end{equation}
Consider for example the cup product of  $1$-cocycles $\phi$ and $\psi$ reresented by the characters
$$
\phi(a_0,a_1)=\int a_0da_1\,,\qquad \psi(b_0,b_1)=\int b_0db_1\,.
$$
(To simplify expressions we omit hats.) One can find 
$$
\begin{array}{l}
\displaystyle (\phi\sqcup\psi)(a_0\otimes b_0, a_1\otimes b_1, a_2\otimes b_2)=\int (a_0\otimes b_0) d( a_1\otimes b_1) d(a_2\otimes b_2)\\[5mm]
\displaystyle=\int (a_0\otimes b_0) (d a_1\otimes b_1+(-1)^{|a_1|}a_1\otimes db_1) (da_2\otimes b_2+(-1)^{|a_2|}a_2\otimes db_2)\\[5mm]
\displaystyle=\int\Big ((-1)^{|b_0|(|a_1|+|a_2|+1)+|a_2|(|b_1|+1)}a_0da_1a_2\otimes b_0b_1db_2\\[5mm]
\displaystyle+(-1)^{|a_1|(|b_0|+1)+(|a_2|+1)(|b_0|+|b_1|+1)}a_0a_1da_2\otimes b_0db_1b_2\Big)\\[5mm]
\displaystyle=\int\Big ((-1)^{|b_0|(|a_1|+|a_2|+1)+|a_2|(|a_0|+|a_1|+|b_1|)}a_2a_0da_1\otimes b_0b_1db_2\\[5mm]
\displaystyle+(-1)^{|a_1|(|b_0|+1)+(|a_2|+|b_2|+1)(|b_0|+|b_1|+1)}a_0a_1da_2\otimes b_2b_0db_1\Big)\\[5mm]
\displaystyle=(-1)^{\epsilon_1}\int a_2a_0da_1\int b_0b_1db_2+(-1)^{\epsilon_2}\int a_0a_1da_2\int b_2b_0db_1\\[5mm]
\displaystyle=(-1)^{\epsilon_1}\phi(a_2a_0,a_1)\psi(b_0b_1,b_2)+(-1)^{\epsilon_2}\phi(a_0a_1,a_2)\psi(b_2b_0,b_1)\,,
\end{array}
$$
where 
$$
\begin{array}{l}
\epsilon_1={(|b_1|+|b_2|+1)(|a_0|+|a_1|+|a_2|+1)+|a_2|(|a_0|+|a_1|+|b_1|)+|a_0||b_0|}\,,\\[5mm]
\epsilon_2=|b_2|(|a_0|+|a_1|+|a_2|+|b_0|+|b_1|)+|a_0|(|b_0|+|b_1|+1)+|a_1||b_1|\,.
\end{array}
$$
Note that it is impossible to define the cup product at the level of cyclic cochains. Cyclicity takes place only for cocycles. 

For reference  we also present the cup product of two cyclic $2$-cocycles 
$$
\phi(a_0,a_1,a_2)\,,\qquad \psi(b_0,b_1,b_2)
$$
for the case of nongraded algebras $A=A_0$ and $B=B_0$. Applying the general formula (\ref{genform}), one can find 
\begin{equation}\label{cup2prod}
(\phi\sqcup\psi)(a_0\otimes b_0,a_1\otimes b_1,a_2\otimes b_2,a_3\otimes b_3,a_4\otimes b_4)
\end{equation}
$$
\begin{array}{c}
  = \phi(a_3a_4a_0,a_1,a_2)\psi(b_0b_1b_2,b_3,b_4)+\phi(a_4a_0a_1,a_2,a_3)\psi(b_1b_2b_3,b_4,b_0)    \\[3mm]
     +\phi(a_1a_2a_3,a_4,a_0)\psi(b_4b_0b_1,b_2,b_3)+\phi(a_0a_1a_2,a_3,a_4)\psi(b_3b_4b_0,b_1,b_2)\\[3mm]
     -\phi(a_4a_0a_1,a_2,a_3)\psi(b_0b_1b_2,b_3,b_4)-\phi(a_0a_1a_2,a_3,a_4)\psi(b_4b_0b_1,b_2,b_3)\\[3mm]
     +\phi(a_0a_1a_2,a_3,a_4)\psi(b_4b_0,b_1b_2,b_3)+\phi(a_4a_0,a_1a_2,a_3)\psi(b_0b_1b_2,b_3,b_4)\\[3mm]
     -\phi(a_4a_0,a_1a_2,a_3)\psi(b_0b_1,b_2b_3,b_4)-\phi(a_0a_1,a_2a_3,a_4)\psi(b_4b_0,b_1b_2,b_3)\,.
\end{array}
$$

As one more example, consider the group $HC^0(\mathrm{Mat}_n(k))\simeq HC^0(k)\simeq k$. The group is clearly generated by the  matrix trace $\mathrm{tr}: \mathrm{Mat}_n(k)\rightarrow k$. Then $A\otimes \mathrm{Mat}_n(k)=\mathrm{Mat}_n(A)$ and one can check that $g\sqcup \mathrm{tr}=\mathrm{cotr}(g)$ for any $g\in C^p(A)$, where the cotrace map is defined by (\ref{cotrc}). 

Suppose now that $A$ is a bialgebra. Then the comultiplication  $\Delta: A\rightarrow A\otimes A$, being  a $k$-algebra homomorphism, induces the map $$\Delta^\ast: HC^\bullet (A\otimes A)\rightarrow HC^\bullet (A)$$ 
(recall that  $HC^\bullet (-)$ is a contravariant functor of algebra). Composing this map with the cup product in cyclic cohomology (\ref{ccup}), we get a new product 
\begin{equation}\label{dc}
    \Delta^\ast\circ \sqcup: HC^\bullet(A)\otimes HC^\bullet(A)\rightarrow HC^\bullet (A)
\end{equation}
that makes $HC^\bullet(A)$ into a graded associative algebra. 
\subsubsection{Periodicity map}\label{PM}

As was mentioned in Sec. \ref{CC},  $HC^{2n}(k)=k$. Let $\sigma\in HC^2(k)$ be a generator in degree $2$ and let $(\Omega,d,\int)$  be a cycle over $k=ke$ such that
$$
1=\sigma(e,e,e)=\int edede\,.
$$

Note that for each $k$-algebra $A$, there is the isomorphism $A\otimes k\simeq A$. Using this isomorphism and the generator $\sigma\in HC^2(k)$, we can define a homomorphism 
\begin{equation}\label{per}
S:HC^{p}(A)\rightarrow HC^{p+2}(A)
\end{equation}
by setting
\begin{equation}\label{u=S}
Sf=\sigma\sqcup f=f\sqcup\sigma\,,\qquad \forall f\in HC^{p}(A)\,.
\end{equation}
The homomorphism $S$ of degree $2$ is called the {\it periodicity map}. For example, applying $S$ to a $1$-cocycle $\phi$, we get
$$
(S\phi)(a_0,a_1,a_2,a_3)=\int (a_0\otimes e)d(a_1\otimes e)d(a_2\otimes e)d(a_3\otimes e)
$$
$$
=\int (a_0\otimes e)(da_1\otimes e+(-1)^{|a_1|}a_1\otimes de)(da_2\otimes e+(-1)^{|a_2|}a_2\otimes de)(da_3\otimes e+(-1)^{|a_3|}a_3\otimes de)
$$
$$
=(-1)^{|a_2|+|a_3|}\int (a_0\otimes e)(da_1\otimes e)(a_2\otimes de)(a_3\otimes de)
$$
$$
+(-1)^{|a_1|+|a_3|}\int (a_0\otimes e)(a_1\otimes de)(da_2\otimes e)(a_3\otimes de)
$$
$$
+(-1)^{|a_1|+|a_2|}\int (a_0\otimes e)(a_1\otimes de)(a_2\otimes de)(da_3\otimes e)
$$
$$
=(-1)^{|a_2|}\int a_0da_1a_2a_3\otimes edede+(-1)^{|a_1|}\int a_0a_1a_2da_3\otimes ededee
$$
$$
=(-1)^{|a_3|+(|a_2|+|a_3|)(|a_0|+|a_1|)}\int a_2a_3da_1\int edede+(-1)^{|a_1|}\int a_0a_1a_2da_3\int edede
$$
$$
=(-1)^{|a_3|+(|a_2|+|a_3|)(|a_0|+|a_1|)}\phi(a_2a_3a_0,a_1)+(-1)^{|a_1|}\phi(a_0a_1a_2,a_3)\,.
$$
The above calculations are simplified if one uses the following identities, which hold for any idempotent $e$:
$$
de=de^2=dee+ede\quad \Rightarrow\qquad edee=0\,.
$$
Actually, treating $k=ke$ as a bialgebra, one can make $HC^\bullet (k)$ into a graded commutative algebra w.r.t. the product (\ref{dc}). 
This algebra appears to be isomorphic to $k[\sigma]$ with the generator $\sigma$ of degree $2$.  Eq. (\ref{u=S}) defines then the action of the algebra $HC^\bullet(k)$ on $HC^\bullet(A)$. This allows us to view $HC^\bullet(A)$ as a module over the graded commutative algebra $HC^\bullet(k)$.

\subsection{\texorpdfstring{Example: the cyclic cohomology of $\Lambda=k[\theta]$}{}}\label{E1}
By way of illustration let us consider a simple, yet important, example of a graded algebra. This is given
by the Grassmann algebra $\Lambda=k[\theta]$ with  the only generator  $\theta$ of degree $1$ subject to the relation $\theta^2=0$.        

Using the normalized Hochschild complex shows that $HH^n(\Lambda,\Lambda^\ast )\simeq k^2$ for all $n\geq 0$. Indeed, since the Hochschild differential vanishes identically upon normalization, one can define a basis of $n$-cocycles 
$\varphi_n^\pm\in \mathrm{Hom}(\Lambda^{\otimes(n+1)},k)$ by the relations
\begin{equation}
    \begin{array}{cc}
    \varphi^+_n(e,\theta,\ldots, \theta)=1 \,,&\qquad \varphi^+_n(\theta ,\theta ,\ldots, \theta )=0\,,\\[5mm]
    \varphi^-_n(e,\theta,\ldots, \theta)=0 \,,&\qquad \varphi^-_n(\theta ,\theta ,\ldots, \theta )=1\,,
    \end{array}
\end{equation}
$e$ being the unit of $\Lambda$. For the same reason, $HH^n(\Lambda,\Lambda)\simeq k^2$ and we can choose the basis $n$-cocycles $\psi^\pm_n\in \mathrm{Hom}(\Lambda^{\otimes n},\Lambda)$  to satisfy the normalization condition and 
\begin{equation}\label{psi-pm}
    \psi^+_n(\theta,\ldots, \theta)=e \,,\qquad \psi^-_n(\theta ,\ldots, \theta )=\theta\,.
\end{equation}
In particular, the equalities 
\begin{equation}
HH^0(\Lambda,\Lambda)=Z(\Lambda)=\Lambda\,,\qquad HH^1(\Lambda,\Lambda)=\mathrm{Der}(\Lambda)
\end{equation}
trivially follow from the graded commutativity of $\Lambda$. Notice that the odd derivation $\psi_1^+$ makes $\Lambda$ into a differential graded algebra  and 
\begin{equation}\label{pcp}
    \psi_n^+\cup\psi_m^+=\psi_{n+m}^+\,,\qquad  \psi_n^-\cup\psi_m^+=\psi_{n+m}^-\,,\qquad \psi^-_n\cup \psi_m^-=0\,,
\end{equation}
where the cup product $\cup$ is defined in Sec. \ref{Gas}. Geometrically, one may regard $\psi_n^\pm$ as $n$-vector fields on a graded manifold with the only (odd) coordinate $\theta$. Then the cup products (\ref{pcp}) become the exterior products of polyvector fields.

The cyclic cohomology of $\Lambda$ is also known \cite[Prop. 3.1.1]{FeiginTsygan}:
\begin{equation}
    HC^{2n}(\Lambda)\simeq k^2\,,\qquad HC^{2n+1}(\Lambda)\simeq k\,,\qquad n=0,1,2,\ldots
\end{equation}
Writing the elements of $\Lambda$ as $a=a^+e +a^-\theta$, where $a^\pm\in k$, we can choose the following basis cocycles:
\begin{equation}\label{lpm}
    \lambda^+_{2n}(a_0,a_1,\ldots,a_{2n})=a_0^+a_1^+\cdots a_{2n}^+\,,\qquad \lambda^-_n(a_0,a_1,\ldots,a_n)=a_0^-a_1^-\cdots a_n^-\,.
\end{equation}
The cyclicity condition (\ref{cyclic}) is obviously satisfied for $\lambda^\pm$'s.

Actually, $\Lambda$ is a bialgebra with the coproduct defined by $\Delta\theta= \theta\otimes e+e\otimes \theta$. 
As explained in Sec. \ref{cpcc}, the space  $HC^\bullet(\Lambda)$ is given the structure of a graded commutative algebra. 
Multiplying the generators (\ref{lpm}) according to (\ref{dc}), one can find \cite[Prop. 5.1.4]{Feigin1987}:
\begin{equation}
    \lambda^+_{2n}\lambda^+_{2m}=\lambda^+_{2m+2n}\,,\qquad \lambda^+_{2k}\lambda^-_m=0\,,\quad k>0\,,\qquad \lambda^-_m\lambda^-_n=0\,. 
\end{equation}
The element $\lambda^+_0$ plays the role of unit in $HC^\bullet(\Lambda)$. It is easy to see, that all $\lambda^-$'s are in the kernel of the homomorphism 
\begin{equation}
    HC^\bullet(\Lambda)\rightarrow HC^\bullet(k)
\end{equation}
induced by the inclusion $k \rightarrow\Lambda$. An earlier discussion of  Hochschild and cyclic  cohomology for Grassmann algebras can be found in \cite{Kast}, \cite{COQUEREAUX1995333}. 

\subsection{K\"unneth theorems  for Hochschild and cyclic cohomology}
Let $N$ be a bimodule over a $k$-algebra $B$ such that all the groups of Hochschild cohomology $HH^p(B,N^\ast)$ are finite-dimensional $k$-vector spaces. Then the cup product (\ref{cup-hoh}) defines a natural isomorphism  
\begin{equation}\label{KunHH}
 HH^n(A\otimes B,M\otimes N)\simeq \bigoplus_{p+q=n} HH^p(A,M)\otimes HH^q(B,N)    
\end{equation}
for any $A$-bimodule $M$. This statement is just the dual version of the K\"unneth formula  for Hochschild homology, see \cite[Ch. X, Thm. 7.4]{MacLane}. 

The cyclic analog of the isomorphism (\ref{KunHH}) is given by the exact sequence
\begin{equation}\label{KS}
 0\rightarrow  HC^\bullet(A)\bigotimes_{HC^\bullet(k)} HC^\bullet(B)\stackrel{\!\!\sqcup}{\rightarrow}  HC^{\bullet}(A\otimes B)\stackrel{\!\alpha}{\rightarrow}                                              \mathrm{Tor}^{HC^\bullet(k)}(HC^\bullet (A), HC^\bullet(B))\rightarrow 0
\end{equation}
under the assumption  that all groups $HC^p(B)$ are finite-dimensional, see \cite[Thm. 1]{Kassel1986}.  Here $\alpha$ is a homomorphism of degree $-1$.  In general, the cup product (\ref{ccup}) defines only an {\it injective} homomorphism from the tensor product of $HC^\bullet(k)$-modules. This homomorphism, however, becomes an isomorphism whenever either of the $HC^\bullet(k)$-modules is torsion free.

\subsection{Relations between various cohomology theories} \label{AR}
The fact that the cyclic complex $C^\bullet_{\mathrm{cyc}}(A)$ is a subcomplex of the Hochschild complex $C^\bullet (A,A^\ast)$ gives rise to the long exact 
sequence in cohomology  
$$
\xymatrix{\cdots\ar[r]& HH^p (A,A^\ast)\ar[r]^-B&HC^{p-1}(A)\ar[r]^-S&HC^{p+1}(A)\ar[r]^-I&HH^{p+1}(A,A^\ast)\ar[r]&\cdots}\,.
$$
The sequence involves the periodicity map (\ref{per}) and is known as Connes' Periodicity Exact Sequences. In many interesting cases it reduces the problem of computation of cyclic cohomology to that of  Hochschild cohomology. The map $I$ is induced by the inclusion $C^\bullet_{\mathrm{cyc}}(A)\rightarrow  C^\bullet (A,A^\ast)$, while the definition of $B$ is more complicated, see \cite{Co}. In low dimensions, Connes' exact sequence takes the form  
\begin{equation}\label{ld}
0 \rightarrow  HC^0\rightarrow HH^0\rightarrow 0\rightarrow  HC^1\rightarrow HH^1\rightarrow HC^0\rightarrow HC^2\rightarrow HH^2\rightarrow \cdots\,,
\end{equation}
hence, $HC^0(A)\simeq HH^0(A,A^\ast)$ (which is clear from the definition of cyclic cohomology).  

Suppose for instance that $HH^p(A,A^\ast)=0$ for all $p>n$. Then applying the exact sequence above, one readily concludes that the periodicity map $S: HC^{q}(A)\rightarrow HC^{q+2}(A)$ is an isomorphism for all $q\geq n$ and an epimorphism for $q=n-1$. If the whole Hochschild cohomology of $A$ is  concentrated in degree $n$, then the nontrivial groups of cyclic cohomology are  
\begin{equation}\label{C=H}
    HC^{n+2k}(A)\simeq HH^{n}(A,A^\ast)\,,\qquad k=0,1,2,\ldots \,.
\end{equation}

Given an associative algebra $A$, denote by $L(A)$ the associated Lie algebra with the Lie bracket given by the commutator in $A$. Any bimodule $M$ over $A$ turns naturally into a (left) module over $L(A)$ by setting $[a,m]=am-(-1)^{|m||a|}ma$ for $m\in M$ and $a\in L(A)$.  
Restricting a Hochschild $p$-cochain   $f: A^{\otimes p}\rightarrow M$
to the subspace of antisymmetric chains $\Lambda^p A\subset A[-1]^{\otimes p}$ gives then a cochain of the Chevalley--Eilenberg complex associated to the Lie algebra $L(A)$, see Sec. \ref{LAC} . Moreover, the restriction appears to be a cochain map, so that 
$$
(\partial f)(a_1\wedge a_2\wedge \ldots\wedge a_{p+1})=(df)(a_0\wedge a_1\wedge\ldots\wedge a_{p+1})
$$
for any $f\in C^p(A,M)$. As a result we have a homomorphism of cohomology groups 
\begin{equation}
    \varepsilon^\ast_p: HH^p(A,M)\rightarrow H^p (L(A),M)
\end{equation}
induced by the inclusion $\varepsilon_p: \Lambda^pA\rightarrow A[-1]^{\otimes p}$. This is known as an {\it antisymmetrization map} \cite[Sec. 1.3.4]{Loday}. 

Similarly, restricting the cyclic $p$-cochains onto the subspace $\Lambda ^{p+1}A\subset A[-1]^{\otimes (p+1)}$, one gets the cochain space $C^{p+1}(L(A))$. 
Again, the restriction map appears to be a morphism of complexes inducing a homomorphism in cohomology,
\begin{equation}
    \varepsilon^\ast_{p+1}: HC^{p}(A)\rightarrow H^{p+1}(L(A))\,.
\end{equation}
Notice that the antisymmetrization map is compatible with cyclicity:
$$
g(a_0\wedge a_1\wedge\ldots\wedge a_p)=(-1)^{\bar a_0(\bar a_1+\cdots+\bar a_p)}g(a_1\wedge \ldots\wedge a_p\wedge a_{0})\,,
$$
cf.(\ref{cyclic}). 

Recall that $\mathrm{Mat}_n(A)$ denotes the associative algebra of $n\times n$-matrices with entries in $A$. When equipped with matrix commutator it becomes a Lie algebra denoted  by $gl_n(A)$.  The same matrix commutator turns the $k$-space $\mathrm{Mat}_n(M)$ into an adjoint module of the Lie algebra $gl_n(A)$ for any $A$-bimodule $M$. 

Composing now the antisymmetrization map with the cotrace of Sec. \ref{cotrmap},  one can define a homomorphism from the Hochschild or cyclic cohomology of $A$ to the cohomology of the Lie algebra $gl_n(A)$:
\begin{equation}\label{adj}
   \phi^\ast= \varepsilon^\ast\circ \mathrm{cotr}^\ast: HH^\bullet (A,M)\rightarrow H^\bullet (gl_n(A), \mathrm{Mat}_n(M))\,,
\end{equation}
\begin{equation}\label{TLQ}
   \varphi^\ast=\varepsilon^\ast\circ \mathrm{cotr}^\ast: HC^{\bullet-1} (A)\rightarrow H^{\bullet} (gl_n(A))\,.
\end{equation}
At the level of cochains the maps are defined by the formulae
\begin{equation}
    \phi(g)(a_1\otimes m_1,\ldots, a_p\otimes m_p)=\sum_{\sigma\in S_p}(-1)^{\epsilon_\sigma}g(a_{\sigma(1)},\ldots, a_{\sigma(p)})\otimes m_{\sigma(1)}\cdots m_{\sigma(p)}\,,
\end{equation}
\begin{equation}
    \varphi(g)(a_0\otimes m_0,\ldots, a_p\otimes m_p)=\sum_{\sigma\in S_p}(-1)^{\epsilon_\sigma}g(a_0,a_{\sigma(1)},\ldots, a_{\sigma(p)})\mathrm{tr}(m_0m_{\sigma(1)}\cdots m_{\sigma(p)})\,.
\end{equation}
Here $(-1)^{\epsilon_\sigma}$ is the Koszul sign caused  by elementary transpositions (\ref{transp}) of arguments.  
(There is no need to antisymmetrise all $p+1$ arguments in the second relation due to cyclicity of $g$ and $\mathrm{tr}$.)

Among important applications of cyclic cohomology is computation of the cohomology of the Lie algebra  $gl(A)$ of `big matrices'. By definition, the algebra $gl(A)$ consists of infinite matrices with only finitely many entries different from zero. Formally, it is defined through the inductive limit $gl(A)=\lim\limits_{\rightarrow}gl_n(A)$ corresponding to the natural inclusions $gl_n(A)\subset gl_{n+1}(A)$ (an $n\times n$-matrix is augmented by zeros). 
A precise  relationship between the cohomology of the Lie algebra of matrices and cyclic  cohomology is established by the following Tsygan--Loday--Quillen theorem \cite{Tsygan_1983}, \cite{LQ}.  
\begin{theorem}\label{A1}
The image of the map (\ref{TLQ}) lies in the indecomposable part of the algebra $H^\bullet (gl_n(A))$ and induces an isomorphism
$$
HC^{p-1}(A)\simeq \mathrm{Indec}\, H^{p}(gl_n(A))
$$
for all $n\geq p$. As an exterior algebra, $H^\bullet(gl(A))$ is freely generated by the graded vector space $HC^{\bullet-1}(A)$. 
\end{theorem}
In other words, the cohomology group $H^p (gl_n(A))$ does not depend on the size of matrices provided it is large enough.  

There is also an analogue of the above theorem for cohomology with coefficients in the adjoint representation, see \cite[Sec. 10.4]{Loday}, \cite{Brodzki}. 
Denoting $\mathrm{Mat}(M)=\lim\limits_{\rightarrow}\mathrm{Mat}_n(M)$, we have 
\begin{theorem} \label{A2} For any $A$-bimodule $M$,
$$
H^\bullet(gl(A),\mathrm{ Mat}(M))\simeq HH^\bullet (A,M)\otimes H^\bullet(gl(A))\,.
$$
\end{theorem}
This theorem says that, viewed as an $H^\bullet(gl(A))$-module, $H^\bullet(gl(A), \mathrm{Mat}(M))$ is freely generated by the spaces of indecomposable elements 
\begin{equation}
\mathrm{Indec}\, H^p (gl(A),\mathrm{Mat}(M))\simeq HH^p (A,M)\,.
\end{equation}

\subsection{\texorpdfstring{Example: the cohomology of $gl(A\otimes \Lambda)$}{Example: cohomology of tensor product of A and Grassmann algebra}}

In studying invariants of a HSGRA associated with a given higher spin algebra $A$ we are interested in cohomology of the Lie algebra $gl(A)$ with coefficients in symmetrized tensor powers of the coadjoint module $gl(A)^\ast$. 
There is a  trick that allows one to incorporate the tensor module  into the structure of a Lie algebra reducing thus the problem to the case of trivial coefficients. The construction goes as follows. 

Let $\Lambda=k[\theta]$ be the Grassmann algebra of Sec. \ref{E1}. Tensoring it with $A$, we define the graded Lie algebra 
\begin{equation}\label{glt}
{gl}(A\otimes \Lambda)=\theta {gl}(A)+\hspace{-1em}\supset{gl}(A) \,.
\end{equation}
Here the commutative ideal $\theta gl(A)$ may be regarded as an adjoint module over the subalgebra $gl(A)$. The semi-direct sum
decomposition (\ref{glt}) leads to the obvious isomorphism 
\begin{equation}\label{HA}
    H^\bullet\big({gl}(A\otimes \Lambda)\big)\simeq H^\bullet\big({gl}(A),S^\bullet({gl}(A)^\ast)\big)
    \end{equation}
for the Lie algebra cohomology groups. Here $S^\bullet$ stands for the symmetric tensor powers of the coadjoint module.
Theorem \ref{A1} states now that the groups on the left form an exterior algebra freely generated by the graded vector space $HC^{\bullet} (A\otimes \Lambda)$.   As to the latter cohomology, we have the following 
\begin{theorem}\label{HCAL} For any $k$-algebra $A$
$$
    HC^p(A\otimes \Lambda)\simeq  HC^p(A)\oplus \left( \bigoplus_{n=0}^p HH^n(A,A^\ast)\right)\,.
$$
\end{theorem}
See  \cite[Prop. 5.1.4]{Feigin1987} and \cite{Kassel1986} for the proof.

Since the groups $HC^p(\Lambda)$ are finite-dimensional, one can apply the K\"unneth exact sequence (\ref{KS}) to establish the isomorphism above. 
Suppose the Hochschild cohomology of $A$ is concentrated in one degree, say $n$. 
Then the cyclic cohomology groups are given by (\ref{C=H}) and $HC^{n+2k}(A)=S^kHC^n(A)$. The periodicity map $S$ defines the action of the algebra $HC^\bullet(k)\simeq k[S]$ on $HC^\bullet(A)$. Clearly, this action gives $HC^\bullet(A)$ the structure of a {\it flat} module over $HC^\bullet(k)$. As a result,  $$\mathrm{Tor}^{HC^\bullet(k)}(HC^\bullet(A), HC^\bullet(\Lambda))=0$$ and the cup product  defines an isomorphism in the short exact sequence (\ref{KS}). 
Taking now the cup product of the elements of $HC^{n}(A)$ with the generators (\ref{lpm})  of the algebra $HC^\bullet(\Lambda)$, we find 
that $HC^p(A\otimes \Lambda)=0$ for all $p<n$ and 
\begin{equation}\label{hchc}
\begin{array}{l}
    HC^{n+2k}(A\otimes \Lambda)\simeq  HH^{n}(A,A^\ast)\oplus HH^{n}(A,A^\ast)\,,\\[5mm]
    HC^{n+2k+1}(A\otimes \Lambda)\simeq  HH^{n}(A,A^\ast)\,,\qquad k=0,1, 2,\ldots
    \end{array}
\end{equation}
This is in line with the statement of Theorem \ref{HCAL}. Therefore, $H^\bullet(gl(A\otimes \Lambda))$ is an exterior algebra freely generated by the graded spaces (\ref{hchc}) above.

\section{The cohomology of the Weyl algebra and its smash products}
\label{app:B}
\subsection{Polynomial Weyl algebras}\label{PWA}
The polynomial Weyl algebra $A_1$ over $k$ is a unital algebra on two generators $q$ and $p$ subject to the relation 
\begin{equation}
    qp-pq=e\,,
\end{equation}
$e$ being the unit of $A_1$. It is known to be a simple Noetherian domain  with  a $k$-basis  consisting of the ordered monomials $q^np^m$, see e.g. \cite{coutinho_1995}. 

The Hochschild cohomology of $A_1$ can easily be computed from the Koszul resolution:
\begin{equation}\label{K}
    \xymatrix{0\ar[r]& A_1\otimes W\otimes A_1\ar[r]^-{\partial_2}& A_1\otimes V\otimes A_1\ar[r]^-{\partial_1}& A_1\otimes A_1\ar[r]^-{m}&A_1\ar[r]&0}\,.
\end{equation}
Here $V$ is the two-dimensional $k$-vector space spanned by $u$ and $v$, $W$ is the one-dimensional space generated by $u\otimes v-v\otimes u$, and the differentials act as follows: 
$$
    m(a_1\otimes a_2)=a_1a_2\,,
    $$
    $$
\partial_1(a_1\otimes u\otimes a_2)=a_1q\otimes a_2-a_1\otimes qa_2\,,\qquad \partial_1(a_1\otimes v\otimes a_2)=a_1p\otimes a_2-a_1\otimes pa_2\,,
$$
$$
\partial_2\big(a_1\otimes (u\otimes v-v\otimes u)\otimes a_2\big)=a_1q\otimes v\otimes a_2-a_1\otimes v\otimes qa_2-a_1p\otimes u\otimes a_2+a_1\otimes u\otimes pa_2\,.
$$
Thus, the exact sequence (\ref{K}) provides us with a resolution of $A_1$ by free $A_1$-bimodules. 
If now $M$ is a unital $A_1$-bimodule, then applying the functor $\mathrm{Hom}_{A_1- A_1}(-,M)$ to the above sequence without the last term $A_1$ yields the complex
$$
\xymatrix{
0\ar[r]&M\ar[r]^-{\partial'_1}&\mathrm{Hom}(V,M)\ar[r]^-{\partial'_2}&\mathrm{Hom}(W,M)\ar[r]&0}
$$
computing the Hochschild cohomology groups $HH^\bullet (A_1, M)$. Here
\begin{equation}\label{dd}
    \partial'_1 m= (qm-mq, pm-mp)\,,\qquad \partial_2'(f(u),f(v))=q f(v)-f(v)q-pf(u)+f(u)p\,.
\end{equation}
As a result one concludes immediately that $HH^p(A_1,M)=0$ for all $p>2$ and any $A_1$-bimodule $M$. 

Let us take for instance $M=A_1$. Then the first equality in (\ref{dd}) identifies $HH^0(A_1,A_1)$ with the centre of $A_1$, which is given by $ke$. 
Introducing now the pair of $k$-linear operators 
\begin{equation}
\begin{array}{ll}
    h_1: A_1\oplus A_1\rightarrow A_1\,,&\qquad h_2: A_1\rightarrow A_1\oplus A_1\,,\\[3mm]
   h_1(q^np^m,q^lp^k)=\frac1{m+1}q^np^{m+1} \,,&\qquad h_2(q^np^m)=\big(0, \frac1{m+1}q^np^{m+1}\big)\,,
    \end{array}
\end{equation}
one can find 
\begin{equation}
    \partial_1'h_1+h_2\partial'_2=1\,,\qquad \partial'_2h_2=1\,.
\end{equation}
Hence, $HH^1(A_1,A_1)=HH^2(A_1,A_1)=0$ and we conclude that 
\begin{equation}\label{HHW}
    HH^p(A_1,A_1)=\left\{\begin{array}{ll} k&\; \mbox{if} \; p=0\,,\\0&\;\mbox{otherwise}\,.\end{array}\right.
\end{equation}
Vanishing of the second cohomology group $HH^2(A_1,A_1)$ means that the first Weyl algebra $A_1$ is rigid, i.e., admits no nontrivial deformation, while the equality $HH^1(A_1,A_1)=0$ says that all derivations of $A_1$ are inner.   

Taking the $n$-fold tensor product $A_1\otimes \cdots \otimes A_1$, one gets the $n$-th Weyl algebra $A_n$. This is generated by  $2n$ elements  $q^i$ and $p_j$ satisfying Heisenberg's  commutation relations 
\begin{equation}\label{qp}
    [q^i,q^j]=0\,,\qquad [p_i,p_j]=0\,,\qquad [q^i,p_j]=\delta^i_je\,.
\end{equation}
The groups (\ref{HHW}) being finite-dimensional, we can apply the K\"unneth theorem (\ref{KunHH}) to get\footnote{Alternatively, one can use the fact that the algebra $A_1$ is Noetherian and apply Theorem 3.1 of \cite[Ch. XI]{CartEil}. }
\begin{equation}\label{HH0}
    HH^\bullet(A_n,A_n)\simeq HH^0(A_n,A_n)\simeq k\,.
\end{equation}
Therefore, all the Weyl algebras $A_n$ are  rigid, have no outer derivations, and their centers are generated by the unit element.  The symmetrized quadratic monomials in $q$'s and $p$'s form
the Lie algebra $sp_{2n}(k)$ w.r.t.  the commutator. 

The Koszul resolution for $A_n$ is constructed in a way similar to above and can be used to establish the isomorphism 
\begin{equation}\label{dual}
HH^p(A_{n}, M)\simeq HH^{2n-p}(A_n, M^\ast)\,,
\end{equation}
which holds for any bimodule $M$. Actually, it is the composition of two isomorphisms:
$$HH_p(A_n, M)^\ast\simeq HH^p(A_n, M^\ast )\,,\qquad HH_{p}(A_{n}, M)\simeq HH^{2n-p}(A_n, M)\,,$$ 
see e.g. \cite[Sec. 4.1]{AFLS}.

\subsection{Skew group algebras and twisted bimodules}

Let $A$ be an associative $k$-algebra and $G\subset \mathrm{Aut}(A)$ a finite group of automorphisms of $A$. 
The {\it skew group algebra} $A\rtimes G$ (aka {\it smash product algebra}) is defined to be the $k$-vector space $A\otimes k[G]$ with multiplication 
\begin{equation}\label{smpr}
    (a_1\otimes g_1)(a_2\otimes g_2)=a_1a_2^{g_1}\otimes g_1g_2\,.
\end{equation}
Here $a^g$ denotes the result of action of $g\in G$ on $a\in A$. 

With any element $g\in G$ one can associate a {\it $g$-twisted bimodule} $Ag$ over $A$. As a $k$-vector space, $Ag$ is isomorphic to $A$ and the action of $A$ on $Ag$ is given by 
\begin{equation}
    a_1 (a) a_2= a_1aa_2^g\,.   
\end{equation}
As is seen, the left action is the usual action of $A$ on itself, while the right action is twisted by the automorphism $g$.

The following statement relates  the Hochschild cohomology of a skew group algebra $A\rtimes G$ with the $G$-invariant cohomology of the $g$-twisted bimodules $Ag$. 

\begin{theorem}\label{thmB1}
If a finite group  $G$ acts by automorphisms on an algebra $A$, then 
$$
HH^\bullet (A\rtimes G, A\rtimes G)\simeq \Big(  \bigoplus_{g\in G} HH^{\bullet}(A,Ag)\Big)^{G}\,.
$$
\end{theorem}
\noindent The poof can be found in \cite[Lemma 9.3]{Etingof}. 

We are going to apply the general constructions above to the $n$-th Weyl algebra $A_n$. The automorphism group of $A_n$ contains a subgroup $Sp_{2n}(k)$ acting by linear transformations on the $2n$-dimensional vector space $V$ spanned by the generators $q$'s and $p$'s. Clearly, the action of $Sp_{2n}(k)$ preserves the commutation relations (\ref{qp}) inducing thus automorphisms of the whole algebra $A_n$. Now let $G$ be a finite subgroup of $Sp_{2n}( k)$ and suppose the ground field $k$ to be algebraically closed, say $\mathbb{C}$. Then $g^{|G|}=1$ for any $g\in G$ and the action of $g$ is diagonalizable in $V$.  Denote by $2\mu_g$ the multiplicity of the eigenvalue $1$ of the operator $g: V\rightarrow V$. The groups of Hochschild cohomology $HH^\bullet(A_n,A_ng)$ were first computed by Alev, Farinati, Lambre and Solotar in \cite{AFLS} (see also \cite{Pinczon}).

\begin{theorem}[AFLS]\label{AFLS1} With the definitions and assumptions above 
$$
HH^\bullet (A_n,A_ng)\simeq HH^{2(n-\mu_g)}(A_n,A_ng)\simeq k\,.
$$
\end{theorem}
When $g=e$, this yields (\ref{HH0}). As is seen  the odd cohomology groups are all zero.  

Let us combine the last theorem with Theorem \ref{thmB1}. The elements of the direct sum $\bigoplus_{g\in G} HH^\bullet(A_n,A_ng)$ are represented by cocycles 
\begin{equation}
    \tau_\gamma=\sum_{g\in G}\gamma(g)\tau_g\,,
\end{equation}
where $\tau_g$ are basis cocycles\footnote{An explicit expression for $\tau_g$ can be found in \cite{Sharapov:2017lxr}.} for nontrivial groups $HH^{2\mu_g}(A_n,A_ng)\simeq k$. 
$G$-invariance implies that $(\tau_\gamma)^h=\tau_\gamma$ for all $h\in G$. On the other hand, it follows from the definition that $(\tau_g)^h=\tau_{hgh^{-1}}$ and $\mu_g=\mu_{hgh^{-1}}$. The latter equality means that the set of all element $g\in G$ with $2\mu_g=p$ is invariant under conjugation. Hence, the $G$-invariance condition amounts to $\gamma(hgh^{-1})=\gamma(g)$, that is, $\gamma: G\rightarrow k$ is a class function on $G$.   In such a way we arrive at the second AFLS theorem. 
\begin{theorem}[AFLS]\label{AFLS2}
Let $n_p(G)$ denote the number of conjugacy classes  of elements $g\in G $ with $2\mu_g=p$, then 
$$
\dim HH^{2n-p}(A_n\rtimes G, A_n\rtimes G)=\dim HH^{p}\big(A_n\rtimes G, (A_n\rtimes G)^\ast\big)=n_p(G)\,.
$$
\end{theorem}

By way of illustration, consider the involutive automorphism  $\varkappa\in \mathrm{Aut}(A_n)$ defined by 
\begin{equation}\label{k}
    (q^i)^\varkappa=-q^i\,,\qquad (p_i)^\varkappa=-p_i\,.
\end{equation}
This equips $A_n$ with the natural  $\mathbb{Z}_2$-grading.  Associated to this $\mathbb{Z}_2$-grading is the supertrace $\mathrm{Str}: A_n\rightarrow k$ defined as a unique homomorphism of unital $k$-algebras (projection on the unit). By definition, the supertrace vanishes on supercommutators, i.e., $\mathrm{Str}\big([a,b]_\varkappa\big)=0$, where 
\begin{equation}
    [a,b]_\varkappa=ab-ba^{\varkappa}\,,\qquad \forall a,b \in A_n\,.
\end{equation}
The associated bilinear form $\kappa(a,b)=\mathrm{Str}(ab)$ is known to be nondegenerate \cite{Pinczon2005}.  This allows one to identify the dual $A_n$-bimodule $A_{n}^\ast$ with the $\varkappa$-twisted bimodule $A_n\varkappa$. Clearly,  $2\mu_\varkappa=0$ and  applying Theorem \ref{AFLS1} yields 
\begin{equation}\label{HH2n}
    HH^\bullet(A_n, A_n^\ast)\simeq HH^{2n}(A_n,A_n^\ast)\simeq k\,.
\end{equation}
The coincidence of the cohomology  groups (\ref{HH0}) and (\ref{HH2n}) is a particular  manifestation of the duality (\ref{dual}).

\subsection{The Feigin--Felder--Shoikhet cocycle}

An explicit formula for a nontrivial $2n$-cocycle $\tau_{2n}$ generating the group (\ref{HH2n}) has been found  by Feigin, Felder, and Shoikhet \cite{FFS} as a consequence of Shoikhet's proof \cite{Shoikhet:2000gw} of Tsygan's formality conjecture.  In order to write  it down we  need a couple of definitions.

First, we identify the elements of the Weyl algebra $A_n$ with polynomials $a(q,p)$ in (commuting) variables $q^i$ and $p_j$ endowed with the Weyl--Moyal product 
\begin{equation}
 a\star b =m \exp \alpha (a\otimes b)\,,     
\end{equation}
where 
\begin{equation}
    \alpha=\frac12\left(\frac{\partial}{\partial p_i}\otimes \frac{\partial}{\partial q^i}-\frac{\partial}{\partial q^i}\otimes \frac{\partial}{\partial p_i}\right)\in \mathrm{End} (A_n\otimes A_n)
\end{equation}
and $m(a\otimes b)=ab$. Next, we introduce the maps 
\begin{equation}
\begin{array}{ccc}
    \alpha_{ij}(a_0\otimes \cdots\otimes a_m) &\displaystyle =\frac12 \Big(a_0\otimes \cdots \otimes \frac{\partial a_i}{\partial p_l}\otimes \cdots\otimes \frac{\partial a_j}{\partial q^l}\otimes \cdots \otimes  a_m&\\[3mm]
    &\displaystyle -a_0\otimes \cdots \otimes \frac{\partial a_i}{\partial q^l}\otimes \cdots\otimes \frac{\partial a_j}{\partial p_l}\otimes \cdots \otimes  a_m\Big)&\in \mathrm{End}(A_n^{\otimes(m+1)})
    \end{array}
\end{equation}
and 
\begin{equation}
    \pi_{2n}(a_0\otimes \cdots \otimes a_{2n})=\sum_{\sigma \in S_{2n}}(-1)^{|\sigma|} a_0\otimes \frac{\partial a_1}{\partial y^{\sigma (1)}}\otimes \frac{\partial a_2}{\partial y^{\sigma (2)}}\otimes \frac{\partial a_{2n}}{\partial y^{\sigma (2n)}}\in \mathrm{End}(A^{\otimes (2n+1)}_n)\,,
\end{equation}
where $y^{2i}=q^i$ and $y^{2i-1}=p_i$, $1\leq i\leq n$. Finally, we define the homomorphisms $\mu_{m}: A_n^{\otimes (m+1)}\rightarrow k$ by
\begin{equation}
    \mu_{m}(a_0\otimes \cdots\otimes a_k)=a_0(0)a_1(0)\cdots a_m(0)\,.
\end{equation}
Here $a(0)$ is the constant term of the polynomial $a(q,p)$. Notice that all the operators introduced are $Sp_{2n}(k)$-invariant. 

The formula for the Feigin--Felder--Shoikhet  cocycle now reads 
\begin{equation}\label{FFS}
    \tau_{2n}(a_0,\ldots,a_{2n})=\mu_{2n}\int_{\Delta_{2n}} du_1\wedge \cdots\wedge du_{2n} \prod_{0\leq i<j\leq 2n} e^{(2u_i-2u_j)\alpha_{ij}} \pi_{2n}(a_0\otimes \cdots\otimes a_{2n})\,.
\end{equation}
Here the integral   is taken over the standard $2n$-simplex: $0=u_0<u_1<\cdots<u_{2n}<1$. The exponential function is to be expanded in the Taylor series and integrated term by term.  

The integral in the definition of $\tau_{2n}$ can be done explicitly. When $n=1$ we have $\pi_2=\alpha_{12}$ and $\tau_2=\mu_2\circ \phi_{FFS}$ where the symbol of the operator $\phi_{FFS}(\alpha_{01},\alpha_{12},\alpha_{02})$ is given by (\ref{FFS-symb}).
As the choice of a representative cocycle is not unique, we are free to impose some additional conditions on it. For example, it follows from the long exact sequence (\ref{ld}) that the natural embedding $I: HC^2(A_1)\rightarrow HH^2(A_1,A_1^\ast)$ is an isomorphism. Therefore, the cocycle $\tau_{2n}$ has to be  cohomologous to a cyclic cocycle $\tau_{2n}^{\mathrm{cyc}}$. An explicit expression for $\tau_{2}^{\mathrm{cyc}}$ is obtained, for example, by mere replacement of the symbol $\phi_{FFS}$ with \eqref{fc}. 
Using the cup product (\ref{ccup}), one can then  obtain similar expressions for all higher Weyl algebras $A_n$. In terms of the so-called $(b,B)$-complex the cyclic cocycles on  Weyl algebras were first constructed in  \cite{PFLAUM20101958}, \cite{WILLWACHER2015277}.

\subsection{\texorpdfstring{Example: the cohomology of $\mathcal{A}=A_1\rtimes \mathbb{Z}_2$ and $\mathcal{A}\otimes \Lambda$}{Example: the cohomology for 3d HSGRA}}

The skew group algebra $\mathcal{A}=A_1\rtimes \mathbb{Z}_2$ is a toy model for and a building block of the higher spin algebra in $4d$ HSGRA. 
Having in mind this application, we specify the ground field $k$ to be $\mathbb{C}$. The group $\mathbb{Z}_2=\{e, \varkappa\}$ acts on $A_1$ by the involution (\ref{k}). Since $2\mu_e=2$ and $2\mu_\varkappa=0$, it follows from Theorem \ref{AFLS2} that all nontrivial groups of Hochschild cohomology are 
\begin{equation}\label{AA02}
   HH^2(\mathcal{A},\mathcal{A})\simeq HH^0(\mathcal{A},\mathcal{A}^\ast)\simeq \mathbb{C}\simeq HH^2(\mathcal{A},\mathcal{A}^\ast)\simeq HH^0(\mathcal{A},\mathcal{A})\,.
\end{equation}
Among other things these isomorphisms tell us that the algebra $\mathcal{A}$ admits a one-parameter deformation and its center is spanned by the unit. 

Since $HH^n(\mathcal{A},\mathcal{A}^\ast)=0$ for $n>2$ it follows from Connes' Periodicity Exact Sequence  that $HC^m(\mathcal{A},\mathcal{A}^\ast)\simeq HC^{m+2}(\mathcal{A},\mathcal{A}^\ast)$ for all $m\geq 2$. In low dimensions, 
 one readily finds from (\ref{ld}) that $HC^0(\mathcal{A})\simeq HH^0(\mathcal{A},\mathcal{A}^\ast)\simeq\mathbb{C}$, $HC^1(\mathcal{A})=0$, and $HC^2(\mathcal{A})\simeq \mathbb{C}^2$. Hence,
$$
HC^{2k-1}(\mathcal{A})=0\,,\qquad 
     HC^0(\mathcal{A})\simeq \mathbb{C}\,,\qquad
HC^{2k}(\mathcal{A})\simeq\mathbb{C}^2\,,\qquad
k=1,2,3,\ldots\,.
$$
We also see that the groups $HC^\bullet(\mathcal{A})$ form a free $HC^\bullet(\mathbb{C})$-module generated (via $S$) by the pair of elements $\phi_0\in HC^0(\mathcal{A})$ and $\phi_2\in HC^2(\mathcal{A})$.  The first one is the trace $\phi_0=\mathrm{Tr}:\mathcal{A}\rightarrow \mathbb{C}$ defined by the projection onto the one-dimensional subspace $\mathbb{C}(e\otimes \varkappa)\subset \mathcal{A}$. We can normalize it by setting $\mathrm{Tr}(e\otimes \varkappa)=1$. 
The second class $\phi_2$ is represented by the cyclic cocycle  with the symbol (\ref{fc}). We refer to $\phi_0$ and $\phi_2$ as {\it primary classes} of cyclic cohomology. 
All the other classes are obtained from these two by successive application of the periodicity operator $S$:
\begin{equation}\label{f'f}
\phi^{^{(1)}}_{2n}=S^n\phi_0\,,\qquad \phi^{^{(2)}}_{2n+2}=S^n\phi_2\,.
\end{equation}
Using the definition  (\ref{u=S}), one can see that 
\begin{equation}
\begin{array}{rcl}
(S^n\mathrm{Tr})(a_0,\ldots, a_{2n})&=&n \mathrm{Tr}(a_0\cdots a_{2n})\,,\\[3mm]
(S\tau^{\mathrm{cyc}}_2)(a_0,a_1,a_2,a_3,a_4)&=&
 \tau^{\mathrm{cyc}}_2(a_0a_1a_2,a_3,a_4)+\tau^{\mathrm{cyc}}_2(a_1a_2a_3,a_4,a_0)\\[3mm]&+&\tau^{\mathrm{cyc}}_2(a_3a_4a_0,a_1,a_2)-\tau^{\mathrm{cyc}}_2(a_0a_1,a_2a_3,a_4)\,,\\[3mm]
 (S^{n}\tau^{\mathrm{cyc}}_2)(a_0,\ldots, a_{n+2})&=&(S^{n-1}\tau^{\mathrm{cyc}}_2)(a_0a_1a_2, a_3,\ldots, a_{n+2})\\[3mm]
&+& \displaystyle \sum_{j=2}^{n+1} \Big[(S^{n-1}\tau^{\mathrm{cyc}}_2)(a_0,\ldots,a_{j-1} a_ja_{j+1},a_{j+2},\ldots, a_{n+2})\\[3mm]
\displaystyle  + \sum_{i=0}^{j-2}(-1)^{j-i+1}(S^{n-1}\tau^{\mathrm{cyc}}_2)&&\hspace{-1.5em}(a_0,\ldots,a_{i} a_{i+1}, a_{i+2},\ldots, a_{j}a_{j+1},a_{j+2},\ldots, a_{n+2})\Big]\,.
 \end{array}
\end{equation}

The $HC^\bullet(\mathbb{C})$-module $HC^\bullet(\mathcal{A})$ being flat, the functor $\mathrm{Tor}^{HC^\bullet(\mathbb{C})}(H^\bullet(\mathcal{A}),-)$ is zero and the K\"unneth sequence (\ref{KS}) yields the isomorphism
\begin{equation}
    HC^\bullet(\mathcal{A})\bigotimes_{HC^\bullet(\mathbb{C})} HC^\bullet(\mathcal{B})\simeq HC^\bullet(\mathcal{A}\otimes \mathcal{B})
\end{equation}
for any algebra $\mathcal{B}$. In case $\mathcal{B}=\Lambda$, this allows us to obtain the cyclic cohomology groups of the tensor product 
$\mathcal{A}\otimes \Lambda$ by the cup product of the primary classes $\phi_0$ and $\phi_2$ with the basis elements (\ref{lpm}) of $HC^\bullet(\Lambda)$. 
Application of the elements  $\lambda^+_{2n}$ reproduces the already known classes (\ref{f'f}) (treated now as elements of $HC^{2n}(\mathcal{A}\otimes \Lambda)$ and $HC^{2n+2}(\mathcal{A}\otimes \Lambda)$, respectively) and the cup product with $\lambda^-_m$ gives two more infinite series of cohomology classes: 
\begin{equation}
    \phi^{^{(3)}}_m=\lambda^-_m\sqcup \phi_0\,,\qquad \phi^{^{(4)}}_{m+2}=\lambda^-_m\sqcup\phi_2\,.
\end{equation}
It is convenient to combine all these classes into a single Table \ref{table1}.

A similar K\"unneth formula 
\begin{equation}
    HH^\bullet(\mathcal{A}\otimes \Lambda,\mathcal{A}\otimes \Lambda )\simeq  HH^\bullet (\mathcal{A},\mathcal{A})\otimes HH^\bullet (\Lambda,\Lambda)
\end{equation}
allows one to compute the Hochschild cohomology groups of $\mathcal{A}\otimes \Lambda$ with coefficients in itself. In view of (\ref{psi-pm}) and (\ref{AA02}), the basis cohomology classes are grouped into the four infinite sequences 
\begin{equation}
    \begin{array}{ll}
        \varphi_n^{^{(1)}} =\psi^+_n\sqcup \varphi_0\,,&\qquad  \varphi_n^{^{(2)}} =\psi^-_n\sqcup \varphi_0\,,\\[3mm]
        \varphi_{n+2}^{^{(3)}} =\psi^+_n\sqcup \varphi_2\,,&\qquad \varphi_{n+2}^{^{(4)}} =\psi^-_n\sqcup \varphi_2\,.
    \end{array}
\end{equation}
where $\varphi_0$ and $\varphi_2$ are basis cohomology classes of  $ HH^0(\mathcal{A},\mathcal{A})$ and $HH^2(\mathcal{A},\mathcal{A})$, respectively.

\begin{table}[h!]
\centering
\begin{tabular}{|c|c|c|c|}
\hline
    $n$ &  $\dim HC^n$ &$\lambda^+$-series &$\lambda^-$-series\tabularnewline
    \hline\hline
0  & 2& $ \lambda^+_0\sqcup \phi_0$ & $\lambda^-_0\sqcup \phi_0$\tabularnewline
\hline
1   & 1&  -- &$\lambda^-_1\sqcup\phi_0 $\tabularnewline
    \hline 
 2   & 4 & $\lambda^+_2\sqcup \phi_0\,, \quad\lambda^+_0\sqcup \phi_2$ & $\lambda^-_2\sqcup\phi_0\,,\quad\lambda^-_0\sqcup\phi_2$ \tabularnewline
    \hline 
  3 &2 &  -- &$\lambda_3^-\sqcup \phi_0\,,\quad\lambda^-_1\sqcup\phi_2$\tabularnewline
    \hline 
   4 & 4& $\lambda^+_4\sqcup\phi_0\,,\quad \lambda^+_2\sqcup\phi_2$ & $\lambda^-_4\sqcup\phi_0\,,\quad \lambda^-_2\sqcup\phi_2$\tabularnewline
    \hline 
  5   &2 &  -- &$\lambda^-_5\sqcup\phi_0\,,\quad \lambda_3^-\sqcup\phi_2$\tabularnewline
    \hline 
6 &4& $\lambda^+_6\sqcup\phi_0\,,\quad \lambda^+_4\sqcup\phi_2$ & $\lambda^-_6\sqcup\phi_0\,,\quad \lambda^-_4\sqcup \phi_2$\tabularnewline
    \hline 
     $\cdots$  & 2, 4 & $\cdots$ &$\cdots$\tabularnewline
   \hline
\end{tabular}
\caption{The basis cohomology classes of $HC^\bullet(\mathcal{A}\otimes \Lambda)$. There are only two primary classes 
$\phi_0\in HC^0(\mathcal{A})$ and $\phi_2\in HC^2(\mathcal{A})$; the classes $\lambda^\pm$'s span $HC^\bullet(\Lambda)$.}\label{table1}
\end{table}

\subsection{The cohomology of (extended) higher spin algebra in four dimensions}\label{EHSA}

By definition, the extended higher spin algebra associated with $4d$ HSGRA is given by the tensor square 
\begin{equation}\label{AA}
\mathfrak{A}=\mathcal{A}\otimes\mathcal{A}=A_2\rtimes (\mathbb{Z}_2\times \mathbb{Z}_2)\,, 
\end{equation}
where $\mathcal{A}=A_1\rtimes \mathbb{Z}_2$ is the complex algebra of the previous subsection. It can also be regarded as the smash product of the second Weyl algebra $A_2$ and the Klein group $\mathbb{Z}_2\times\mathbb{Z}_2$ acting on $A_2$ by symplectic reflections. The associated Lie algebra $L(\mathfrak{A})$ contains a finite-dimensional subalgebra $sp_4(\mathbb{R})$ generated by the homogeneous quadratic polynomials  in $q$'s and $p$'s over reals. The well known isomorphism $sp_4(\mathbb{R})\simeq so(3,2)$ relates this algebra with the group of isometries   of $4d$ anti-de Sitter space, thereby explaining the relevance of the algebra (\ref{AA}) to $4d$ HSGRA. 

By the K\"unneth formula (\ref{KunHH}) for  Hochschild cohomology
$$
HH^n(\mathfrak{A}, \mathfrak{A})=\bigoplus_{q+p=n}HH^q(\mathcal{A},\mathcal{A})\otimes HH^p(\mathcal{A},\mathcal{A})
$$
we find that 
\begin{equation}\label{HHAA}
    HH^0(\mathfrak{A},\mathfrak{A})\simeq \mathbb{C}\,,\qquad HH^2(\mathfrak{A},\mathfrak{A})\simeq \mathbb{C}^2\,,\qquad HH^4(\mathfrak{A},\mathfrak{A})\simeq\mathbb{C}\,,
\end{equation}
and the other groups vanish. 
Then the standard interpretation of the second and third groups of Hochschild cohomology suggests that the algebra $\mathfrak{A}$ admits a two-parameter family of formal  deformations.  We could also arrive at the above isomorphisms by applying Theorem \ref{AFLS2}, which also gives  
\begin{equation}
     HH^0(\mathfrak{A},\mathfrak{A}^\ast)\simeq \mathbb{C}\,,\qquad HH^2(\mathfrak{A},\mathfrak{A}^\ast)\simeq \mathbb{C}^2\,,\qquad HH^4(\mathfrak{A},\mathfrak{A}^\ast)\simeq\mathbb{C}\,.
\end{equation}

Now the cyclic cohomology of $\mathfrak{A}$ can easily be computed from  the Connes exact sequence.  Moving from left to right, we find 
$$
\begin{array}{ccccccccccccccccc}
     0&\rightarrow &HC^0&\rightarrow&HH^0&\rightarrow&0&\rightarrow&HC^1&\rightarrow& HH^1&\rightarrow&HC^0&\rightarrow&HC^2&\rightarrow&HH^2\\
      &            &||  &           &||  &            & &           &||&            &  || &           & || &           &||&             &||\\
      &            &\mathbb{C}&\Leftarrow &\mathbb{C}&&  & &0&\Leftarrow&0&&\mathbb{C}&\Rightarrow &\mathbb{C}^3&\Leftarrow&\mathbb{C}^2\\[5mm]
      &\rightarrow &HC^1&\rightarrow & HC^3&\rightarrow &HH^3&\rightarrow & HC^2&\rightarrow &HC^4&\rightarrow &HH^4&\rightarrow&HC^3&\rightarrow &HC^5\\
      &&||&&||&&||&&||&&||&&||&&||&&||\\
      &&0&\Rightarrow&0&\Leftarrow&0&&\mathbb{C}^3&\Rightarrow&\mathbb{C}^4&\Leftarrow&\mathbb{C}&&0&\Rightarrow&0\\[5mm]
      &\rightarrow&HH^5&\rightarrow &HC^4&\rightarrow&HC^6&\rightarrow &HH^6&\rightarrow&HC^5&\rightarrow&HC^7&\rightarrow&HH^7&\rightarrow&\cdots\\
      &           &||&&||&&||&&||&&||&&||&&||&&\\
      &\Leftarrow&0&&\mathbb{C}^4&\Rightarrow &\mathbb{C}^4&\Leftarrow&0&&0&\Rightarrow&0&\Leftarrow&0&&.
\end{array}
$$
Therefore,
\begin{equation}\label{HHHH}
\begin{array}{lll}
     HC^0(\mathfrak{A})\simeq\mathbb{C}\,,&\qquad HC^2(\mathfrak{A})\simeq \mathbb{C}^3\,, &  \\[5mm]
     HC^{4+2k}(\mathfrak{A})\simeq \mathbb{C}^4\,,&\qquad  HC^{2k+1}(\mathfrak{A})=0\,,&\quad k=0,1,2\ldots \,.
\end{array}
\end{equation}
On the other hand, we know that $HC^\bullet(\mathcal{A})$ is a free $HC^\bullet(\mathbb{C})$-module of rank $2$.  This leads to the isomorphism 
\begin{equation}\label{HCHC}
    HC^\bullet(\mathcal{A})\bigotimes_{HC^\bullet(\mathbb{C})} HC^\bullet(\mathcal{A})\simeq HC^\bullet(\mathfrak{A})
\end{equation}
meaning  that the cyclic cohomology groups on the right are obtained by taking cup products of classes (\ref{f'f}). As the tensor product on the left is over  $HC^\bullet(\mathbb{C})$, we may fix either factor to be a primary cocycle. This yields the four infinite series:
\begin{equation}\label{Phi4}
\begin{array}{ll}
    \Phi^{{}^{(1)}}_{2n}=S^n\phi_0\sqcup \bar \phi_0\,,&\qquad \Phi^{^{(2)}}_{2n+4}=S^n\phi_2\sqcup \bar\phi_2\,,\\[5mm]
    \Phi^{^{(3)}}_{2n+2}=S^n\phi_0\sqcup \bar \phi_2\,,&\qquad \Phi^{^{(4)}}_{2n+2}=S^n\phi_2\sqcup\bar\phi_0\,.
    \end{array}
\end{equation}
Here $\phi_{0,2}$ and $\bar \phi_{0,2}$ denote the primary cohomology classes of the left and right factors in (\ref{HCHC}). 
Combining the classes of these series according to their degree, we re-derive the isomorphisms (\ref{HHHH}) above.

Rels. (\ref{Phi4}) also say that $HC^\bullet(\mathfrak{A})$ is a free $HC^\bullet(\mathbb{C})$-module of rank $4$ generated by the products
$\phi_0\sqcup\bar\phi_0$, $\phi_0\sqcup\bar\phi_2$, $\phi_2\sqcup\bar\phi_0$, and $\phi_2\sqcup\bar\phi_2$. Again, 
this allows one to compute the cyclic cohomology 
of the tensor  product $\mathfrak{A}\otimes \Lambda$ from the K\"unneth isomorphism
\begin{equation}
    HC^\bullet(\mathfrak{A}\otimes \Lambda)\simeq  HC^\bullet(\mathcal{A})\bigotimes_{HC^\bullet(\mathbb{C})} HC^\bullet(\mathcal{A})\bigotimes_{HC^\bullet(\mathbb{C})} HC^\bullet(\Lambda)\,.
\end{equation}
The result of application of the generators (\ref{lpm}) to the primary cohomology classes is presented in Table \ref{table2}.  
The same complex dimensions of the cohomology groups $HC^\bullet(\mathfrak{A}\otimes\Lambda)$ follow directly from Theorem  \ref{HCAL}. 

When constructing the HSGRA equations of motion it is important to know some of the Hochschild cohomology groups 
$HH^\bullet(\mathfrak{A}\otimes \Lambda,\mathfrak{A}\otimes \Lambda )$ that control interaction. These can easily be obtained from  (\ref{HHAA}) by the K\"unneth formula 
\begin{equation}
    HH^\bullet(\mathfrak{A}\otimes \Lambda,\mathfrak{A}\otimes \Lambda )\simeq  HH^\bullet (\mathfrak{A},\mathfrak{A})\otimes HH^\bullet (\Lambda,\Lambda)\,.
\end{equation}
Writing $\varphi_0$ $\varphi_2$, $\bar \varphi_2$, and $\varphi_4$ for the basis elements of (\ref{HHAA}), we can produce eight infinite 
series of cohomology classes that span  $HH^\bullet(\mathfrak{A}\otimes \Lambda,\mathfrak{A}\otimes \Lambda )$ by taking cup product with the classes $\psi_n^\pm$ defined by (\ref{psi-pm}). The result is presented in Table \ref{table3}, where these classes are grouped according to their degrees. Among these classes only the elements of the series $\psi^+_n\sqcup \varphi_2$ and $\psi_n^+\sqcup\bar\varphi_2$ are actually relevant to deformation of  the free equations of motion\footnote{Upon identification these cohomology classes with interaction vertices one should take into account an extra grading in the Lie algebra cohomology (\ref{HA}) coming from the splitting (\ref{glt}). }.

Now we can use the above results on the cyclic and Hochschild cohomology of the algebra $\mathfrak{A}\otimes \Lambda$ 
for  the description of the Lie algebra cohomology of  `big matrices' $gl(\mathcal{A}\otimes \Lambda)$. 
It is these cohomology groups that control the structure of HSGRA interactions and  the corresponding characteristic cohomology. 
By straightforward 
adaptation  of Theorems \ref{A1} and \ref{A2}, we obtain the following two statements. 

\begin{table}[h!]
\centering
\begin{tabular}{|c|c|w{8cm}|}
\hline
    $n$ &  $\dim HH^n$ &$\psi^\pm$-series \tabularnewline
    \hline\hline
0  & 2& $ \psi^\pm_0\sqcup \varphi_0 $ \tabularnewline
\hline
1   & 2& $ \psi^\pm_1\sqcup \varphi_0 $  \tabularnewline
    \hline 
 2   & 6 & $\psi^\pm_2\sqcup \varphi_0\,, \quad\psi^\pm_0\sqcup \varphi_2\,, \quad\psi^\pm_0\sqcup \bar\varphi_2$\tabularnewline
    \hline 
  3 & 6&  $\psi^\pm_3\sqcup \varphi_0\,, \quad \psi^\pm_1\sqcup \varphi_2\,, \quad\psi^\pm_1\sqcup \bar\varphi_2 $ \tabularnewline
    \hline 
   4 & 8& $\psi^\pm_4\sqcup \varphi_0\,, \quad \psi^\pm_2\sqcup \varphi_2\,, \quad\psi^\pm_2\sqcup \bar\varphi_2\,,\quad  \psi^\pm_0\sqcup \varphi_4$\tabularnewline
    \hline 
  5   & 8&  $\psi^\pm_5\sqcup \varphi_0\,, \quad \psi^\pm_3\sqcup \varphi_2\,, \quad\psi^\pm_3\sqcup \bar\varphi_2\,,\quad  \psi^\pm_1\sqcup \varphi_4$ \tabularnewline
    \hline 
     $\cdots$  & 8 & $\cdots$ \tabularnewline
   \hline
\end{tabular}
\caption{The basis cohomology classes of $HH^\bullet(\mathfrak{A}\otimes \Lambda, \mathfrak{A}\otimes \Lambda )$. }\label{table3}
\end{table}

\begin{proposition}\label{P1}
The Lie algebra cohomology $H^\bullet \big(gl(\mathfrak{A}\otimes \Lambda)\big)$ is an exterior algebra of the graded vector space spanned by the classes of Table \ref{table2}. 
\end{proposition}

\begin{proposition}
The Lie algebra cohomology $H^\bullet \big(gl(\mathfrak{A}\otimes \Lambda), gl(\mathfrak{A}\otimes \Lambda)\big)$ is a free $H^\bullet \big(gl(\mathfrak{A}\otimes \Lambda)\big)$-module generated by the elements of Table  \ref{table3}. 
\end{proposition}

We conclude with some remarks on the relation of the extended higher spin algebra $\mathfrak{A}$ to the algebra $\mathfrak{A}_0=A_2^e\rtimes \mathbb{Z}_2$ underlying $4d$ HSGRA, see Sec. \ref{4dim}. Let us consider first the algebra $$A_2\rtimes \mathbb{Z}_2=A_1\otimes (A_1\rtimes \mathbb{Z}_2)\,,$$ where  $\mathbb{Z}_2$ acts on $A_1$ by the symplectic reflection (\ref{k}). This action obviously commutes with the full reflection in $A_2$ defined by the same formula (\ref{k}). The algebra $\mathfrak{A}_0$ is now identified with the subalgebra $(A_2\rtimes \mathbb{Z}_2)^{\mathbb{Z}_2}\subset A_2\rtimes \mathbb{Z}_2 $ of elements that are invariant under the full reflection. It is generated by the even polynomials in $q$'s and $p$'s. 
Since both the algebras $A_2\rtimes \mathbb{Z}_2$ and $(A_2\rtimes \mathbb{Z}_2)^{\mathbb{Z}_2}$ are simple and the full reflection is an outer automorphism, the algebra  $\mathfrak{A}_0=(A_2\rtimes \mathbb{Z}_2)^{\mathbb{Z}_2}$ is Morita equivalent to the skew group algebra $(A_2\rtimes \mathbb{Z}_2)\rtimes \mathbb{Z}_2$, see \cite[Thm. 2.5]{S.Montgomery}. On the other hand,
\begin{equation}
    (A_2\rtimes \mathbb{Z}_2)\rtimes \mathbb{Z}_2\simeq A_2\rtimes (\mathbb{Z}_2\times \mathbb{Z}_2)=\mathfrak{A}\,,
\end{equation}
as the actions of $\mathbb{Z}_2$ groups commute to each other. This means the Morita equivalence $\mathfrak{A}_0\sim \mathfrak{A}$ and the isomorphisms of the corresponding cyclic, Hochschild, and Lie cohomology groups:
\begin{equation}
\begin{array}{c}
    CH^\bullet(\mathfrak{A})\simeq CH^\bullet(\mathfrak{A}_0)\,,\qquad  HH^\bullet(\mathfrak{A}, \mathfrak{A}^\ast)\simeq HH^\bullet(\mathfrak{A}_0,\mathfrak{A}_0^\ast)\,,\\[3mm]
    H^\bullet\big(gl(\mathfrak{A}\otimes\Lambda) \big)\simeq H^\bullet\big(gl(\mathfrak{A}_0\otimes\Lambda)\big)\,.
    \end{array}
\end{equation}
As a consequence, all our computations collected in Table \ref{table2} and summarized in Proposition \ref{P1} hold true for the algebra $\mathfrak{A}_0$.

\begin{table}[h!]
\centering
\begin{tabular}{|c|c|w{5cm}|w{5cm}|}
\hline
    $n$ &  $\dim HC^n$ &$\lambda^+$-series &$\lambda^-$-series\tabularnewline
    \hline\hline
0  & 2& $ \lambda^+_0\sqcup \phi_0\sqcup \bar\phi_0$ & $\lambda^-_0\sqcup \phi_0\sqcup\bar\phi_0$\tabularnewline
\hline
1   & 1&   &$\lambda^-_1\sqcup\phi_0 \sqcup\bar\phi_0$\tabularnewline
    \hline 
 2   & 6&  $\lambda^+_0\sqcup \phi_2\sqcup\bar\phi_0$, 
 $ \lambda^+_0\sqcup \phi_0\sqcup\bar \phi_2$,
 
 $\lambda^+_2\sqcup \phi_0\sqcup\bar\phi_0$
  & 
$\lambda^-_0\sqcup\phi_2\sqcup \bar\phi_0$,
$ \lambda^-_0\sqcup\phi_0\sqcup\bar\phi_2$, 
 
  $\lambda^-_2\sqcup\phi_0\sqcup\bar\phi_0$
 \tabularnewline
    \hline 
  3 &3 &   &
  
  $\lambda^-_1\sqcup\phi_2\sqcup\bar\phi_0$, $ \lambda^-_1\sqcup\phi_0\sqcup\bar\phi_2$,

  $\lambda_3^-\sqcup \phi_0\sqcup\bar\phi_0$
  \tabularnewline
    \hline 
   4 & 8& $\lambda^+_4\sqcup\phi_0\sqcup\bar\phi_0$,
    $ \lambda^+_2\sqcup\phi_2\sqcup\bar\phi_0$, 
   
   $\lambda^+_2\sqcup\phi_0\sqcup\bar\phi_2$,
   $\lambda^+_0\sqcup\phi_2\sqcup\bar\phi_2$
   & $\lambda^-_4\sqcup\phi_0\sqcup\bar\phi_0$, $\lambda^-_2\cup\phi_2\sqcup\bar\phi_0$, 
   
   $\lambda^-_2\sqcup\phi_0\sqcup\bar\phi_2$, 
   $\lambda^-_0\sqcup\phi_2\sqcup\bar\phi_2$
   \tabularnewline
    \hline 
  5   &4 &  &  $\lambda^-_5\sqcup\phi_0\sqcup\bar\phi_0$,
  $ \lambda^-_3\sqcup\phi_2\sqcup\bar\phi_0$,
   
   $\lambda^-_3\sqcup\phi_0\sqcup\bar\phi_2$,
   $\lambda^-_1\sqcup\phi_2\sqcup\bar\phi_2$
   \tabularnewline
    \hline 
6 &8& $\lambda^+_6\sqcup\phi_0\sqcup\bar\phi_0$,
$\lambda^+_4\sqcup\phi_2\sqcup\bar\phi_0$, 
   
   $\lambda^+_4\sqcup\phi_0\sqcup\bar\phi_2$,
   $\lambda^+_2\sqcup\phi_2\sqcup\bar\phi_2$&
   $\lambda^-_6\sqcup\phi_0\sqcup\bar\phi_0$,
   $\lambda^-_4\sqcup\phi_2\sqcup\bar\phi_0$, 
   
   $\lambda^-_4\sqcup\phi_0\sqcup\bar\phi_2$,
   $\lambda^-_2\sqcup\phi_2\sqcup\bar\phi_2$
   \tabularnewline
    \hline 
     $\cdots$  & 4, 8 & $\cdots$ &$\cdots$\tabularnewline
   \hline
\end{tabular}
\caption{The cyclic cohomology classes of the algebra $\mathfrak{A}\otimes \Lambda$. As is seen, the dimensions of the groups $HC^n(\mathfrak{A}\otimes\Lambda)$ stabilize starting from degree four.}  \label{table2}
\end{table}

\footnotesize
\providecommand{\href}[2]{#2}\begingroup\raggedright\endgroup

\end{document}